\documentclass[fleqn,usenatbib]{mnras}
\usepackage{newtxtext,newtxmath}
\usepackage[T1]{fontenc}
\usepackage{xcolor}

\DeclareRobustCommand{\VAN}[3]{#2}
\let\VANthebibliography\thebibliography
\def\thebibliography{\DeclareRobustCommand{\VAN}[3]{##3}\VANthebibliography}
\defcitealias{esposito22}{E22}

\usepackage{graphicx}	
\usepackage{amsmath}	
\usepackage{booktabs}   
\usepackage{caption}
\usepackage{subcaption}
\usepackage[flushleft]{threeparttable}
\usepackage{comment}
\usepackage{array}

\DeclareMathOperator\erf{erf}
\DeclareGraphicsExtensions{.jpg,.png}

\title[GMC modelling]{
Modelling molecular clouds and CO excitation 
in AGN-host galaxies}

\author[F. Esposito et al.]{
Federico Esposito$^{1,2}$\thanks{E-mail: federico.esposito7@unibo.it},
Livia Vallini$^{2}$,
Francesca Pozzi$^{1,2}$,
Viviana Casasola$^{3}$,
\newauthor{
Almudena Alonso-Herrero$^{4}$,
Santiago García-Burillo$^{5}$,
Roberto Decarli$^{2}$,
Francesco Calura$^{2}$,
}
\newauthor{
Cristian Vignali$^{1,2}$,
Matilde Mingozzi$^{6}$,
Carlotta Gruppioni$^{2}$,
and Dhrubojyoti Sengupta$^{1,2}$
}
\vadjust{\vspace{3pt}} \\
$^{1}$Dipartimento di Fisica e Astronomia, Università degli Studi di Bologna, Via P. Gobetti 93/2, I-40129 Bologna, Italy\\
$^{2}$Osservatorio di Astrofisica e Scienza dello Spazio (INAF–OAS), Via P. Gobetti 93/3, I-40129 Bologna, Italy\\
$^{3}$INAF – Istituto di Radioastronomia, Via P. Gobetti 101, 40129, Bologna, Italy\\
$^{4}$Centro de Astrobiología (CAB), CSIC-INTA, Camino Bajo del Castillo s/n, E-28692 Villanueva de la Cañada, Madrid, Spain\\
$^{5}$Observatorio de Madrid, OAN-IGN, Alfonso XII, 3, E-28014 Madrid, Spain\\
$^{6}$Space Telescope Science Institute, 3700 San Martin Drive, Baltimore, MD 21218, USA
}

\date{Accepted XXX. Received YYY; in original form ZZZ}

\pubyear{2023}

\begin{document}
\label{firstpage}
\pagerange{\pageref{firstpage}--\pageref{lastpage}}
\maketitle

\begin{abstract}
We present a new physically-motivated model 
for estimating the molecular line
emission in active galaxies.
The model takes into account 
(\textit{i}) the internal density
structure of giant molecular clouds (GMCs),
(\textit{ii}) the heating associated both to 
stars and to the active galactic nuclei (AGN), 
respectively producing photodissociation regions
(PDRs) and X-ray dominated regions (XDRs)
within the GMCs, and (\textit{iii}) 
the mass distribution of GMCs
within the galaxy volume.
The model needs, as input parameters,
the radial profiles of molecular mass, 
far-UV flux and X-ray flux
for a given galaxy,
and it has two free parameters:
the CO-to-H$_2$ conversion factor
$\alpha_{\text{CO}}$, and the
X-ray attenuation column density $N_{\text{H}}$.
We test this model on a sample
of 24 local ($z \leq 0.06$)
AGN-host galaxies,
simulating their carbon monoxide
spectral line energy distribution (CO SLED).
We compare the results with 
the available observations and calculate, 
for each galaxy, 
the best ($\alpha_{\text{CO}}$, $N_{\text{H}}$)
with a Markov chain Monte Carlo algorithm,
finding values consistent with 
those present in the literature.
We find a median 
$\alpha_{\text{CO}} = 4.8$
M$_{\odot}$ (K km s$^{-1}$ pc$^{2}$)$^{-1}$
for our sample.
In all the modelled galaxies,
we find the XDR component of the CO SLED
to dominate the CO luminosity 
from $J_{\text{upp}} \geq 4$.
We conclude that, once a detailed
distribution of molecular gas density
is taken into account,
PDR emission at mid-/high-$J$
becomes negligible with respect to XDR.
\end{abstract}



\section{Introduction}
\label{sec:introduction}

The molecular gas in galaxies
is the fuel for star formation (SF) 
and is affected by active galactic nuclei (AGN)
feedback processes
\citep{fabian12, kennicutt12, cattaneo19},
such as accretion
\citep{combes13, izumi16, farrah22}
and/or outflows
(the review by \citealt{veilleux20}, and more recently
\citealt{garciabernete21, ramosalmeida22, alonsoherrero23}).
Most of the molecular gas resides in
Giant Molecular Clouds (GMCs) that are the birthplaces of stars 
and, for this reason, influence the overall chemical and dynamical
evolution of a galaxy 
\citep{hoyle53, tan00, ballesterosparedes20}.

From observational and theoretical works
\citep[see][for an extensive review]{elmegreen04},
it is well established that
GMCs have an internal hierarchical 
structure constituted by clumps
resulting from the supersonic compression 
produced by the turbulence
\citep{mckee07, hennebelle12}.
Furthermore,
turbulence in GMCs provides 
global support against the gravitational collapse
\citep{federrath13}.
Several studies have
collected catalogues of GMCs through the galaxy structure,
finding that their distribution in mass and size 
is well described by a power-law function
\citep{fukui10, heyer15, brunetti21, rosolowsky21}.

The molecular gas in GMCs is mainly observed through
the carbon monoxide (CO) rotational lines. 
The CO(1--0) luminosity is the most used proxy
for the total molecular gas mass in galaxies,
which can be obtained assuming the so-called
CO-to-H$_2$ conversion factor
\citep[][and references therein]{solomon05, bolatto13}.
Moreover, the CO Spectral Line Energy Distribution
(CO SLED), i.e. the luminosity of CO rotational lines
as a function of the rotational quantum number $J$,
is a very powerful diagnostic for the physical
conditions of the molecular interstellar medium
\citep[ISM,][]{narayanan14, kamenetzky18, vallini19, valentino21, wolfire22, farrah22b}.

Two radiation sources
dominate the molecular gas heating in active galaxies: 
OB stars, emitting far-ultraviolet (FUV, $6-13.6$ eV) radiation,
and the AGN, through hard X-rays ($E > 1$ keV) emission.
FUV and X-ray photons create the so-called photo-dissociation regions
\citep[PDRs,][]{hollenbach97, wolfire22}
and X-ray dominated regions
\citep[XDRs,][]{maloney96, wolfire22}, respectively.
Usually, low-$J$ CO lines trace FUV-heated gas
within PDRs, while high-$J$ lines are XDR tracers \citep{wolfire22}.
This is due to the increasing critical density 
and excitation temperatures
of the CO rotational transitions 
as a function of their quantum number $J$
($n_{\text{crit}} \propto (J+1)^3$)
and to the fact that X-rays can penetrate
at larger column densities with
respect to FUV photons
\citep{maloney96, meijerink05}, thus keeping the dense gas warmer.

Many works in literature exploit the 
observed CO SLED of active galaxies 
\citep[e.g.][]{vanderwerf10, pozzi17, mingozzi18, esposito22} 
to infer the global molecular gas properties 
(e.g. density, temperature) and the heating mechanism 
acting on the molecular gas 
(FUV from star formation, and/or X-ray from AGN).
This is done by searching for the best-fit PDR and XDR models
reproducing the CO SLED. PDR and XDR models are radiative transfer
calculations that compute the line emissivity given the incident
radiation field, the gas density, the metallicity, 
and other free parameters set by the user
\citep[see][for a recent review]{wolfire22}. 
This approach is obviously a simplification, 
because the gas density and the heating mechanism 
inferred from the CO SLED fitting are global average values 
over the whole galaxy. In fact, most of the PDR and XDR models 
do not account for the spatial distribution of GMCs 
in the galaxy and/or for their internal structure.

In \citealt[][]{esposito22}
(hereafter \citetalias{esposito22})
we used alike single-density models
to interpret the CO SLED
of a sample of 35 local AGN-host galaxies,
for which we gathered a wealth
of multiwavelength observations.
We found that PDR models could reproduce
the high-$J$ CO lines only by
using very high density ($n > 10^5$ cm$^{-3}$), 
while XDR models need more moderate 
($n\sim 10^{3}\, \rm cm^{-3}$) densities.

With the aim of improving the characterization 
of the molecular gas properties in AGN-host galaxies 
we build a new, physically-motivated, model
that couples PDR and XDR radiative transfer (RT) calculation with
the internal structure of GMCs, and their observed
distribution within the galaxy disc.
This is done by integrating single-density
single-flux RT predictions, performed with \textsc{Cloudy} 
\citep{ferland17}, into a more complex model 
able to catch the gas distribution properties
of a galaxy, building upon previous analysis by
\cite{vallini17, vallini18, vallini19}. 
We apply the model to the \citetalias{esposito22} sample, 
but the general goal is that of providing a flexible modelling tool
exploitable for any AGN-host source.

This paper is structured as follows:
in Section~\ref{sec:GMC_model} we describe how we model 
(\textit{i}) the internal structure of GMCs, 
(\textit{ii}) the RT within the GMCs,
and (\textit{iii}) the mass distribution of GMCs
within a galaxy. We test the model
on a sample of AGN-host galaxies,
described in Section~\ref{sec:data},
and in Section~\ref{sec:results} we present
the model results. We discuss them in 
Section~\ref{sec:discussion}
and we draw our conclusions 
in Section~\ref{sec:conclusions}.
For all the revelant calculations,
we assume a $\Lambda$CDM cosmology with
$H_0 = 70$ km s$^{-1}$ Mpc$^{-1}$,
$\Omega_m = 0.3$ and $\Omega_{\Lambda} = 0.7$.


\begin{figure}
    \includegraphics[width=\columnwidth]{
    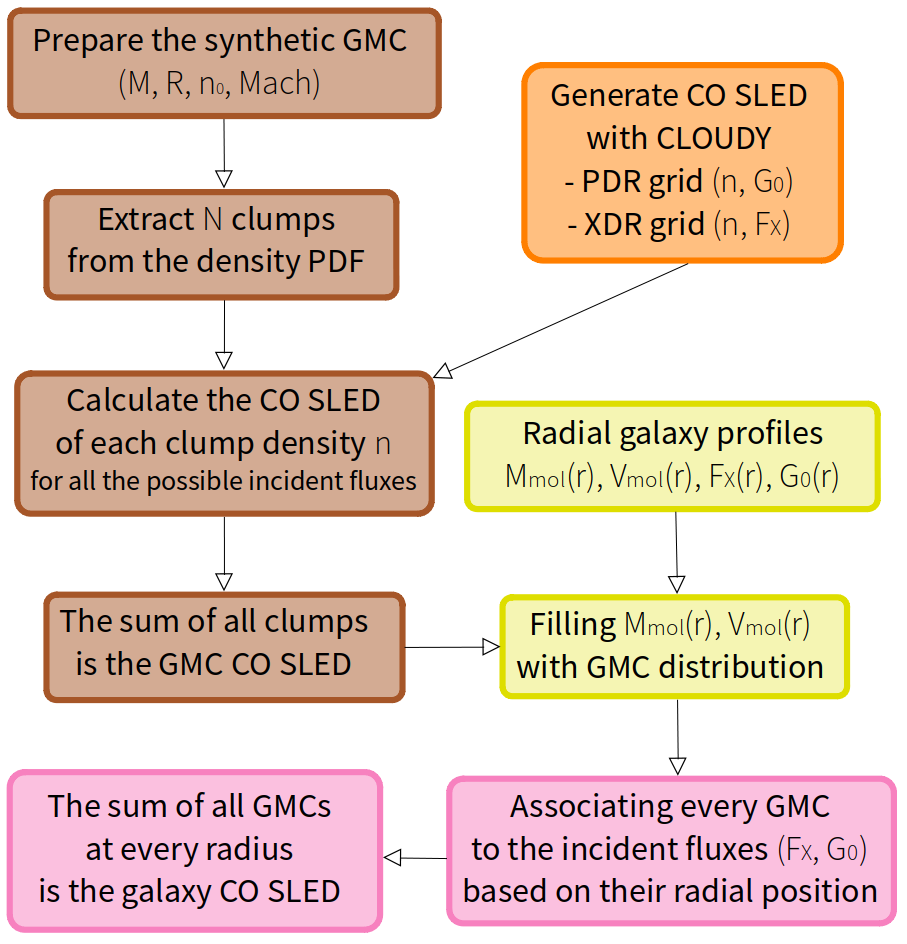}
    \caption{A sketch of the workflow followed
    in this paper. First, we define a synthetic GMC
    with four physical variables 
    ($M, R, n_0, \mathcal{M}$), which define
    the density distribution of its molecular clumps
    (as described in Section~\ref{sec:clumpsdistribution}).
    We combine \textsc{Cloudy} results 
    to generate the CO SLED of each clump
    and of the synthetic GMC
    (see Section~\ref{sec:cosled_clump}).
    Having collected the four radial profiles
    for a real galaxy 
    ($M_{\text{mol}}(r), V_{\text{mol}}(r), 
    G_0(r), F_{\text{X}}(r)$), we filled it with
    the GMC distribution described
    in Equation~(\ref{eq:dNdM}),
    we associate every GMC with the corresponding
    incident fluxes, and eventually we
    have the CO SLED of the studied galaxy.}
    \label{fig:sketch}
\end{figure}


\section{Model outline}
\label{sec:GMC_model}

We build a physically motivated model\footnote{
We publicly release the code, called 
galaxySLED, on GitHub 
(\url{https://federicoesposito.github.io/galaxySLED/})
along with a Jupyter notebook tutorial.
}
that decribes the interaction of the 
radiation from AGN and star formation 
with the molecular gas in galaxies.
This model takes the
\cite{vallini17, vallini18, vallini19} works
- focused on the far-infrared (FIR) and molecular
line emission from GMCs in high-$z$ galaxies - as starting point.
In those works, single GMCs were modelled
as collections of clumps in a
log-normal density distribution,
their CO SLEDs were computed
with radiative transfer calculations,
and a galaxy was filled with a uniform
distribution of identical GMCs.
The aim of the present work is to 
extend such analysis,
including different GMCs in single galaxies
(following a mass distribution),
and illuminating them with a differential
radiative flux (following 
observed radial profiles).

Figure~\ref{fig:sketch} shows a sketch 
that summarizes its modular structure,
which from the sub-pc
scales of clumps within GMCs
(see Sec. \ref{sec:clumpsdistribution})
progressively zooms-out to the kpc scales of 
the gas distribution within galaxies 
(see Sec. \ref{sec:galaxy_model}). 
More precisely, Section~\ref{sec:clumpsdistribution}
deals with the analytical description 
of the internal structure and mass distribution of GMCs,
Section~\ref{sec:cosled_clump} 
outlines the RT modelling
implemented to compute the CO emission, 
whereas in Sec.~\ref{sec:galaxy_model}
we present our assumptions concerning 
the molecular gas distribution on kpc-scales 
and the resulting total CO emission from galaxies.


\subsection{GMC internal structure and mass distribution}
\label{sec:clumpsdistribution}

We characterize GMCs with 
four physical parameters: the total
mass ($M$), the GMC radius ($R$), the
mean density ($\rho_0$) 
and the Mach number ($\mathcal{M}$).
The density structure of GMCs 
\citep[see e.g.][]{hennebelle12}
is well described by a log-normal distribution 
\citep{vazquezsemadeni94, padoan11, federrath11, vallini17}.
The volume-weighted
probability distribution function (PDF) 
of gas density $\rho$, in a supersonically turbulent, 
isothermal cloud of mean density $\rho_0$, is
\begin{equation} \label{eq:PDF}
p_s \, ds = \frac{1}{\sqrt{(2 \pi \sigma_s^2)}}
\exp \left[ -\frac{1}{2} \left( 
\frac{s-s_0}{\sigma_s} \right)^2 \right] \, ds \; \; ,
\end{equation}
where $ s = \ln (\rho / \rho_0) $
is the logarithmic density, with mean value
$ s_0 = -\sigma_s^2 / 2 $. 
The standard deviation of the distribution, $\sigma_s$, 
depends on $\mathcal{M}$
and on the $b$ factor, 
which parametrizes the kinetic energy injection mechanism 
(often referred to as forcing) driving the turbulence
\citep{molina12}:
\begin{equation} \label{eq:sigma_s}
\sigma_s^2 = \ln \left( 1 + b^2 \mathcal{M}^2 \right)
\; \; .
\end{equation}
We assume $b = 0.3$,
which is the value for purely solenoidal forcing
\citep{federrath08, molina12}.


\begin{figure}
    \includegraphics[width=\columnwidth]{
    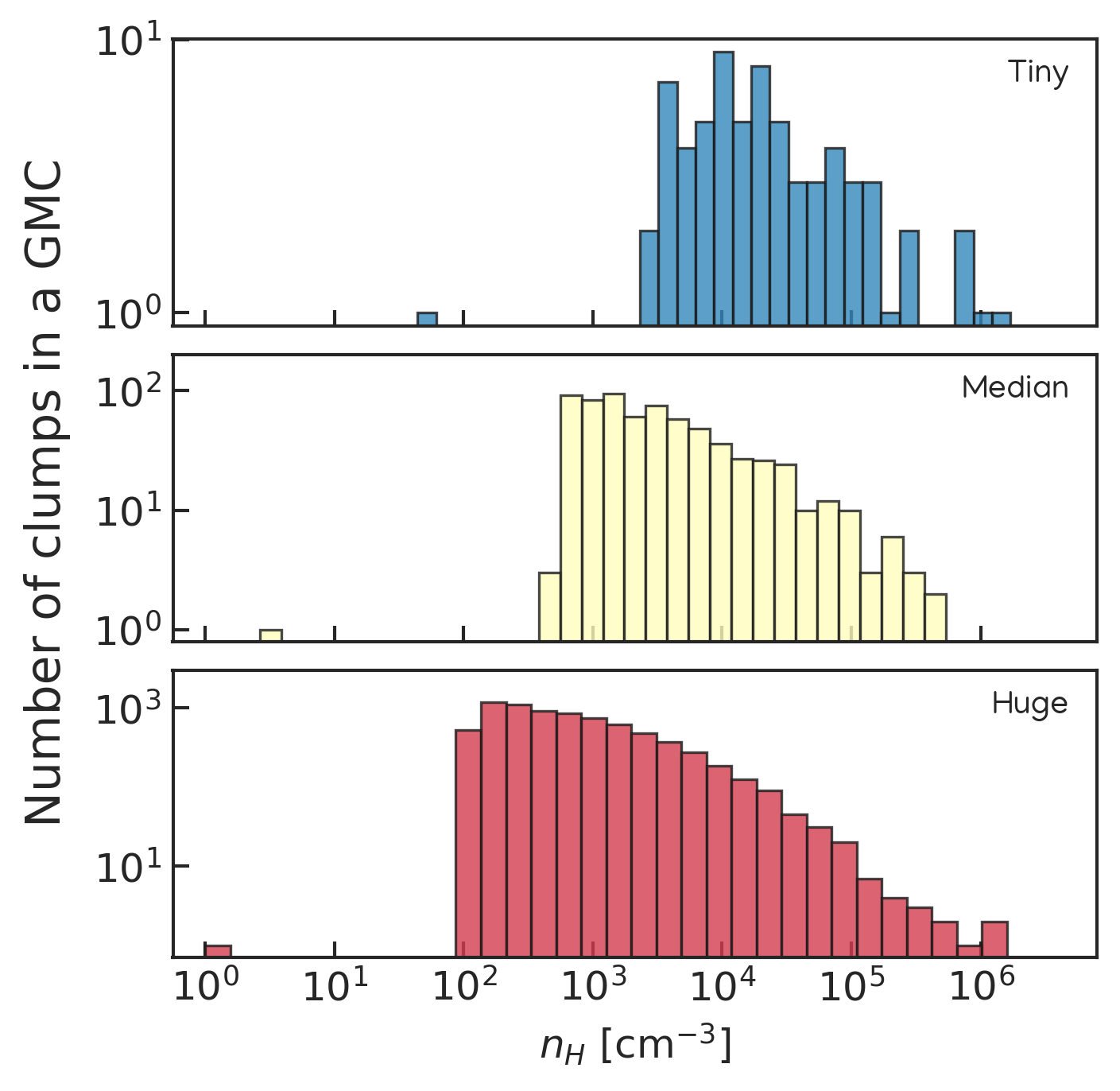}
    \caption{
    Histograms of clumps density distributions
    within the three GMCs 
    listed in Table~\ref{tab:GMCproperties}.
    The shapes are set from Equation~\ref{eq:PDF},
    and mainly depend on each GMC
    mean number density $n_0$.
    }
    \label{fig:GMC_clumps}
\end{figure}

\begin{table}
    \begin{center}
    \begin{tabular}{ccccc}
    \textbf{Name} & \textbf{M} [M$_{\odot}$] 
    & \textbf{R} [pc]
    & \textbf{n\textsubscript{0}} [cm$^{-3}$]
    & \textbf{N$_{\mbox{clumps}}$} \\
    \hline
    Tiny & $1.3 \times 10^3$ & $1.5$ & $2755.6$ & $69$ \\
    Median & $3.2 \times 10^4$ & $7.7$ & $549.8$ & $674$ \\
    Huge & $7.9 \times 10^5$ & $38.6$ & $109.7$ & $7644$ \\
    \end{tabular}
    \caption{Main properties of the 
    smallest, median and largest
    GMCs in the mass distribution described
    in Section~\ref{sec:clumpsdistribution}.
    The columns are the mass, radius,
    mean number density and number of 
    extracted clumps.
    }
    \label{tab:GMCproperties}
    \end{center}
\end{table}

\begin{figure*}
    \includegraphics[width=\textwidth]{
    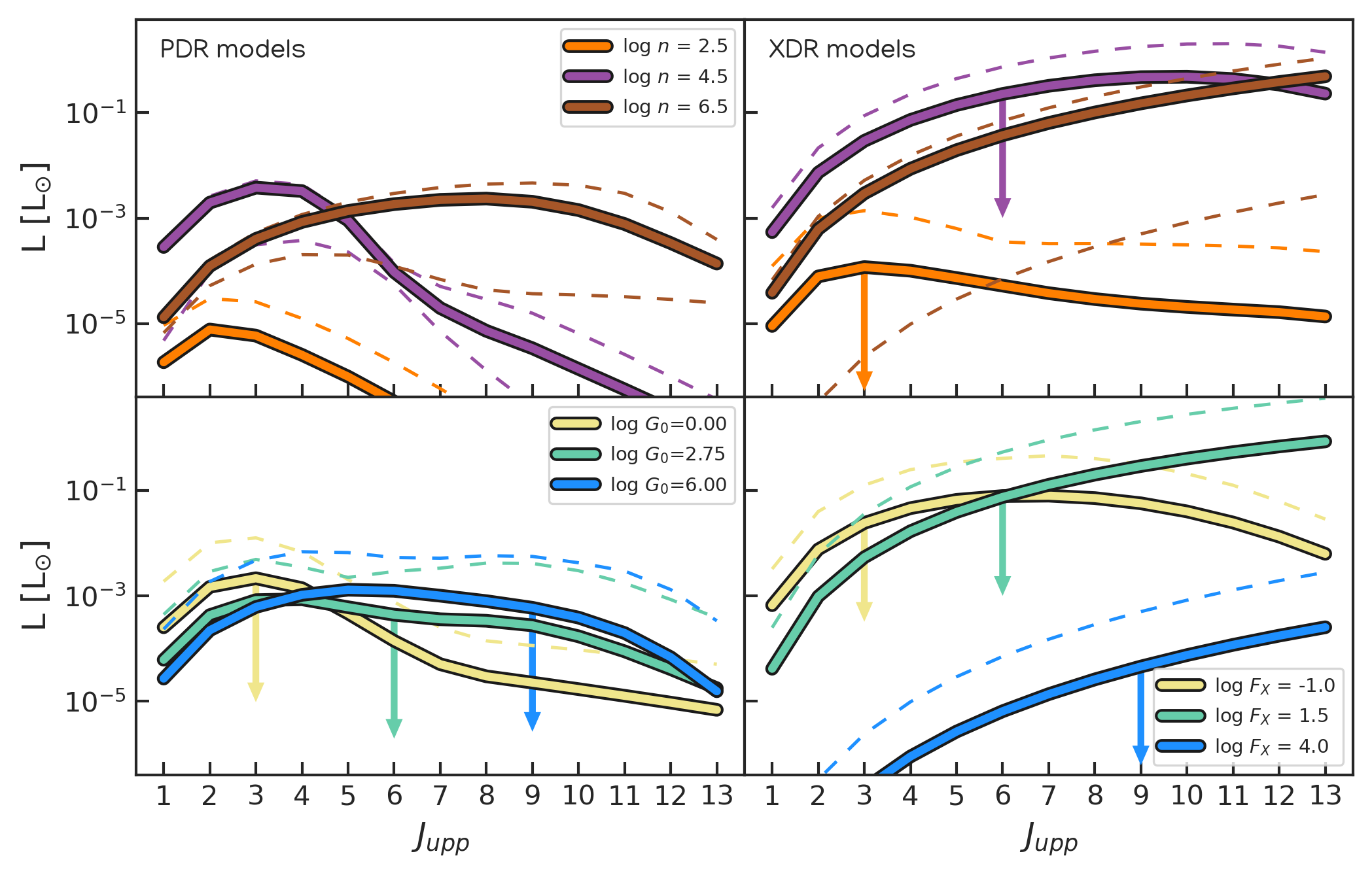}
    \caption{Synthetic CO SLEDs
    from CO($1-0$) to CO($13-12$)
    for spherical molecular clumps of radius 
    $R = R_J$ (Equation~\ref{eq:jeans}).
    The clump fluxes have been converted
    to solar units.
    Left and right panels are PDR and XDR models, respectively.
    In the upper panels the orange, purple, and brown lines represent
    the mean CO SLED for $\log n = 2.5,\,3.5,\,4.5 \, \rm{cm^{-3}}$, 
    respectively, with the incident flux
    ($G_0$ for PDRs, $F_{\text{X}}$ for XDRs)
    left free to vary over the ranges $10^0 - 10^6$ and
    $10^{-1} - 10^4$ erg s$^{-1}$ cm$^{-2}$, respectively. 
    In the lower panels the yellow, green, and blue lines represent 
    the median CO SLED for $\log G_0 = 0.0,\,2.75,\,6.0$
    (bottom-left panel) and for 
    $\log F_{\text{X}} = -1.0,\,1.5,\,4.0 \, \rm{erg s^{-1} cm^{-2}}$
    (bottom-right panel), respectively,
    while the volume density is left free to vary over the range
    $10^0 - 10^{6.5} \rm{cm^{-3}}$.
    In all panels the solid lines represent the mean CO SLED,
    while the dashed lines with the same colors show the minimum 
    and maximum CO SLEDs at the given density or incident flux.
    If the minimum luminosity
    of a given SLED is out of the plot, 
    then the median curve is plotted with a downward arrow
    to highlight the presence of very low luminosities.
    }
    \label{fig:clumpSLED}
\end{figure*}


We define a GMC as a spherical collection of clumps,
whose densities are distributed following the PDF in Equation~(\ref{eq:PDF}). 
Accounting for the presence of clumps within the GMCs 
is one of the fundamental features of this work,
as dense clumps emit the bulk of the CO luminosity
\citep[$n_{\text{crit}} > 10^4$
cm$^{-3}$ for $J \geq 2$,][]{carilli13}.
We then randomly extract 
clumps with $\rho \geq \rho_0$ from the PDF
and for each of them we
calculate its Jeans mass, 
$M_J = (4 \pi / 3) R_J^3 \mu m_p n$, 
and the Jeans radius:
\begin{equation} \label{eq:jeans}
R_J(n) = \sqrt{\frac{\pi c_s^2}{G \rho}} \; \; .
\end{equation}
Here $c_s = \sqrt{\gamma k_B T / (\mu m_p)}$
is the sound speed, that depends,
after we set the adiabatic index to $\gamma = 5/3$ and
the mean molecular weight to $\mu = 1.22$,
only on the gas temperature $T$
and the clump number density $n$.
We assume a fixed temperature 
of $T = 10$ K
for all our clumps,
which is the typical value for dense clumps
in molecular clouds
\citep{spitzer78, hughes16, elia21}.
We proceed with the clumps extraction
until we fill the entire GMC mass
(with a $10\%$ tolerance).
Following \citet{vallini17},
we distribute the leftover mass through the whole GMC, 
accounting for what we define
\emph{intraclump} medium (ICM).

We consider 15 different GMC models,
with masses in the range 
$M=10^3-10^6\, \rm M_{\odot}$ (0.2 dex steps),
and a fixed surface density 
$\Sigma = 170 \, \rm M_{\odot}\, pc^{-2}$ \citep{mckee07}.
This translates into GMCs radii in the range $R=1-40\, \rm pc$,
and mean number densities, $n_0 = 3M/(\mu m_p 4 \pi R^3)$, 
in the range $n_0 = 10^2- 3 \times 10^3 \, \rm cm^{-3}$.
The Mach number is set to $\mathcal{M} = 10$
for all the GMCs in this work
\citep{krumholz05}.
In Figure~\ref{fig:GMC_clumps} we plot
the distribution in density of the clumps
within the three GMCs listed
in Table~\ref{tab:GMCproperties}:
different GMCs (shown with different colors)
have a different peak in the clumps
density distribution, since they have
a different mean density $n_0$.
The ICM appears, in Figure~\ref{fig:GMC_clumps},
in single clumps at low-end densities.

The observed GMC mass distribution 
in galaxies is well approximated by a power-law
\citep[see][ for a recent review]{chevance22}.
In particular,
we follow \cite{romanduval10} and \cite{dutkowska22} 
and we set the mass distribution as follows:
\begin{equation} 
\label{eq:dNdM}
\frac{dN}{dM} \propto M^{-1.64}.
\end{equation}
Equation~(\ref{eq:dNdM}) has been derived 
by fitting the observed GMC mass distribution
in the Milky Way with a power-law, 
albeit only in the mass range 
$M=10^5-10^6\, \rm  M_{\odot}$ 
due to observational limits 
that make it hard to sample the distribution 
towards the low-mass end
\citep[][]{romanduval10}.
\citet{dutkowska22} 
extrapolate the relation down to 
$M=10^4\, \rm M_{\odot}$, 
but in this work we extrapolate 
Equation~(\ref{eq:dNdM}) down to $M=10^3 \rm \, M_{\odot}$, 
because we expect the majority of GMCs in galaxies 
to be small and with low masses 
\citep[e.g.][]{fukui10, colombo14, rosolowsky21}.

We verified that the choice of different power-law exponents
for Equation~(\ref{eq:dNdM}), namely 
1.39, 1.89 and 2.5, corresponding to
the lowest and highest values in 
\cite{romanduval10}
and the highest value from 
\cite{chevance22} does not greatly affect our results.
In Table~\ref{tab:GMCproperties}, we list
the two extremes in mass, labelled as 
"Tiny" and "Huge", 
respectively, and the "Median" GMCs, extracted
from the distribution.


\begin{figure*}
    \includegraphics[width=\textwidth]{
    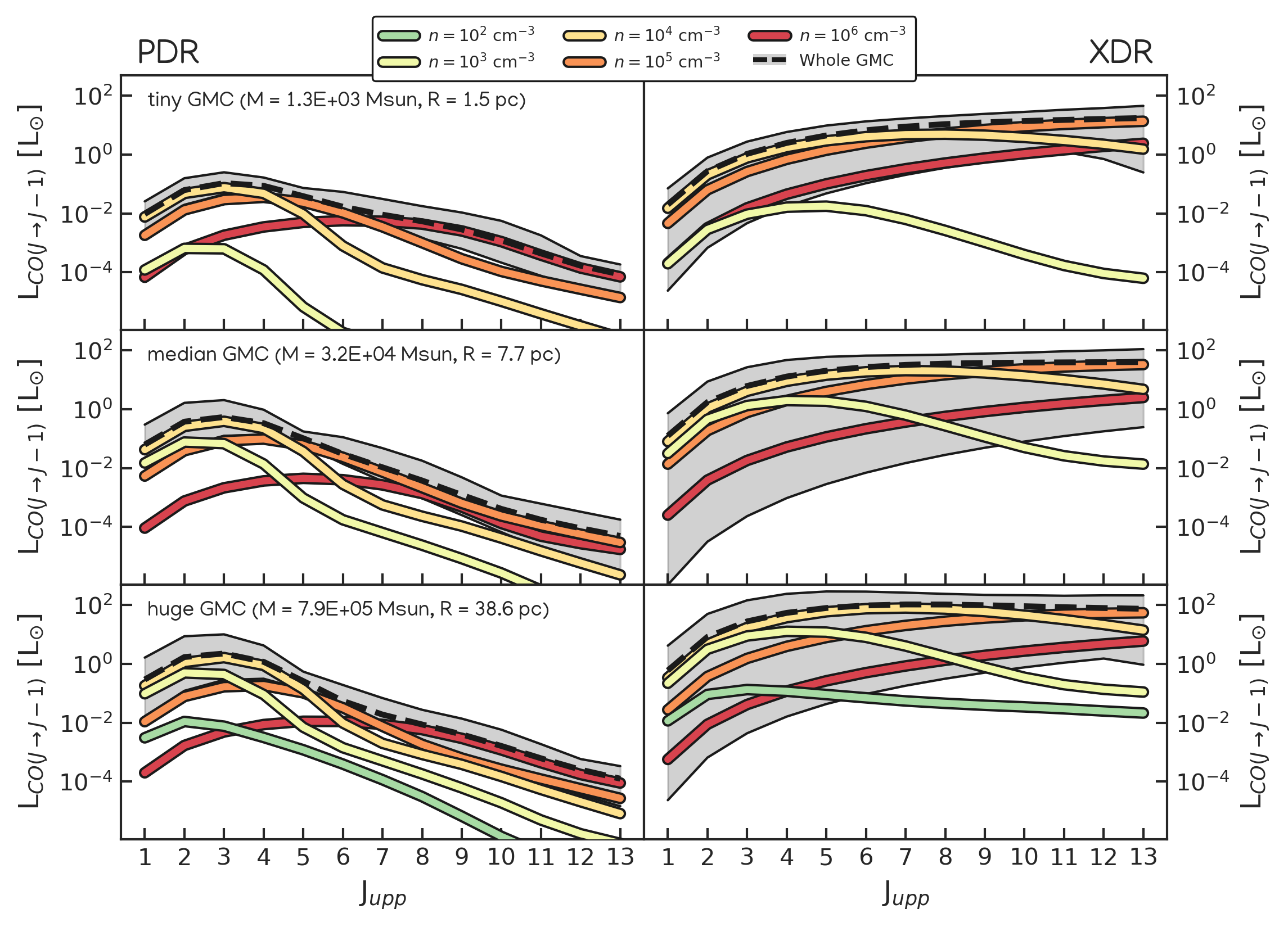}
    \caption{CO SLED of the three GMCs 
    listed in Table~\ref{tab:GMCproperties}.
    The dashed black lines represent the CO SLED
    of the tiny (upper panels), median (middle panels), 
    huge (lower panels) GMC models, 
    while the grey shaded area highlights 
    the variation in the GMC CO SLED 
    produced by changing the incident flux. 
    Left panels are for PDR models
    (i.e. the incident flux range
    is $10^0 \leq G_0 \leq 10^6$),
    right panels are for XDR models
    (i.e. the incident flux range
    is $10^{-1} \leq F_{\text{X}} / (\rm{erg \, s^{-1} \, cm^{-2}}) 
    \leq 10^4$). 
    The colored lines highlight the contribution 
    of the clumps within each GMC,
    as a function of their density $n$.}
    \label{fig:gmcSLED}
\end{figure*}


\subsection{Radiative transfer modelling}
\label{sec:cosled_clump}

We use \textsc{Cloudy} 
\citep[version 17.02, ][]{ferland17}
to model the CO line emission
from a 1-D gas slab with constant density $n$
and irradiated by a stellar or AGN incident flux.
In what follows we label a \textsc{Cloudy}
run as "PDR-model" if the incident far-ultraviolet (FUV) photons flux 
is produced by a stellar population. 
As it is a standard in PDR studies \citep{wolfire22}
the FUV flux ($G_0$) is parametrized in terms of the Habing field 
$1.6 \times 10^{-3}\, \rm erg\, s^{-1}\, cm^{-2}$ \citep{habing68},
which equates to $G_0 = 1$.
Conversely, we label a run as
"XDR-model" if the incident flux
$F_{\text{X}}$ (in the $1-100\rm \, keV$ range)
comes from an AGN.

For PDR models we adopt the
Spectral Energy Distribution (SED) 
of the incident radiation from the 
stellar population synthesis
code \textsc{Starburst99} \citep{leitherer99},
assuming a continuous star formation model 
with age $t=10\, \rm Myr$ and solar metallicity,
and $\log(G_0) = 0 - 6$.
In the XDR models, we use 
$\log[F_{\text{X}} / (\text{erg s}^{-1} \text{ cm}^{-2})]
= -1 - 4$,
and the SED is set
with the \textsc{Cloudy} command \texttt{table xdr}, 
that generates a truncated
AGN X-ray SED, $f_{\nu} \propto h \nu^{-0.7}$, in the 1--100 keV range
\citep[see][for details]{maloney96}.

The second free parameter in the PDR and XDR models
is the gas density $n$.
To cover the density range of clumps and ICM in GMCs
(see Section \ref{sec:clumpsdistribution}), 
we consider a grid of \textsc{Cloudy} runs 
spanning $\log(n/\text{cm}^{-3}) = 0 - 6.5$.
The ranges of $n$, $G_0$ and $F_{\text{X}}$
have logarithmic steps of 0.25 dex.
All our \textsc{Cloudy} runs assume
solar metallicity for the gas, and 
elemental abundances from \cite{grevesse10}. 
Moreover, we account for the Cosmic Microwave Background
($T_{\rm CMB} = 2.7$ K), the Milky Way
cosmic ray ionization rate,
$\zeta = 2 \times 10^{-16}$ s$^{-1}$
\citep{indriolo07}, and we set 
the turbulence velocity $v=1.5\rm \, km\,s^{-1}$
\citep[see][]{pensabene21}.

\textsc{Cloudy} solves the RT through the gas slab 
by dividing it into a large number of thin layers. 
Starting at the illuminated face of the slab,
it computes the cumulative emergent flux at every layer. 
While in this work we are interested in the CO lines 
from CO($1-0$) to CO($13-12$), \textsc{Cloudy} also computes
other molecular and atomic lines
(e.g. HCN, HCO$^+$, [CII]) that we store in our database.
We compute the emergent line emission up to a
total gas column density of
$\log(N_{\text{H}} / \rm cm^{-2}) = 24.5$:
this choice allows us to fully sample
the molecular part of the irradiated gas clouds,
typically located at $N_{\text{H}} > 2 \times 10^{22}$ cm$^{-2}$
\citep{mckee07}.
The CO emission from a GMC is obtained by 
interpolating the 
\textsc{Cloudy} outputs 
at the density $n_i$ and radius $R_i$ 
for each $i$-th clump and then summing up all the clumps luminosities.

\subsubsection{The CO SLED of single clumps and GMCs}

In what follows, we first discuss
the resulting CO SLED from clumps 
of different densities within a given GMC, 
noting that the luminosity of each
$i$-th clump is 
$L = 4 \pi R_i^2 F_{CO}(R_i)$,
where $F_{CO}(R_i)$ is the flux 
computed by \textsc{Cloudy}. 
This is shown in Figure~\ref{fig:clumpSLED}.

As shown in the upper panel of Figure \ref{fig:clumpSLED} 
the CO SLEDs for XDR clump models are overall brighter 
at a given gas density,
with respect to the CO SLEDs of PDR clump models.
Leaving the incident flux free to vary
(upper panels of Figure~\ref{fig:clumpSLED})
has a limited effect in the CO cooling energy output
of PDR models (top-left panel).
Varying the incident flux seems more
important in XDR models (top-right panel), 
since the derived CO SLED luminosities 
can decrease by more than four orders
of magnitudes
(as pointed by the downward arrows
in the top-right panel of
Figure~\ref{fig:clumpSLED}).

Leaving the density free to vary while
fixing the incident flux, however,
could make the CO SLED luminosity
range over several orders of magnitude
for any possible value of incident flux,
both for PDR and XDR models
(bottom panels of Figure~\ref{fig:clumpSLED}).
The very low-luminosity SLEDs
correspond to the lowest densities computed
($\log n/{\rm cm^{-3}} = 0$).
It is interesting to note that
the highest $F_{\text{X}}$ values, 
which correspond to clumps closer to the AGN,
return very faint CO SLEDs:
this is because high X-ray photon fluxes 
lead to the CO dissociation
\citep[see][for a recent review of XDR processes]{wolfire22}.

\begin{figure*}
    \includegraphics[width=\textwidth]{
    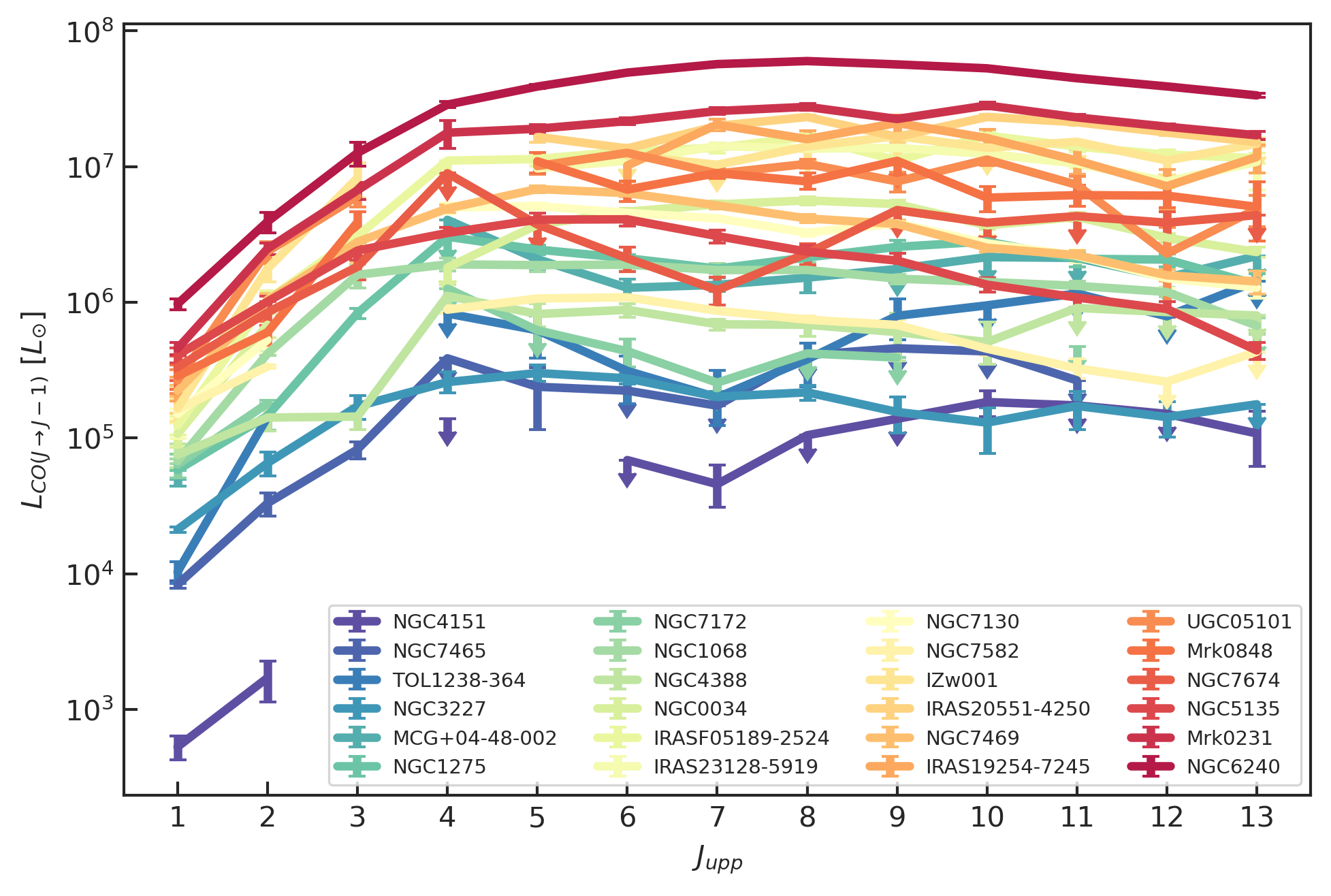}
    \caption{Observed CO SLED of our AGN sample,
    composed of 24 galaxies.
    The order of the objects depends
    on their molecular mass
    within a radius of $2 r_{\text{CO}}$
    (blue is low, red is high,
    with the range being
    [$5 \times 10^7 - 9 \times 10^{10}$]
    M$_{\odot}$ when calculated
    with a Milky-Way $\alpha_{\text{CO}}$).
    Downward arrows are upper limits.}
    \label{fig:sampleSLED}
\end{figure*}

As detailed in Section~\ref{sec:clumpsdistribution}, 
for a given modelled cloud
we have a list of extracted clumps.
Once the contribution from the various clumps 
in the GMC is accounted for, by summing up their 
CO luminosity, the global CO SLED of a GMC 
depends on the incident flux only. 
In Figure~\ref{fig:gmcSLED},
we show the CO SLED for the three GMCs listed in 
Table~\ref{tab:GMCproperties}.
The variation in the GMC CO SLED
introduced by spanning the whole $G_0$ and $F_{\text{X}}$ ranges
is highlighted by a grey shaded area.
We note that varying $G_0$
has a limited effect
on the predicted CO SLEDs
(left~panels~in~Figure~\ref{fig:gmcSLED}),
as opposed to varying $F_{\text{X}}$ (right~panels).

In Figure~\ref{fig:gmcSLED} we plot
the contribution of clumps
of different densities
to the global GMC CO SLED.
PDR models
(left panels of Figure~\ref{fig:gmcSLED})
are dominated by clumps with density
$n = 10^4,  10^5, 10^6\, \rm cm^{-3}$
in the low ($J \leq 4$),
medium ($5 \leq J \leq 7$)
and high-$J$ ($J \geq 8$)
transitions, respectively. 
In the XDR case
(right panels of Figure~\ref{fig:gmcSLED}), 
instead, the very high-density clumps
($n \sim 10^6$ cm$^{-3}$)
never dominate the CO SLED:
their contribution is always at least
one order of magnitude lower than
clumps with $n = 10^4-10^5\, \rm cm
^{-3}$.
However, as can be noticed in 
Figure~\ref{fig:clumpSLED},
the very high-density clumps can 
sustain a high XDR emission only
when a large incident flux
($F_{\text{X}} \sim 10^4$ erg s$^{-1}$ cm$^{-2}$) is present.
This is because at high volume and, thus,
column densities
($N_{\text{H}} = n R_J \propto n^{1/2}$),
the molecular gas is efficiently shielded
from external radiation,
so that only a large incident flux
can penetrate the gas.
The same argument applies to PDRs. 
 

\subsection{Galaxy radial profiles}
\label{sec:galaxy_model}

In this Section we describe how we 
distribute the GMCs
throughout the galaxy volume, and how we associate to each
GMC an incident flux ($G_0$ or $F_{\text{X}}$)
depending on the position within the galaxy.
To do so, we need the radial profiles
of the molecular mass $M_{\text{mol}}(r)$, 
the molecular volume $V_{\text{mol}}(r)$,
the FUV flux $G_0(r)$,
and the X-ray flux $F_{\text{X}}(r)$.
Spatially resolved observations 
of nearby galaxies have shown that 
low-$J$ CO emission is well described by 
an exponential profile along the 
galactocentric radius $r$
\citep{boselli14, casasola17, casasola20},
with scale factor $r_{\text{CO}}=0.17 r_{25}$,
where $r_{25}$ is the radius of the galaxy 
at the isophotal level 25 mag arcsec$^{-2}$ 
in the B-band.
The CO($1-0$) emission can be converted into the
molecular mass by adopting a CO-to-H$_2$
conversion factor $\alpha_{\text{CO}}$
\citep[e.g.][for a review]{bolatto13}.
We adopt the Milky Way value of
$\alpha_{\text{CO}} = 4.3$
M$_{\odot}$ (K km s$^{-1}$ pc$^2$)$^{-1}$
(which includes the helium contribution)
as our initial fiducial value, 
but we will discuss the effect of relaxing this assumption
in Section~\ref{sec:alphaCO}.
We thus obtain the following
cumulative molecular mass radial profile:
\begin{equation} 
\label{eq:M_r}
M_{\text{mol}} (r) = 2.08 \times 10^4
\, \alpha_{\text{CO}} \, L_{\text{CO,tot}}
\, [1 - e^{-r/r_{\text{CO}}} (r/r_{\text{CO}} + 1)]
\end{equation}
where $L_{\text{CO,tot}}$ is 
the CO($1-0$) luminosity of the whole galaxy
in L$_{\odot}$,
$M_{\text{mol}}$ is in M$_{\odot}$.
This molecular gas mass is also confined within
a volume radial profile $V(r)$.
We approximate the galaxy as a disc,
with half-height $z_{\text{CO}} = 0.01 r_{25}$ 
\citep{boselli14,casasola20},
in its external part, while in its inner part
($r<1.5 \, z_{\text{CO}}$)
we use the equation for a sphere
to model a bulge:
\begin{equation} \label{eq:V_r}
V_{\text{mol}} (r) = 
\begin{cases}
(4/3) \pi r^3 & 
\text{for $r \leq 1.5 \, z_{\text{CO}}$} \\
2 \pi z_{\text{CO}} r^2 & 
\text{for $r \geq 1.5 \, z_{\text{CO}}$}
\end{cases}
\end{equation}
The profile $V(r)$ changes at $r=1.5 z_{\text{CO}}$
to ensure a smooth transition between the spherical
and the disc-like regions.
Equations~(\ref{eq:M_r})~$-$~(\ref{eq:V_r})
imply that the molecular gas volume density
$\rho_{\text{mol}}(r)$ increases towards small radii.
The gas surface density $\Sigma(r)$ 
is instead roughly constant in the central part
(i.e. at $r \lesssim r_{\text{CO}}$),
and decreases exponentially as
$\Sigma_{\text{mol}}(r) = 
\Sigma_{\text{mol}}(0) e^{-r/r_{\text{CO}}}$.

In our model the molecular gas --
distributed according to 
Equations~(\ref{eq:M_r})~$-$~(\ref{eq:V_r}) -- is irradiated
by a FUV or X-ray flux that depend 
on the galactocentric radius.
Following \citetalias{esposito22}, we model
$G_0(r)$ with a Sersic function \citep{sersic63}:
\begin{equation} \label{eq:G0_r}
G_0 (r) = G_0 (r_e)
\exp \left\{ -b_n \left[ \left(
\frac{r}{r_e} \right)^{1/n} -1 \right] \right\}
\end{equation}
which is characterized by three parameters:
the effective radius $r_e$, 
the shape parameter $n$ and
the normalization $G_0(r_e)$.
We assume a minimum $G_0=1$ at every radius,
which is equal to the Milky Way 
interstellar radiation field (ISRF).

The X-ray profile, $F_{\text{X}}(r)$,
is derived from the intrinsic
X-ray luminosity $L_{\text{X}}$, 
in the $1-100\, \rm keV$ range.
Given the importance of the 
attenuation of the X-ray flux
due to obscuring gas,
usually measured in terms of
gas column density $N_{\text{H}}(r)$,
we use the following formula 
\citep{maloney96, galliano03}:
\begin{equation} \label{eq:FX_r}
F_{\text{X}}(r) = \frac{L_{\text{X}}}{4 \pi r^2} N_{22}(r)^{-0.9}
\end{equation}
where $N_{22}(r) = N_{\text{H}}(r)/10^{22}\rm \, cm^{-2}$.

From the derived radial profiles 
of the molecular mass and volume, 
we can easily compute the column density 
radial profile as
\begin{equation} \label{eq:NH_r}
N_{\text{H}} (r) = 
\int_0^r n_{\text{H}}(x) \, dx
= \int \frac{M_{\text{mol}}(x)}{\mu m_p V_{\text{mol}}(x)} 
\, dx \; \; \; .
\end{equation}
Whenever the gas column density
$N_{\text{H}}(r) < 10^{22}\rm \, cm^{-2}$,
due to the marginal impact on the 
X-rays attenuation
\citep[e.g.][]{hickox18},
we use the classical definition
$F_{\text{X}}(r) = L_{\text{X}} / (4 \pi r^2)$.
We term this "Baseline model", 
as it is also possible to fit the 
average $N_{\text{H}}$ value 
from the observed CO SLEDs (see Section~\ref{sec:results}).
We will discuss in detail the effect of
$N_{\text{H}}(r)$ in Section~\ref{sec:logNH}.

\section{The dataset}
\label{sec:data}

In \citetalias{esposito22}, we gathered a wealth of observational data
for a sample of 35 local ($z \leq 0.15$)
AGN-host galaxies.
For each source 
we collected: CO line luminosities
from CO($1-0$) to CO($13-12$),
mostly coming from \textit{Herschel} observations;
the optical radius $r_{25}$
(from the 
HyperLeda\footnote{\url{http://leda.univ-lyon1.fr}} 
database);
the total IR luminosity
$L_{\text{IR}}$ ($8-1000$ $\mu$m),
mostly from IRAS observations;
the intrinsic X-ray luminosity $L_{\text{X}}$ 
($2-10\, \rm keV$) and
the obscuring column density
of the X-ray photons, $N_{\text{H, X-ray}}$,
from a variety of X-ray observatories;
and the FIR flux 
(observed by \textit{Herschel})
from which we derived the
FUV field radial profile, $G_0 (r)$
(i.e. the three parameters
of Equation~\ref{eq:G0_r}:
$G_0(r_e)$, $r_e$, and $n$).
Any change in the data with respect
to \citetalias{esposito22} is reported
in Appendix~\ref{sec:update}.

First of all, we select a sub-sample
of the 35 galaxies analyzed in \citetalias{esposito22}.
We adopt the following selection criteria:
\textit{(i)} the availability
of the CO($1-0$) detection,
necessary for the derivation of
the molecular gas mass of a galaxy;
\textit{(ii)} the availability of both
incident flux radial profiles, 
i.e. $G_0(r)$ and $F_{\text{X}}(r)$;
\textit{(iii)} an apparent optical radius
$r_{25} \leq 250 ''$,
since larger objects would have
a $r_{\text{CO}}$ larger
than the CO beams;
\textit{(iv)} at least three CO
detections (i.e. without counting the upper limits)
between CO($1-0$) and CO($13-12$),
to ensure we have enough data points
to make a meaningful comparison
with the model output.
These selection criteria reduce our sample 
to 24 active galaxies.
We list, in Table~\ref{tab:sample_results},
the $r_{\text{CO}}$, the median $G_0(r)$,
the $L_{\text{X}}$ and the $N_{\text{H, X-ray}}$
of each galaxy in the sample.

The parent sample was selected
considering the availability 
of \textit{Herschel} data,
and being their intrinsic X-ray luminosity
$L_{\text{X}} \geq 10^{42}$ erg s$^{-1}$ 
in the $2-10$ keV range.
The sample of 24 objects
considered in this work is
composed by nearby
($z \leq 0.062$)
moderately powerful
($L_{\text{X}} \leq 10^{44}$ erg s$^{-1}$)
AGN-host galaxies, but at the same time
they display a wide variety of physical
properties.
Four galaxies of our sample
(IRAS 19254$-$7245, IRAS 23128$-$5919, Mrk 848, 
and NGC 6240) are clear mergers, while
other six (IRAS F05189$-$2524,
IRAS 20551$-$4250, Mrk 231, NGC 34, NGC 1275, and UGC 5101)
show some signs of interaction as tidal tails.
The rest of the sample has a spiral-like
morphology.
Half of the galaxies (12/24) are
luminous infrared galaxies
(LIRGs: $10^{11} \leq L_{\text{IR}} < 10^{12}$ L$_{\odot}$),
while 5 are ultra-LIRGs 
(ULIRGs: $L_{\text{IR}} \geq 10^{12}$ L$_{\odot}$).
Almost all of the galaxies (22/24)
have their nuclear activity classified
as Seyfert, with the exceptions
of Mrk 848 and IRAS 20551$-$4250
which are classified as 
low-ionization nuclear emission-line regions
(LINERs).
NGC 1275 (also known as 3C 84, Perseus A) 
is the only radio-galaxy of our sample,
and it is also the central dominant galaxy in the
Perseus Cluster.
Two of the furthest galaxies
(Mrk 231 and I Zw 1) are also
commonly classified as quasi-stellar objects
(QSOs).
We refer to \citetalias{esposito22} for further details 
on the galaxy sample and data collection.
The CO SLEDs of all the objects
in our sample
are shown in Figure~\ref{fig:sampleSLED}.


\setlength{\extrarowheight}{1mm}
\begin{table*}
\centering
\caption{Properties and results on the sample of 24 AGN.}
\label{tab:sample_results}
\begin{threeparttable}
\begin{tabular}{lrrrrrrrrrrr}
\toprule
{} &  $r_{\text{CO}}$ &  $\log M_{\text{mol}}$ &  $\log G_0$ &  $\log L_{\text{X}}$ &  $\log N_{\text{H,X}}$ &  $f_{\text{PDR}}^{\text{(1-0)}}$ &  $f_{\text{XDR}}^{\text{(1-0)}}$ &  $f_{\text{PDR}}^{\text{(4-3)}}$ &   $f_{\text{XDR}}^{\text{(4-3)}}$ &  $\alpha_{\text{CO}}$ &  $\log N_{\text{H}}$ \\[-0.5mm]
Name & (kpc)  & (M$_{\odot}$) & {}   & (erg s$^{-1}$) & (cm$^{-2}$) & {} & {} & {} & {} & {} & (cm$^{-2}$) \\
\midrule
I Zw 1             & 2.78 &  10.36 &  1.64 &  43.6 &  $-$   &  0.76 &  0.20 &  0.24 &  0.80 &  $6.6^{+1.7}_{-1.7}$ &  $23.50^{+0.37}_{-0.28}$ \\
IRAS F05189$-$2524 & 2.15 &   9.74 &  3.83 &  43.2 &  22.86 &  0.17 &  0.02 &  0.83 &  0.98 &  $2.1^{+0.4}_{-0.2}$ &  $22.09^{+0.12}_{-0.04}$ \\
IRAS 19254$-$7245  & 3.83 &  10.55 &  3.94 &  42.8 &  23.58 &  0.57 &  0.11 &  0.43 &  0.89 &  $7.3^{+1.5}_{-0.9}$ &  $22.11^{+0.10}_{-0.05}$ \\
IRAS 20551$-$4250  & 2.92 &  10.46 &  4.22 &  42.3 &  23.69 &  0.64 &  0.14 &  0.36 &  0.86 &  $7.2^{+0.9}_{-0.6}$ &  $22.04^{+0.04}_{-0.02}$ \\
IRAS 23128$-$5919  & 4.18 &  10.15 &  3.71 &  43.2 &  $-$   &  0.46 &  0.07 &  0.54 &  0.93 &  $5.5^{+1.0}_{-0.4}$ &  $22.08^{+0.10}_{-0.03}$ \\
MCG +04$-$48$-$002 & 1.45 &   9.85 &  2.54 &  43.1 &  23.86 &  0.81 &  0.27 &  0.19 &  0.73 &  $7.3^{+1.3}_{-1.3}$ &  $23.96^{+0.10}_{-0.12}$ \\
Mrk 231            & 5.99 &  10.95 &  4.77 &  42.5 &  22.85 &  0.86 &  0.34 &  0.14 &  0.66 &  $9.4^{+0.5}_{-0.5}$ &  $22.02^{+0.01}_{-0.02}$ \\
Mrk 848            & 2.62 &  10.26 &  3.72 &  42.3 &  23.93 &  0.66 &  0.15 &  0.34 &  0.85 &  $3.1^{+0.4}_{-0.3}$ &  $22.05^{+0.04}_{-0.03}$ \\
NGC 34             & 2.33 &  10.20 &  2.93 &  42.1 &  23.72 &  0.77 &  0.21 &  0.23 &  0.79 &  $7.1^{+0.5}_{-0.8}$ &  $22.07^{+0.03}_{-0.04}$ \\
NGC 1068           & 2.46 &   9.88 &  3.69 &  42.4 &  24.70 &  0.80 &  0.24 &  0.20 &  0.76 &  $5.4^{+0.9}_{-0.5}$ &  $22.40^{+0.10}_{-0.04}$ \\
NGC 1275           & 3.89 &   8.78 &  3.60 &  44.0 &  21.68 &  0.25 &  0.02 &  0.75 &  0.98 &  $0.5^{+0.1}_{-0.0}$ &  $22.11^{+0.17}_{-0.08}$ \\
NGC 3227           & 1.62 &   8.94 &  3.00 &  42.1 &  20.95 &  0.74 &  0.16 &  0.26 &  0.84 &  $2.0^{+0.2}_{-0.3}$ &  $22.53^{+0.07}_{-0.09}$ \\
NGC 4151           & 1.01 &   7.30 &  2.28 &  42.3 &  22.71 &  0.66 &  0.11 &  0.34 &  0.89 &  $1.8^{+0.7}_{-0.5}$ &  $22.66^{+1.02}_{-0.47}$ \\
NGC 4388           & 4.74 &   9.16 &  2.90 &  42.6 &  23.50 &  0.87 &  0.30 &  0.13 &  0.70 &  $0.9^{+0.1}_{-0.1}$ &  $22.02^{+0.01}_{-0.02}$ \\
NGC 5135           & 3.42 &  10.27 &  3.11 &  42.0 &  24.47 &  0.86 &  0.32 &  0.14 &  0.68 &  $2.2^{+0.2}_{-0.2}$ &  $22.03^{+0.06}_{-0.03}$ \\
NGC 6240           & 5.51 &  10.60 &  4.10 &  43.6 &  24.20 &  0.54 &  0.07 &  0.46 &  0.93 &  $2.0^{+0.1}_{-0.1}$ &  $22.05^{+0.02}_{-0.02}$ \\
NGC 7130           & 2.60 &  10.05 &  3.68 &  42.3 &  24.10 &  0.76 &  0.19 &  0.24 &  0.81 &  $3.6^{+0.5}_{-0.5}$ &  $22.08^{+0.08}_{-0.05}$ \\
NGC 7172           & 2.28 &   9.75 &  2.67 &  42.8 &  22.91 &   $-$ &   $-$ &   $-$ &   $-$ &                  $-$ &                      $-$ \\
NGC 7465           & 0.74 &   8.91 &  2.48 &  42.0 &  21.46 &  0.96 &  0.64 &  0.04 &  0.36 &  $4.7^{+0.7}_{-0.7}$ &  $23.94^{+0.65}_{-0.48}$ \\
NGC 7469           & 2.34 &  10.39 &  3.30 &  43.2 &  20.53 &  0.83 &  0.29 &  0.17 &  0.71 &  $5.3^{+0.5}_{-0.7}$ &  $23.54^{+0.05}_{-0.09}$ \\
NGC 7582           & 3.82 &   9.58 &  3.25 &  42.5 &  24.20 &  0.46 &  0.06 &  0.54 &  0.94 &  $1.1^{+0.2}_{-0.1}$ &  $22.29^{+0.10}_{-0.07}$ \\
NGC 7674           & 3.32 &  10.27 &  4.06 &  43.6 &  $-$   &  0.94 &  0.54 &  0.06 &  0.46 &  $2.7^{+0.4}_{-0.4}$ &  $24.08^{+0.11}_{-0.12}$ \\
TOL 1238-364       & 1.44 &   9.14 &  3.45 &  43.4 &  24.95 &  0.83 &  0.28 &  0.17 &  0.72 &  $6.5^{+1.2}_{-1.1}$ &  $24.09^{+0.14}_{-0.14}$ \\
UGC 5101           & 4.78 &  10.58 &  4.36 &  43.1 &  24.08 &  0.78 &  0.22 &  0.22 &  0.78 &  $6.9^{+1.3}_{-1.2}$ &  $22.51^{+0.10}_{-0.11}$ \\
\bottomrule
\end{tabular}
\begin{tablenotes}
\item \textbf{Notes.}
The molecular masses $M_{\text{mol}}$
are calculated within a $2 \, r_{\text{CO}}$ radius and
by using the best-fit $\alpha_{\text{CO}}$
(except for NGC 7172, for which we used
$\alpha_{\text{CO}} = 4.3$).
$G_0$ is the median value of the $G_0(r)$ profile.
$N_{\text{H,X}}$ is the column density
derived from the X-ray SED.
The best-fit
$f_{\text{PDR}}^{\text{(1-0)}}$,
$f_{\text{XDR}}^{\text{(1-0)}}$,
$f_{\text{PDR}}^{\text{(4-3)}}$, and
$f_{\text{XDR}}^{\text{(4-3)}}$ 
are the fractions to the total
CO($1-0$) and CO($4-3$) luminosities, respectively,
due to PDR and XDR emission, respectively.
Best-fit $\alpha_{\text{CO}}$ is in units of
M$_{\odot}$ (K km s$^{-1}$ pc$^2$)$^{-1}$.
For both $\alpha_{\text{CO}}$ and $\log N_{\text{H}}$,
we report the median values of the marginalized posterior
distributions, with $1\sigma$ width as errors.
\end{tablenotes}
\end{threeparttable}
\end{table*}

\section{Results}
\label{sec:results}

The model takes into account
the radial profiles
defined in Equations~(\ref{eq:M_r})~$-$~(\ref{eq:NH_r})
to spatially distribute the 15 GMCs
described in Section~\ref{sec:clumpsdistribution}
in a real galaxy.
The profiles for all the galaxies 
in the \citetalias{esposito22} sample are presented
in Figure~\ref{fig:diagnostics}.
We divide these profiles
into logarithmically-spaced radial bins.
Based on some tests, we adopt
a 0.05 dex bin size,
starting from $r_{\text{min}} = 10^{-3}$ kpc,
and up to $r_{\text{max}} = 2 r_{\text{CO}}$.
First, since
for every bin we know the molecular mass
and the molecular volume,
we extract random GMCs from the distribution
(Equation~\ref{eq:dNdM}), until we fill up
the entire mass and/or volume of the bin.
Then, we associate each GMC,
in each radial bin,
with the matching incident fluxes
$G_0(r)$, $F_{\text{X}}(r)$,
and finally produce the expected CO SLED of each galaxy.

\subsection{The different models}

The molecular mass of a galaxy
is highly dependent
on the choice of the CO-to-H$_2$
conversion factor $\alpha_{\text{CO}}$, 
while the X-ray flux can
be strongly attenuated by
gas clouds between the AGN and the GMCs
which we parametrize in terms of
a hydrogen column density profile $N_{\text{H}}(r)$.

The model discussed in
Section~\ref{sec:GMC_model} assumes a default 
Milky-Way value of
$\alpha_{\text{CO}} = 4.3$
M$_{\odot}$ (K km s$^{-1}$ pc$^2$)$^{-1}$,
and the intrinsic $N_{\text{H}}(r)$
profile from Equation~(\ref{eq:NH_r}).
We label this "Baseline model".

\begin{figure}
    \includegraphics[width=\columnwidth]{
    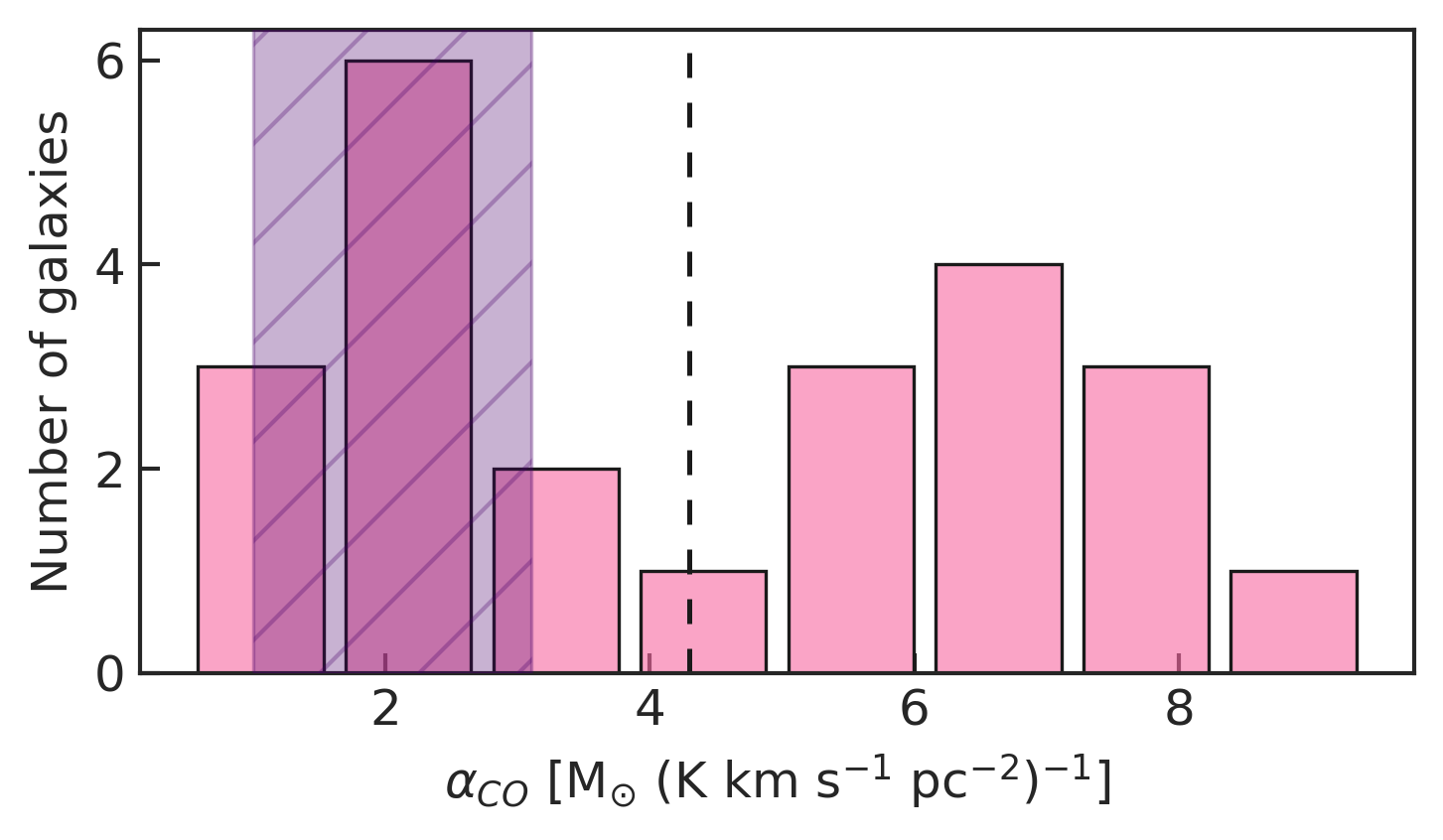}
    \caption{
    Histogram representing the
    best-fit CO-to-H$_2$ conversion
    factor $\alpha_{\text{CO}}$,
    selected through the minimization of
    the $\chi_{\nu}^2$,
    for our sample of 24 galaxies.
    The dashed line is for the Milky Way value
    \citep{bolatto13},
    while the shaded hatched area highlights 
    the values associated to ULIRGs
    \citep{pereztorres21}.}
    \label{fig:alphaCO_hist}
\end{figure}

\begin{figure}
    \includegraphics[width=\columnwidth]{
    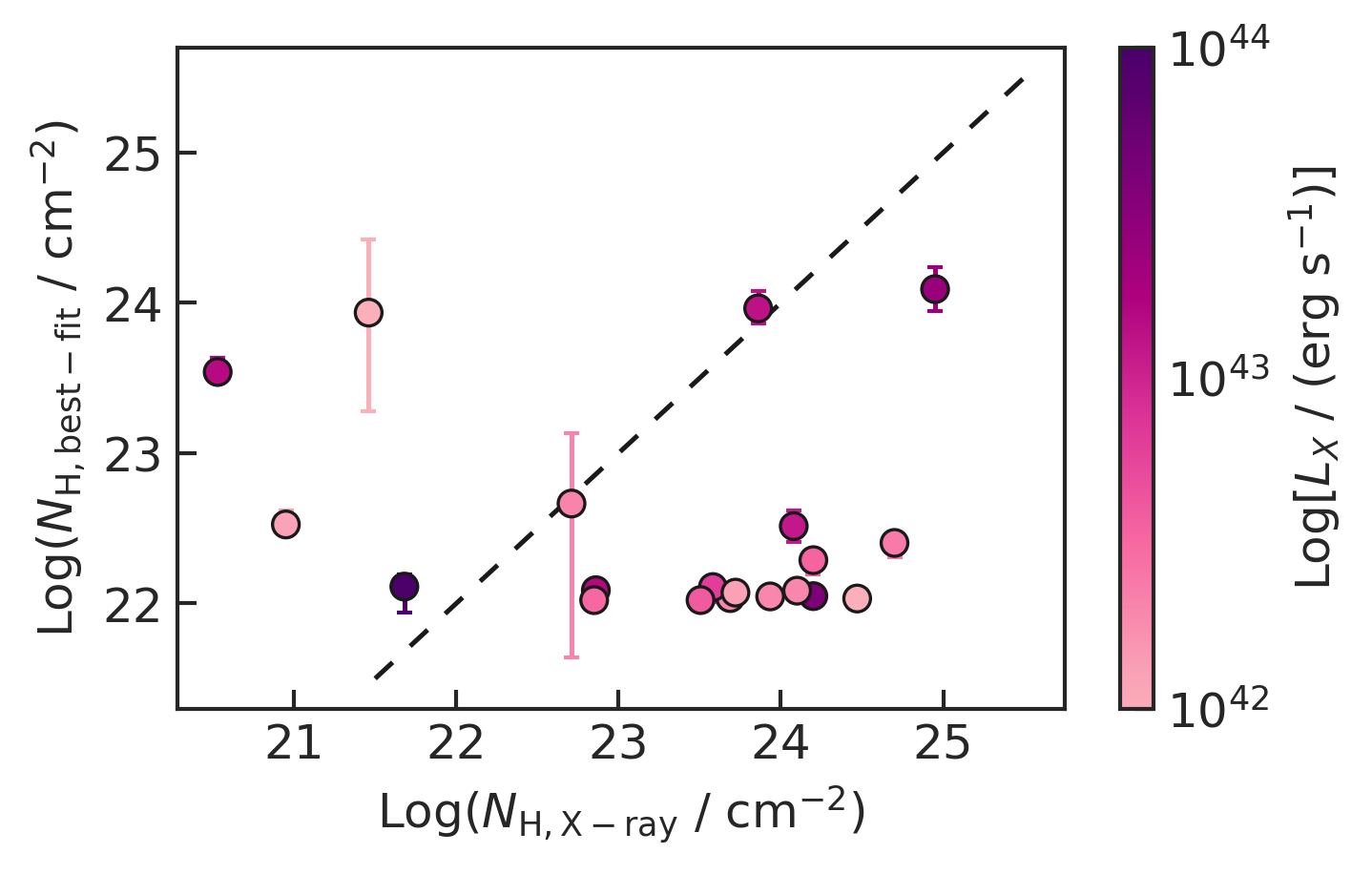}
    \caption{A comparison of the column
    densities $N_{\text{H}}$ from our fitting
    procedure (on the $y$-axis)
    and the $N_{\text{H, X-ray}}$
    derived by modelling the 
    absorption of X-rays (on the $x$-axis).
    The colour of each circle
    (i.e. of each galaxy)
    depends on the intrinsic X-ray
    luminosity $L_{\text{X}}$ ($2-10$ keV).
    The dashed line is the 1:1 line.}
    \label{fig:logNH_scatter}
\end{figure}

For each galaxy we search also for the "Best-fit model"
following the Bayesian Markov chain Monte Carlo
(MCMC) method: we use the $\chi^2$ likelihood function
to fit the observed CO SLED and determine the 
posterior probability distribution of the model
parameters, i.e. $\alpha_{\text{CO}}$ and $N_{\text{H}}$,
with uniform prior distributions
$\alpha_{\text{CO}}=[0.43,\, 43]$ and
$N_{\text{H}} = [10^{22}$, $10^{25}$] cm$^{-2}$.
In this model, $N_{\text{H}}$ has a constant value
at every radius, and it acts as an average
$N_{\text{H}}$ seen by the GMCs.
To run the MCMC algorithm, we use
the open-source Python package \texttt{emcee}
\citep{foremanmackey13},
which implements the Goodman and Weare’s 
Affine Invariant MCMC Ensemble sampler
\citep{goodman10}.
In this way we are able to fully characterize
any degeneracy between our model parameters, while
also providing the 1$\sigma$ spread of the posterior
distribution for each of them.
To be able to include also upper limits
in the likelihood function,
we follow the approach described in \cite{sawicki12}
and \cite{boquien19}, splitting the $\chi^2$ formula in 
two sums:
\begin{equation} \label{eq:chisquare}
\chi^2 = \sum_{j=1}^{13} \left( \frac{f_j - m_j}{\sigma_j} \right)
- 2 \sum_{j=1}^{13} \ln \left\{ \frac{1}{2} \left[ 1 + \erf
\left( \frac{f_{\text{ul},j} - m_j}{\sqrt{2} \sigma_j}
\right) \right] \right\}
\end{equation}
where the first sum contains the detections $f_j$
and their errors $\sigma_j$ ($j$ covers the first 
13 lines of the CO SLED),
the second one containing the $3\sigma$ upper limits
$f_{\text{ul},j}$ (whereas we used the $1\sigma$ upper limit
as the measured error $\sigma_j$),
and in both sums $m_j$ are the model values.

Finally, we produce a third set of synthetic 
CO SLEDs by keeping 
the default $\alpha_{\text{CO}} = 4.3$
M$_{\odot}$ (K km s$^{-1}$ pc$^2$)$^{-1}$
but using the column density $N_{\text{H, X-ray}}$ 
derived from the absorption of X-rays along
the line of sight 
\citep[][and Salvestrini in prep.]{brightman11, ricci17, marchesi19, lacaria19}.
We label these CO SLED predictions
"$N_{\text{H, X-ray}}$ model". 
Also in this case we set $N_{\text{H, X-ray}}$
constant at every radius.

\subsection{Best-fit model results}
\label{sec:BFresults}

We run \texttt{emcee} with
10 walkers exploring the 
parameter space for $10^4$ chain steps.
The chains have been initialized by
distributing the walkers around
$\alpha_{\text{CO}} = 4.3$
M$_{\odot}$ (K km s$^{-1}$ pc$^2$)$^{-1}$
and the median $N_{\text{H}}(r)$
for each galaxy (if larger than $10^{22}$ cm$^{-2}$,
$10^{22}$ cm$^{-2}$ otherwise).
For each walker,
we first run the algorithm with a burn-in chunk
of 5000 steps which we later discard.
The procedure gives a mean autocorrelation length $\tau = 65$
for both $\alpha_{\text{CO}}$ and $N_{\text{H}}$,
and a mean acceptance fraction of 0.59.
In the following we report the
median values of the marginalized posterior
probability distributions for the two parameters,
with a 68\% confidence interval as errors.
The MCMC code did not converge for NGC 7172,
due to the dominance of upper limits
(7, with only 3 detections):
in the subsequent paragraphs and sections, 
we only consider the remaining 23 galaxies of our sample.

The best-fit procedure returns median values
$0.5 \leq \alpha_{\text{CO}} < 9.5$
M$_{\odot}$ (K km s$^{-1}$ pc$^2$)$^{-1}$,
as shown in Figure~\ref{fig:alphaCO_hist}.
The median CO-to-H$_2$ conversion factor
for our galaxy sample is
$\alpha_{\text{CO}} = 4.7^{+2.4}_{-2.8}$,
where the lower and upper errors
are the $16^{\text{th}}$ and 
$84^{\text{th}}$ percentiles
of the sample distribution,
respectively.
This value is comparable to that of the Milky Way,
and the range agrees well with 
the available literature
\citep[e.g.][]{bolatto13, leroy15, 
mashian15, accurso17, seifried17, casasola17, dunne22}.

The best-fit values of $N_{\text{H}}$
are shown in Figure~\ref{fig:logNH_scatter}
against $N_{\text{H, X-ray}}$.
We find values 
$10^{22.0} \leq N_{\text{H}} \leq 10^{24.1}$ cm$^{-2}$
with a median of
$\log (N_{\text{H}} / \rm{cm}^{-2}) = 22.1^{+1.6}_{-0.1}$.
By comparison, the 21/24 galaxies for which we have
$N_{\text{H, X-ray}}$ have a median
$\log (N_{\text{H, X-ray}}/ \rm{cm}^{-2}) = 23.7^{+0.5}_{-1.8}$.

The maximum fitted obscuration of our sample
is in TOL 1238-364
($N_{\text{H}} = 10^{24.1} \text{ cm}^{-2}$), which also has
the maximum $N_{\text{H, X-ray}} = 10^{25}$ cm$^{-2}$.
As shown in Figure~\ref{fig:logNH_scatter}, however,
the $N_{\text{H}}$ values, derived from the fit of the CO SLED
and from the fit of the X-ray spectrum, 
do not correlate. This, anyway, does not 
constitute a critical issue of our modelling, 
as will be explained in Section~\ref{sec:logNH}.

\begin{figure}
    \includegraphics[width=\columnwidth]{
    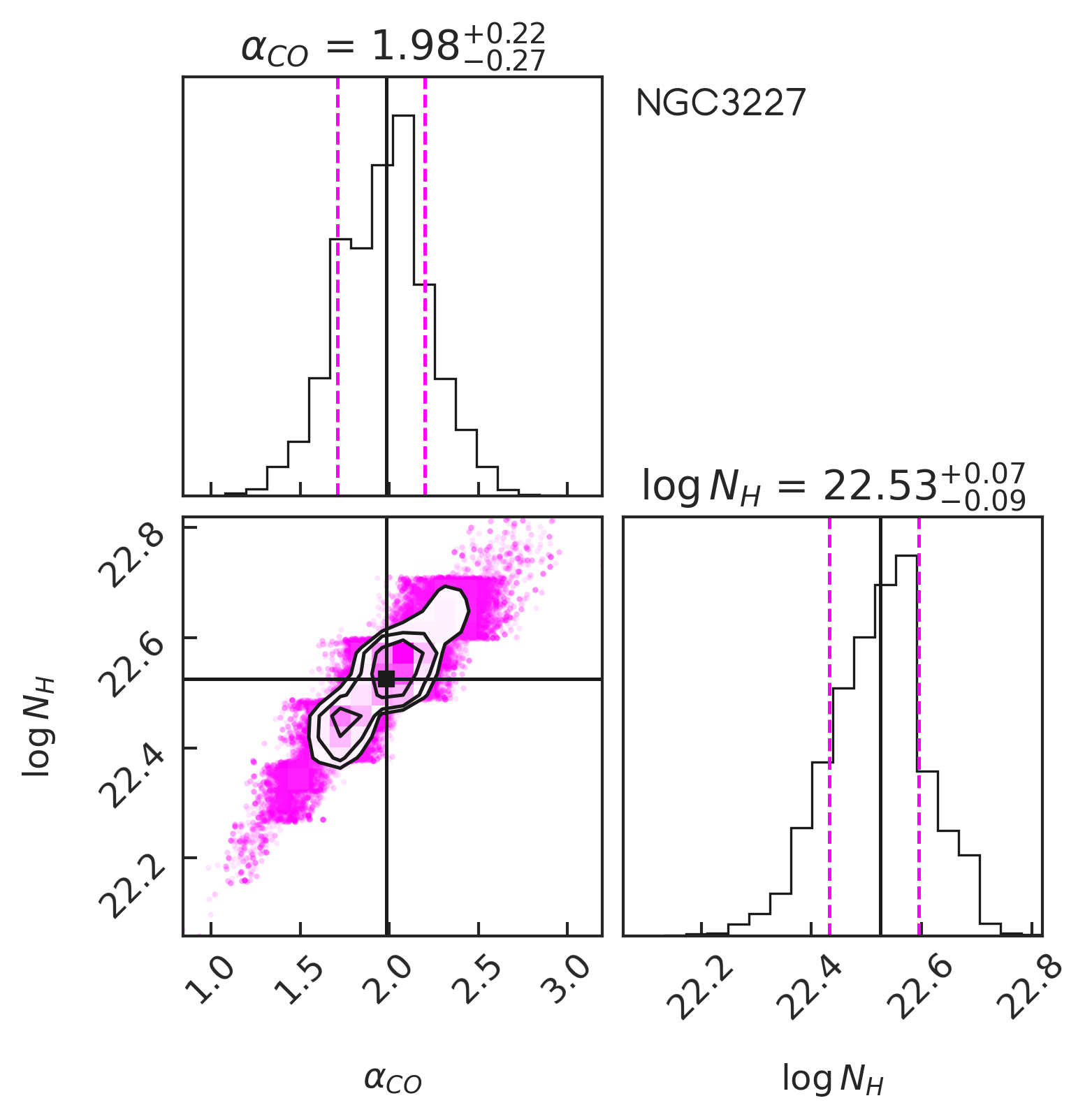}
    \caption{
    Corner plot showing the marginalized posterior
    distributions of $\alpha_{\text{CO}}$
    (in M$_{\odot}$ (K km s$^{-1}$ pc$^2$)$^{-1}$)
    and $N_{\text{H}}$ (in cm$^{-2}$) 
    for the galaxy NGC 3227.
    The contours represent $(1, 1.5, 2) \sigma$
    levels for the 2D distribution.
    The best-fitting parameters and the 
    $16^{\text{th}}$ and $84^{\text{th}}$ percentiles
    are plotted with solid black lines and square and 
    dashed magenta lines, respectively.
    }
    \label{fig:corner}
\end{figure}

We find a degeneracy between the
$\alpha_{\text{CO}}$ and $N_{\text{H}}$
sampled values in almost all the galaxies
(see e.g. Figure~\ref{fig:corner} for NGC 3227).
This is not unexpected, since it is known
that $\alpha_{\text{CO}}$ depends on 
the optical depth \citep{bolatto13, teng23},
which in turn depends on $N_{\text{H}}$.
Increasing $\alpha_{\text{CO}}$ in a galaxy means
more molecular mass within the same volume,
hence $N_{\text{H}}$ has to increase accordingly.
The galaxies for which the sampled parameters
do not show degeneracy are also among the ones
for which our model works worse
(e.g. Mrk 231, NGC 4151, NGC 4388,
see Appendix~\ref{sec:appendix}).

\begin{figure}
    \includegraphics[width=\columnwidth]{
    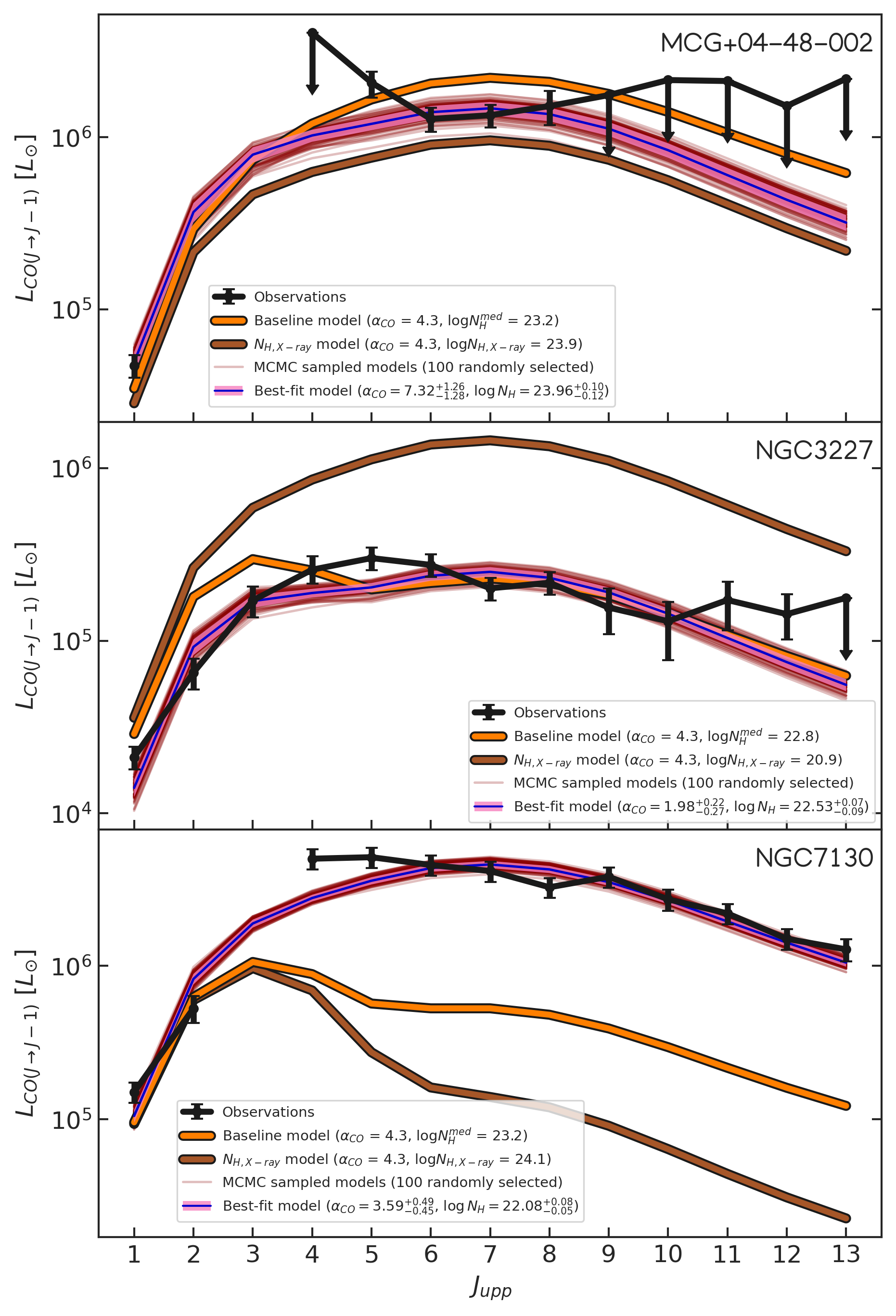}
    \caption{The modelled CO SLEDs of 3 galaxies
    (from the top to the bottom panel:
    MCG+04$-$48$-$002, 
    NGC 3227 and NGC 7130)
    of our sample, in physical units of $L_{\odot}$.
    For each panel, the black line is the observed
    CO SLED (with downward arrows indicating
    censored data points), and the 
    orange, brown, red, and pink lines
    are the modelled CO SLEDs:
    our baseline model without fitting any free parameter 
    and with a negligible $N_{\text{H}}(r)$,
    the baseline model with a constant
    $N_{\text{H,X-ray}}$ derived by modelling 
    the absorption of X-rays,
    100 MCMC modelled SLEDs randomly picked
    from the parameters posterior distributions,
    and the best-fit model covering the $1\sigma$
    spread of such distributions, 
    respectively.
    The $x$-axis represents the upper rotational
    quantum number $J$ of each CO line,
    from CO($1-0$) to CO($13-12$).
    }
    \label{fig:comparison_SLED}
\end{figure}

In Appendix~\ref{sec:appendix},
together with observed and modelled CO SLEDs
for all the galaxies, we also plot the 
relative (i.e. normalized to
the best-fit model)
residuals between observed data
and best-fit model.
We find relative residual values $\geq 2$ 
for at least two detected CO lines in 4 galaxies:
IRAS 20551$-$4250, Mrk 231, NGC 4151, and NGC 4388.
Apart from NGC 4388, which has a very peculiar CO SLED
shape that our model is not able to reproduce,
for the other three galaxies our model fails
to reach such high luminosities for the high-$J$ lines.
This may be due to an additional source of excitation
(other than FUV and X-ray flux), as shocks or
cosmic rays.
In fact, IRAS 20551$-$4250 and Mrk 231 are known
to host powerful CO outflows, with velocities
up to 500 and 700 km s$^{-1}$, respectively
\citep{lutz20}.
NGC 4388 has also a detected CO outflow,
reaching 150 km s$^{-1}$
\citep{dominguezfernandez20},
while for NGC 4151 we only have a detected
H$_2$ outflow travelling at
300 km s$^{-1}$ \citep{may20}.
It is tempting to intepret these results
as evidences of AGN mechanical feedback
on CO excitation, but this is 
beyond the purpose of this work.

The best-fit $\alpha_{\text{CO}}$ and $N_{\text{H}}$,
together with the $1\sigma$ spread of their
posterior distributions,
are listed, for each galaxy,
in Table~\ref{tab:sample_results},
together with the molecular mass
within a $2 r_{\text{CO}}$ radius
(recalculated with the best-fit $\alpha_{\text{CO}}$).

\subsection{Three examples of modelled galaxies}

In Figure~\ref{fig:comparison_SLED} we show
the comparison between the observed CO SLED 
and that resulting from 
the "Baseline", "Best-fit",
and "$N_{\text{H, X-ray}}$" modelling,
where the "Best-fit" model is calculated with
the $\pm 1\sigma$ values of the 
($\alpha_{\text{CO}}$, $\log N_{\text{H}}$) 
posterior distributions.
We show three sample galaxies:
MCG+04$-$48$-$002, NGC 3227 and NGC 7130,
chosen to be representative
(due to the spread of their best-fit parameters)
of the results obtained by 
means of our fitting procedure,
with best-fit median values
$\alpha_{\text{CO}} = [7.32,\, 1.98,\, 3.59]$
M$_{\odot}$ (K km s$^{-1}$ pc$^2$)$^{-1}$,
respectively, and 
$N_{\text{H}} = [10^{23.96},\, 10^{22.53},\, 10^{22.08}]$
cm$^{-2}$, respectively.
In Appendix~\ref{sec:appendix}
we gather the observed and theoretical CO SLEDs 
for the whole galaxy sample.

MCG+04$-$48$-$002
(top panel of Figure~\ref{fig:comparison_SLED})
has very similar "$N_{\text{H, X-ray}}$",
"Baseline" and "Best-fit" CO SLEDs,
due to their similar $N_{\text{H}}$ values.
The effect of a high $N_{\text{H}}$
on the CO SLED is visible especially at high $J$.
The effect of changing $\alpha_{\text{CO}}$
(by a factor of $\sim 80 \%$)
is evident by comparing the 
"$N_{\text{H, X-ray}}$",
and "Best-fit" CO SLEDs.

In NGC 3227
(middle panel of Figure~\ref{fig:comparison_SLED})
we can notice the difference between the
"Baseline" and "Best-fit" CO SLEDs
due to different $\alpha_{\text{CO}}$ and $N_{\text{H}}$:
a higher $\alpha_{\text{CO}}$ would boost
the luminosity of all the CO lines,
but a higher $N_{\text{H}}$ decreases
the high-$J$ lines, exposing a typical
PDR bump in the low-$J$s
(cfr. Figures~\ref{fig:clumpSLED}
and \ref{fig:gmcSLED}).
The "$N_{\text{H, X-ray}}$" SLED is very bright
due to its low $N_{\text{H, X-ray}}$,
which boosts the XDR emission.

NGC 7130
(bottom panel of Figure~\ref{fig:comparison_SLED})
has instead three very different modelled CO SLEDs.
In the "Baseline" and "$N_{\text{H, X-ray}}$"
models we can clearly see the PDR bump
at low-$J$, due to the high $N_{\text{H}}$
which absorbs the X-rays.
The higher $N_{\text{H, X-ray}}$ makes its SLED
to dramatically decrease towards the high-$J$ lines,
where the XDR emission is dominant.
The "Best-fit" CO SLED has instead a lower PDR
component due to its low
$\alpha_{\text{CO}} = 3.59$,
which better reproduces the observed CO SLED.


\begin{figure}
    \includegraphics[width=\columnwidth]{
    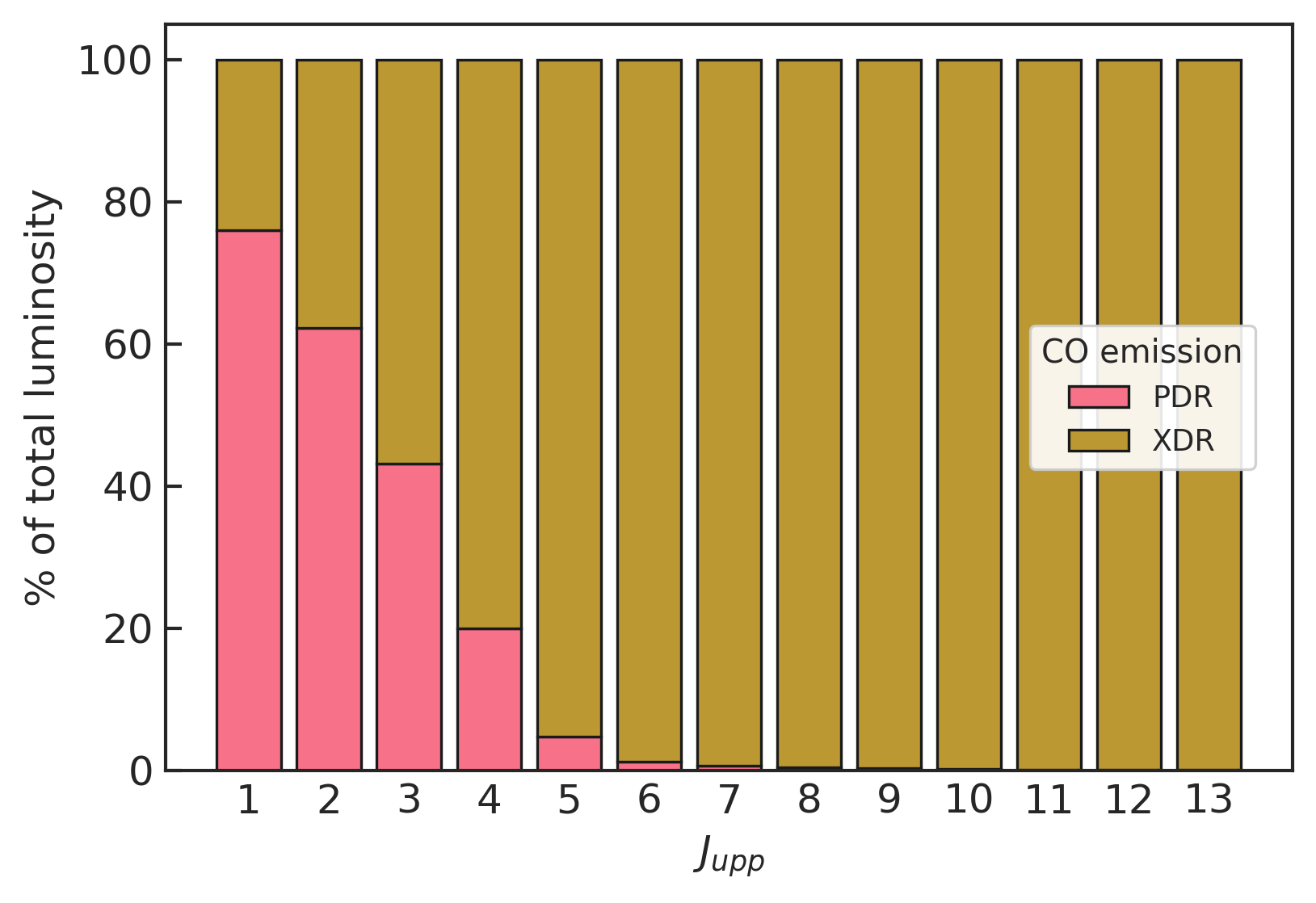}
    \caption{Bar plot showing the relative
    percentage to the total modelled luminosity
    of PDR (pink side) and XDR (brown side)
    emission for each analyzed CO line, 
    from CO($1-0$) to CO($13-12$).
    }
    \label{fig:violin}
\end{figure}

\begin{figure}
    \includegraphics[width=\columnwidth]{
    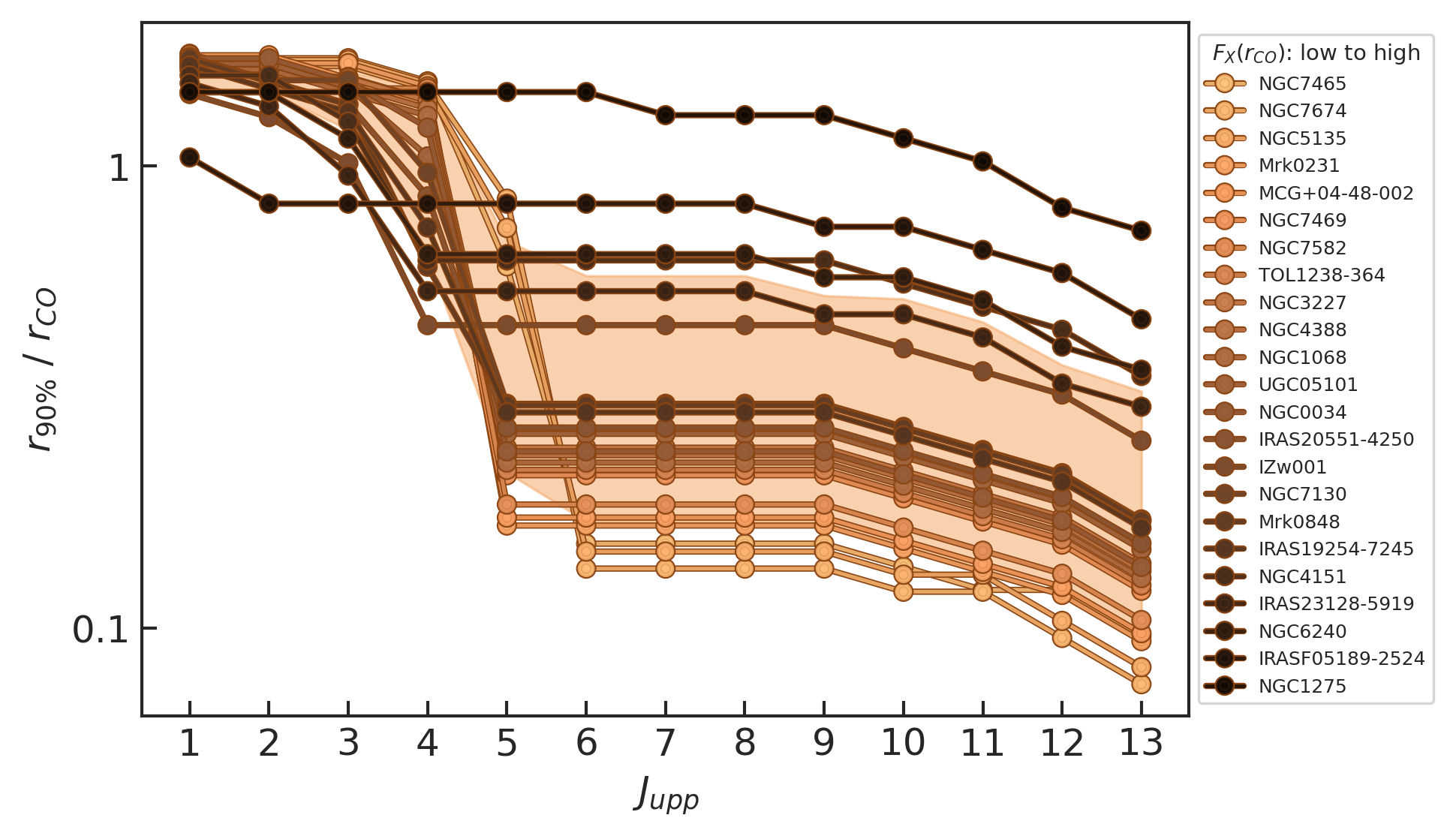}
    \caption{
    Galactocentric radius at which the luminosity
    of a given CO transition reaches $90 \%$
    of the total value for our sample of galaxies,
    divided by their $r_{\text{CO}}$.
    The filled circles, linked with
    solid lines, mark the $r_{90 \%}$ for 
    every galaxy, color-coded according to
    their $F_{\text{X}}(r_{\text{CO}})$,
    while the brown shaded area represents the
    values between $16\%$ and $84\%$ of the 
    radii distributions for our sample.
    }
    \label{fig:CObuildup}
\end{figure}


\subsection{PDR vs. XDR emission}

In Figure~\ref{fig:violin} we plot
the relative importance of PDR and XDR
emission as a function of $J$
for our galaxy sample as resulting from the "Best-fit" model.
It is remarkable the fact that
from CO($4-3$) going upwards
the CO luminosity is almost all due to
XDR emission, even after taking into account
the effect of X-ray flux attenuation
with a best-fit $N_{\text{H}}$.
This confirms what is shown in 
Figure~\ref{fig:clumpSLED}
for single molecular clumps,
and in Figure~\ref{fig:gmcSLED}
for single GMCs, 
i.e. the XDR models dominate
the overall CO luminosity,
with the exception of the low-$J$ lines.
These results are extensively
discussed in Section~\ref{sec:xdr_domination}.

\subsection{The CO SLED radial build-up}

Since we modelled the GMC distribution
and their CO emission as a function of 
galactocentric radius,
we can study the spatial distribution of different
CO lines.
In Figure~\ref{fig:CObuildup} we plot
the radius at which the luminosity
of a given CO transition reaches $90 \%$
of the total value for our sample of galaxies,
$r_{90 \%}$.
It is immediate to see that there is a separation
between low-$J$ lines, dominated by PDR emission
(Figure~\ref{fig:violin}),
which emit up to several kpc
(i.e. through the whole star-forming galaxy disc),
and mid- and high-$J$ lines, dominated by XDR emission,
which emit most of their luminosity
within $r=1$ kpc.
The spread in radii for a given CO line
is due to the different sizes
of the studied galaxies,
but also, at least for the mid-/high-$J$,
to the different incident $F_{\text{X}}$:
a higher $F_{\text{X}}$ indeed produces
a larger XDR emission
\citep[as in][]{meijerink05}.


\begin{figure*}
    \includegraphics[width=\textwidth]{
    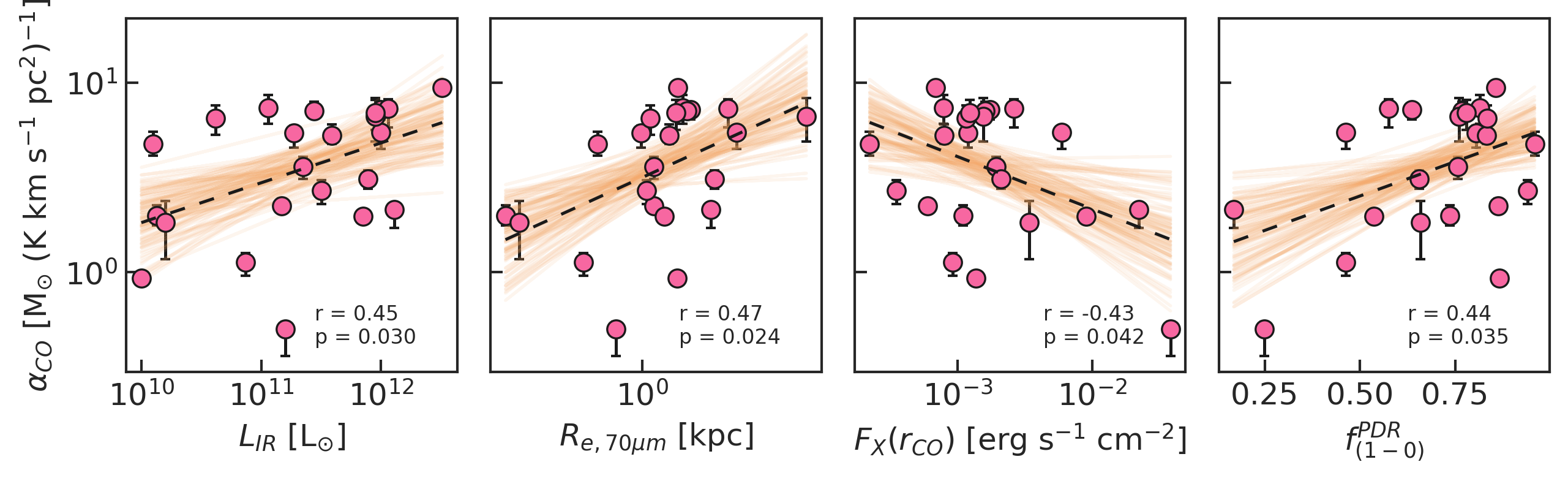}
    \caption{
    Scatter plot of best-fit $\alpha_{\text{CO}}$
    (on the $y$-axis) against, from left to right,
    the total IR luminosity, the effective radius
    of the 70 $\mu$m maps, the PDR fraction of the 
    total CO($1-0$) modelled emission, and 
    the X-ray flux at $r_{\text{CO}}$,
    for the whole galaxy sample.
    The dashed black line is the regression fit
    between the quantities on the $x$ and $y$ axes,
    and the brown lines are 100 bootstrapped fits,
    which highlight the confidence interval of
    the regression fit.
    The Pearson correlation coefficient $r$
    and the $p$-value appear
    on every panel on the bottom right.
    }
    \label{fig:alphaCO_regress}
\end{figure*}


\section{Discussion}
\label{sec:discussion}

In this section we discuss
the implications
of the derived values of
the CO-to-H$_2$ conversion
factor $\alpha_{\text{CO}}$
and the gas column densities
$N_{\text{H}}$ for our galaxy sample.
Finally, we discuss about the relative
importance of PDRs and XDRs
to the CO luminosity.

\subsection[The CO-to-H2 conversion factor
alphaCO]{The CO-to-H\textsubscript{2} 
conversion factor $\alpha_{\text{CO}}$}
\label{sec:alphaCO}

The best-fit procedure
returns reasonable values
for $\alpha_{\text{CO}}$.
We find a median value of
$\alpha_{\text{CO}} = 4.8^{+2.4}_{-2.8}$
M$_{\odot}$ (K km s$^{-1}$ pc$^2$)$^{-1}$,
which is similar to the Milky-Way adopted value.
Less than a third (8/24)
of our sample has
a ULIRG-like $\alpha_{\text{CO}} = 1.8^{+1.3}_{-0.8}$
M$_{\odot}$ (K km s$^{-1}$ pc$^2$)$^{-1}$
\citep{herreroillana19, pereztorres21},
even though only one of them
(IRAS F05189$-$2524) is a ULIRG,
and four (Mrk 848, NGC 5135, NGC 6240, and NGC 7674)
are LIRGs.

For some targets we have independent measures
of $\alpha_{\text{CO}}$ from the literature
to compare with.
By modelling the gas dynamics,
\cite{fei23} found, for I Zw 1, 
$\alpha_{\text{CO}} = 1.6 \pm 0.5$
M$_{\odot}$ (K km s$^{-1}$ pc$^2$)$^{-1}$,
which is lower than our result
($\alpha_{\text{CO}} = 6.6 \pm 1.7$
M$_{\odot}$ (K km s$^{-1}$ pc$^2$)$^{-1}$);
however, their analysis is limited to the central
1 kpc, whereas our model extends up to
$2 r_{\text{CO}}$ (i.e. 5.6 kpc for I Zw 1).
By comparing gas and dust emission,
\cite{garciaburillo14} found
values in the range $0.43-1.43$
M$_{\odot}$ (K km s$^{-1}$ pc$^2$)$^{-1}$
for the central 700 pc of NGC 1068 
(whereas we find 
$\alpha_{\text{CO}} = 5.4^{+0.9}_{-0.5}$
M$_{\odot}$ (K km s$^{-1}$ pc$^2$)$^{-1}$
up to a diameter of 4.9 kpc).
In NGC 6240, the quiescent gas 
and the outflowing one have different 
$\alpha_{\text{CO}} = 3.2 \pm 1.8$
M$_{\odot}$ (K km s$^{-1}$ pc$^2$)$^{-1}$
and $\alpha_{\text{CO}} = 2.1 \pm 1.2$
M$_{\odot}$ (K km s$^{-1}$ pc$^2$)$^{-1}$,
respectively \citep{cicone18}:
our value for this galaxy
($\alpha_{\text{CO}} = 2.0 \pm 0.1$
M$_{\odot}$ (K km s$^{-1}$ pc$^2$)$^{-1}$)
is closer to the outflowing gas,
which is where the bulk of the molecular
mass is, according to \cite{cicone18}.
The circumnuclear region
(up to a 175 pc radius) of NGC 7469
has a $\alpha_{\text{CO}}$ between
1.7 and 3.4 M$_{\odot}$ (K km s$^{-1}$ pc$^2$)$^{-1}$
according to \cite{davies04}.
Our model gives a higher result
($\alpha_{\text{CO}} = 5.3^{+0.5}_{-0.7}$
M$_{\odot}$ (K km s$^{-1}$ pc$^2$)$^{-1}$)
but it is derived for the molecular gas
extending up to a 4.7 kpc radius.

Surprisingly, we find a positive,
moderate 
(Pearson coefficient $r=0.45$, $p=0.03$),
correlation between $\alpha_{\text{CO}}$
and the total IR luminosity $L_{\text{IR}}$,
as shown in the first panel
of Figure~\ref{fig:alphaCO_regress}.
The value of $\alpha_{\text{CO}}$
is predicted to decrease for higher
gas temperatures, hence for starburst
galaxies as (U)LIRGs \citep{bolatto13}.
However, high gas and column densities
are expected to increase $\alpha_{\text{CO}}$:
for example, \cite{teng23} found $\alpha_{\text{CO}}$
to increase in the central region of
local barred galaxies due to the high
CO optical depth, which they find to be
a more dominant driver than the gas temperature.
A limit of the present analysis
is the fact that we used a single
$\alpha_{\text{CO}}$ for every radius,
while it is likely to vary 
with different local conditions,
as metallicity, temperature, density
and optical depth \citep{bolatto13, teng23}.
Another important point to make is that
our (U)LIRGs (17/24 galaxies have 
$L_{\text{IR}} \geq 10^{11}$ L$_{\odot}$)
are AGN, so that a fraction of the IR luminosity
is due to AGN activity rather than SF
\citep[e.g.][]{nardini10, alonsoherrero12}.
For example, the largest value we derive is
$\alpha_{\text{CO}} = 9.41$
M$_{\odot}$ (K km s$^{-1}$ pc$^2$)$^{-1}$
for Mrk 231, the only quasar of our sample,
with $L_{\text{IR}} = 10^{12.5}$ L$_{\odot}$,
the highest of our sample.

We also find $\alpha_{\text{CO}}$ to correlate
with the effective radius $R_{e,70\mu m}$
of the 70 $\mu$m emission
($r=0.47$, $p=0.02$,
second panel of Figure~\ref{fig:alphaCO_regress}).
These radii result from the Sersic fit
on the \textit{Herschel} maps analyzed in 
\citetalias{esposito22}.
No significant correlation ($r=-0.12$, $p=0.58$)
shows up instead
between $\alpha_{\text{CO}}$ and $r_{\text{CO}}$:
this means that the optical and molecular radii
($r_{\text{CO}} = 0.17 \, r_{25}$)
do not affect the CO-to-H$_2$ conversion factor,
while it seems influenced by
the size of the star-forming region
(traced by the 70 $\mu$m emission in our analysis).
We do also find $L_{\text{IR}}$ and $R_{e,70\mu m}$
to strongly correlate between each other 
($r=0.72$, $p=8 \times 10^{-5}$),
so it is possible that only one of the two
correlations is actually meaningful.
There is no significant correlation 
($r=0.13$, $p=0.56$)
between $\alpha_{\text{CO}}$ and the median $G_0$ 
listed in Table~\ref{tab:sample_results},
or the $G_0(R_{e,70\mu m})$, so it
seems $G_0$ is not the main driver
of the $\alpha_{\text{CO}}$ variation.

Some clues can come from the way we
implemented $\alpha_{\text{CO}}$ in our model,
i.e. as a normalization factor of the
"Best-fit model". A bright IR galaxy
has also bright CO emission at every $J$
\citep[e.g. ][]{kamenetzky16}.
However, there is complex interplay
between the combined effects of $\alpha_{\text{CO}}$
and $N_{\text{H}}$ in our model.
For example, MCG+04$-$48$-$002
(top panel of Figure~\ref{fig:comparison_SLED})
has a high $\alpha_{\text{CO}}=7.32$
M$_{\odot}$ (K km s$^{-1}$ pc$^2$)$^{-1}$,
even though the "Baseline model", made
with a lower $\alpha_{\text{CO}}$,
is brighter for some CO lines,
and this is due to the high $N_{\text{H}}$
that we derived for this galaxy,
which suppresses the XDR emission.
We do not find 
($r=0.28$, $p=0.20$)
a convincing correlation between
the two best-fit parameters
($\alpha_{\text{CO}}$, $N_{\text{H}}$),
but we do find (third panel of
Figure~\ref{fig:alphaCO_regress})
an anti-correlation 
($r=-0.43$, $p=0.04$) between $\alpha_{\text{CO}}$
and $F_{\text{X}}$ measured at a
distance $r_{\text{CO}}$ from the AGN, and
obscured with the best-fit $N_{\text{H}}$:
a high $F_{\text{X}}$ is indeed expected
to boost the XDR CO emission,
hence lowering $\alpha_{\text{CO}}$.

In the last panel of Figure~\ref{fig:alphaCO_regress},
we show a positive, 
moderate ($r=0.44$, $p=0.04$),
correlation between
$\alpha_{\text{CO}}$ and the PDR fractions
to the total CO($1-0$) modelled emission.
For some galaxies 
(e.g. MCG+04$-$48$-$002, which has a 
high $f^{(1-0)}_{\text{PDR}} = 0.81$
-- see Table~\ref{tab:sample_results})
this is due to the combined effect of
$\alpha_{\text{CO}}$ and $N_{\text{H}}$:
a high $N_{\text{H}}$ suppresses the XDR emission,
so that a high $\alpha_{\text{CO}}$ is needed
to boost the otherwise low PDR emission
and increase the likelihood.

We stress that the characterization of a class
of objects (local AGN as in our study) 
or even more of individual galaxies
by constraining galaxy properties such as 
$\alpha_{\text{CO}}$, is in line with 
some recent studies
\citep[e.g.,][]{ellison21, casasola22, salvestrini22, thorp22}.
These works highlight the heterogeneity of
properties of the local galaxies, 
in addition to provide observational constraints 
to theoretical models and
for high-redshift studies.

\subsection{The X-ray attenuation column density}
\label{sec:logNH}

Taking into account the $1\sigma$ uncertainty,
we find non-negligible 
(i.e. $N_{\text{H}} > 10^{22}$ cm$^{-2}$,
see Equation~\ref{eq:FX_r}) X-ray attenuation for most 
(14/23 galaxies)
of our sample.
The median is
$\log(N_{\text{H}} / \text{cm}^{-2}) = 22.1^{+1.6}_{-0.1}$,
and the range [$10^{22}, 10^{24}$] cm$^{-2}$.
By comparison the column density
derived from the X-ray SED, for 20/23 galaxies,
has median $\log (N_{\text{H, X-ray}}/\text{cm}^{-2})$
$= 23.7^{+0.5}_{-2.0}$
and range [$10^{20.5}, 10^{25}$] cm$^{-2}$.

Although our best-fit $N_{\text{H}}$ 
approximately has the same range
of $N_{\text{H, X-ray}}$,
the two values do not correlate
(Figure~\ref{fig:logNH_scatter}).
This uncorrelation 
can be due to the different locations
of the X-ray absorbers:
in the case of $N_{\text{H, X-ray}}$,
the X-ray flux is attenuated by gas clouds
along the line of sight only,
while the $N_{\text{H}}$ we derive is due to clouds
between the AGN and the GMCs
in the whole galaxy volume
(i.e. considering multiple lines of sight).

Lately, more complex models of X-ray absorption
are leading to different estimates of $N_{\text{H}}$,
measuring the obscuration along the line of sight
plus a reflective component from the nuclear torus
\citep{yaqoob12, esparzaarredondo21}.
Several works targeting local AGNs 
show that these different $N_{\text{H}}$,
both estimated from the X-ray SED,
fail to correlate between each other
\citep{torresalba21, sengupta23}.

\cite{garciaburillo21} compared
instead different estimates of $N_{\text{H}}$, from
X-ray absorption, and from high-resolution
(7-10 pc) ALMA CO observations, 
for a sample of 19 local Seyferts, 
finding similar median values but with a large
dispersion around the 1:1 line.
As our estimates of Figure~\ref{fig:logNH_scatter},
they also find less absorbed galaxies
(with $N_{\text{H,X-ray}} < 10^{22}$ cm$^{-2}$)
to systematically lie above the 1:1 line,
and vice versa.
This could be expected if,
in less absorbed galaxies,
$N_{\text{H,X-ray}}$
is dominated by gas clouds
smaller than our spatial resolution,
which is dictated by the diameter
of our smallest modelled GMC (3 pc, 
see Table~\ref{tab:GMCproperties}).

It is believed that, at least in local AGNs,
most of the obscuration is due to the
nuclear torus rather than gas
on galactic scales
(see \citealt{hickox18} for a recent review,
and \citealt{gilli22} for high-redshift systems).
However, the orientation, filling factor 
and opening angle of the torus are
all important quantities that affect
the measurement of $N_{\text{H, X-ray}}$.
With our analysis we give a
way to measure the attenuation of AGN obscuration
on the molecular gas in all the possible
directions.

\subsection{PDR vs. XDR emission}
\label{sec:xdr_domination}

As shown in Figures
\ref{fig:clumpSLED} and \ref{fig:gmcSLED},
the CO luminosity is dominated,
for our modelled clumps and GMCs,
by XDR emission,
at least at $J > 3$.
We find the same by fitting the
observed CO SLEDs (Figure~\ref{fig:violin}).
The fact that high-$J$ lines
need X-rays to be radiatively excited
is well known and studied
\citep{bradford09, vanderwerf10, rosenberg15, kamenetzky17},
however this is the first time
a detailed study of CO emission,
coming from different gas sizes (clumps, GMCs),
produces this same result.
The majority of the studies have found,
so far, CO excitation to be consistent
with PDRs up to CO($6-5$)
\citep{vanderwerf10, haileydunsheath12, pozzi17}.
One notable exception is the work by
\cite{mingozzi18}, where their fiducial
model for the CO SLED of NGC 34
has a XDR component which starts
dominating the luminosity at $J_{\text{upp}}=3$.

All these studies used a combination
of two or three PDR and XDR components,
each with a single density and incident flux.
To be able to reproduce PDR emission 
consistent with observations
at mid-/high-$J$, these models need
very high gas densities 
($n_{\text{H}} \gtrsim 10^5$ cm$^{-3}$),
more typical of high-redshift, turbulent star-forming regions 
\citep{calura22}.
The mass fraction of such dense clumps 
within the modelled GMCs is $\sim 2\%$,
being most of the molecular mass in the range
$[10^2, 10^4]$ cm$^{-3}$, 
where only XDRs are able to emit
significant high-$J$ CO emission
(see also Figure~\ref{fig:gmcSLED}).
This also confirms the findings of \citetalias{esposito22},
i.e. that only unlikely extreme gas densities
PDR models can reproduce the CO emission.

As a caveat, we note that 
our model does not include
(\textit{i}) variations in the
cosmic ray ionisation rate (CRIR),
and (\textit{ii}) the effect of shocks.
It is known that both
high cosmic-ray fluxes
\citep{vallini19, bisbas23}
and shocks 
\citep{meijerink13, mingozzi18, bellocchi20}
influence the CO SLED,
especially the high-$J$ lines.
Further developments of our model
may be able to include a varying CRIR
and the treatment of large scale shocks, 
as in the case of galaxies with winds and outflows
\citep{garciaburillo14, morganti15, fiore17, speranza22}
or mergers \citep{ueda14, larson16, ellison19}.


\section{Conclusions}
\label{sec:conclusions}

We have presented a new 
physically-motivated model for 
computing the CO SLED in
AGN-host galaxies,
which takes into account
the internal structure of GMCs,
the radiative transfer of
external FUV and X-ray flux on
the molecular gas,
and the mass distribution of GMCs
within a galaxy volume.
The assumptions are that
the molecular clumps in a GMC
follow a log-normal density distribution,
the clumps size is equal to their Jeans radius, 
and that the GMC masses, in a galaxy,
follow a power-law distribution.
The model can predict the CO SLED
of a galaxy from the radial profiles
of molecular mass $M_{\text{mol}}(r)$,
FUV flux $G_0(r)$ and X-ray flux $F_{\text{X}}(r)$.
We also set two free parameters:
the CO-to-H$_2$ conversion factor
$\alpha_{\text{CO}}$, and the
X-ray attenuation column density $N_{\text{H}}$.
We have used \textsc{Cloudy} to predict
the CO lines emission of
clumps and GMCs, through a combination
of PDR and XDR simulations,
and by taking into account 
the observed radial profiles.

To test the validity of our model,
we extracted, from the galaxy sample of \citetalias{esposito22},
a sub-sample of 24 
X-ray selected
AGN-host galaxies
with a well-sampled CO SLED.
The sample represents a good variety of local
($z \leq 0.06$) moderately luminous
($10^{42} \leq L_{\text{X}} [\text{erg s}^{-1}]
\leq 10^{44}$, $10^{10} \leq L_{\text{IR}} 
[\text{L}_{\odot}] \leq 10^{12.5}$) AGNs.
We compared the CO SLEDs
produced by our model with the observed ones,
and we selected the best-fit 
$\alpha_{\text{CO}}$ and $N_{\text{H}}$ 
with a MCMC analysis for each galaxy.
Our main results are here summarized:
\begin{itemize}
\item the best-fit procedure returned,
for our galaxy sample, median values of
$\alpha_{\text{CO}} = 4.8^{+2.4}_{-2.8}$
M$_{\odot}$ (K km s$^{-1}$ pc$^2$)$^{-1}$ and
$\log (N_{\text{H}}/\text{cm}^{-2}) = 22.1^{+1.6}_{-0.1}$,
where the lower and upper errors
are the $16^{\text{th}}$ and 
$84^{\text{th}}$ percentiles, respectively;
we find $\alpha_{\text{CO}}$ consistent with 
the Galactic value, and $N_{\text{H}}$ that do not
correlate with $N_{\text{H, X-ray}}$;
\item XDRs are fundamental
to reproduce observed CO SLEDs
of AGN-host galaxies,
particularly for $J_{\text{upp}} \gtrsim 3$.
\end{itemize}
We argue that the larger importance
of X-rays to the
CO luminosity, with respect to other
works in the literature, 
is mainly due to 
the fact that the majority of 
molecular gas mass in a galaxy is at 
$n_{\text{H}} \sim 10^2 - 10^4$ cm$^{-3}$.
This conclusion comes from a
physically-motivated 
structure for the molecular gas,
rather than from single-density
radiative transfer models.

The model predictions can be
used to estimate
$\alpha_{\text{CO}}$ in AGN-host galaxies,
by constraining both the relative and 
absolute intensity of several CO lines.
They can also be useful to estimate
the effective attenuation of X-ray photons
on the molecular gas in a spatially-resolved
way, rather than only on the line-of-sight
(i.e., $N_{\text{H, X-ray}}$).

Moving forward, 
the model is able
to predict the emission of other molecular
species, as HCN, HCO$^+$, H$_2$, H$_2$O.
These molecules have been increasingly used
in detailed studies of local AGN
\citep{butterworth22, eibensteiner22, huang22}.
The model is also built to exploit
a wealth of observed galaxy data,
and compare different molecular lines
spatially resolved by ALMA
\citep[as done with NGC 1068 in][]{garciaburillo19, nakajima23}
with our model predictions.

\section*{Acknowledgements}
We thank the anonymous referee for suggestions 
and comments that helped to improve 
the presentation of this work.
We acknowledge use of
Astropy \citep{astropy1, astropy2},
\texttt{corner.py} \citep{corner},
Matplotlib \citep{matplotlib},
NumPy \citep{numpy},
Pandas \citep{pandas},
Python \citep{python3},
Seaborn \citep{seaborn},
SciPy \citep{scipy}.
We acknowledge the usage of the 
HyperLeda database 
(\url{http://leda.univ-lyon1.fr}).
The authors acknowledge the use of 
computational resources from the parallel
computing cluster of the 
Open Physics Hub 
(\url{https://site.unibo.it/openphysicshub/en})
at the Physics and Astronomy 
Department in Bologna.
FE and FP acknowledge support from grant PRIN MIUR 2017- 20173ML3WW$\_$001.
SGB  acknowledges support from the research
project PID2019-106027GA-C44 
of the Spanish Ministerio de Ciencia e Innovación. 
FE, LV, FP, VC, FC and CV acknowledge funding 
from the INAF Mini Grant 2022 program 
“Face-to-Face with the Local Universe: 
ISM’s Empowerment (LOCAL)”.
AAH acknowledges support from research grant 
PID2021-124665NB-I00 funded by 
MCIN/AEI/10.13039/501100011033 
and by ERDF "A way of making Europe".

\section*{Data Availability}

We publicly release the code, called 
galaxySLED, on GitHub 
(\url{https://federicoesposito.github.io/galaxySLED/}).
The data generated in this research
will be shared on reasonable request 
to the corresponding author.

\bibliographystyle{mnras}
\bibliography{GMCmodelAGN}

\clearpage

\appendix

\section{Update of observed fluxes for our sample}
\label{sec:update}

Reviewing the data contained in \citetalias{esposito22},
we have found a better source for
one CO line luminosity in one galaxy.
The CO($2-1$) of NGC 1275 was taken, in \citetalias{esposito22},
from \cite{salome11}, which was converted
to $303^{+8}_{-8} \times 10^2$ L$_{\odot}$.
Since this flux was observed, in \cite{salome11},
from a very small nuclear region of NGC 1275,
we wanted to improve the measurement
by finding another observation with
a larger beam.
\cite{lazareff89} observed the CO($2-1$)
emission with the IRAM-30m telescope,
which has a full-width at half maximum
(FWHM) of 10\farcs5, finding
a larger CO($2-1$) luminosity of 
$15^{+3}_{-3} \times 10^4$ L$_{\odot}$.
Throughout this paper
we use this value, since it is the one
with the larger beam we can find for this galaxy;
in this way the CO($2-1$) beam is
almost equal to the projected $r_{\text{CO}}$
of the galaxy: $r_{\text{CO}} = $ 10\farcs9.

\section{Radial profiles of galaxies}

\begin{figure}
    \includegraphics[width=\columnwidth]{
    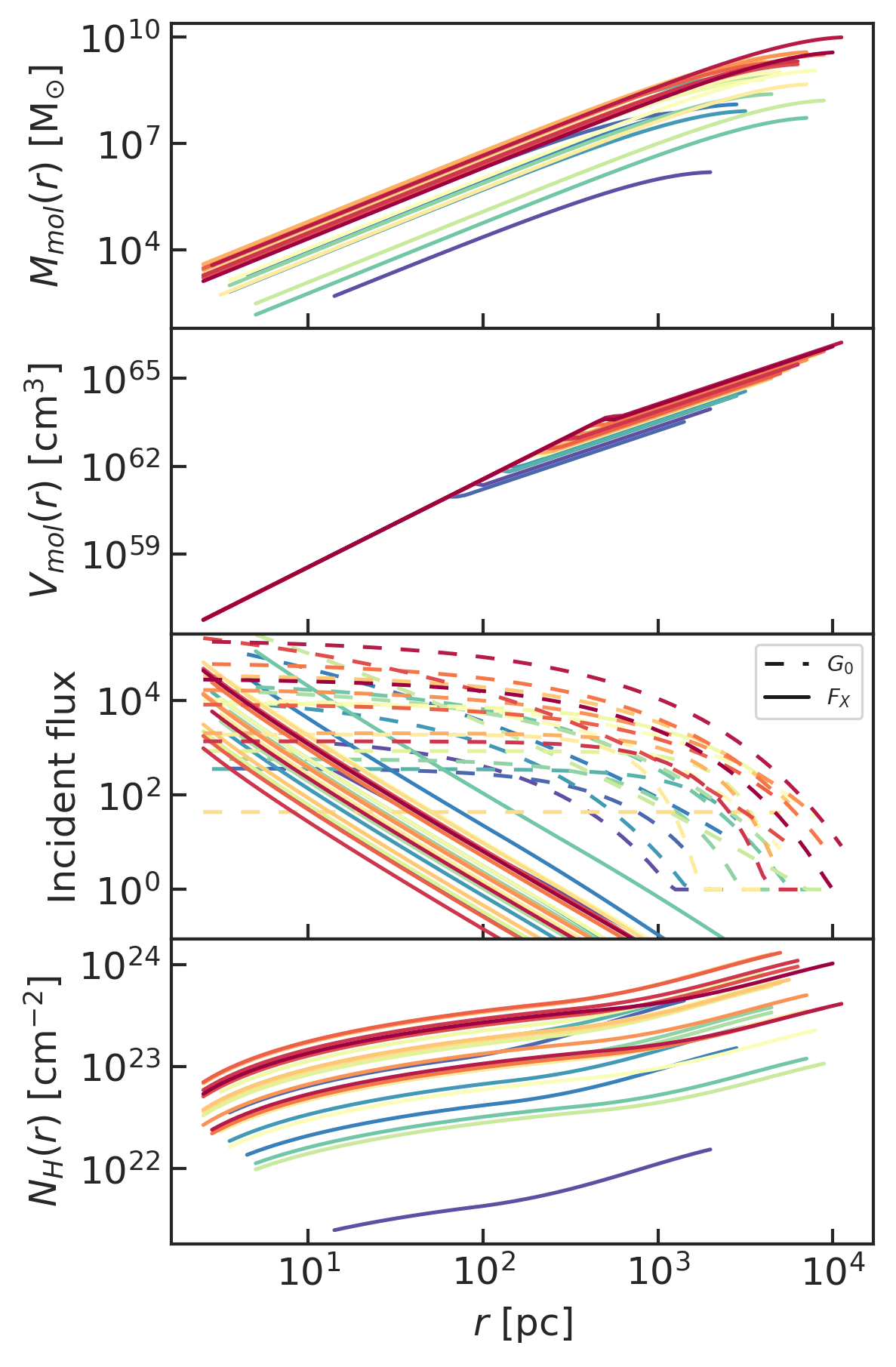}
    \caption{
    Radial profiles of our sample of
    24 galaxies, sorted by their total
    molecular mass when measured with a Milky Way
    $\alpha_{\text{CO}}$.
    The four panels show, from top to bottom,
    the molecular mass $M_{\text{mol}}(r)$
    (recalculated with the best-fit $\alpha_{\text{CO}}$),
    the molecular volume $V_{\text{mol}}(r)$,
    the incident fluxes $G_0(r)$ and $F_{\text{X}}(r)$,
    and the "Baseline model" 
    column density $N_{\text{H}}(r)$.
    $G_0(r)$ and $F_{\text{X}}(r)$ are in $G_0$
    and erg s$^{-1}$ cm$^{-2}$ units and they are
    plotted with dashed and solid lines, respectively.
    Colors are the same as
    Figure~\ref{fig:sampleSLED}.
    }
    \label{fig:diagnostics}
\end{figure}

In Figure~\ref{fig:diagnostics} we show the radial profiles
of the molecular mass $M_{\text{mol}}(r)$,
the volume $V_{\text{mol}}(r)$,
the incident fluxes $G_0(r)$ and $F_{\text{X}}(r)$,
and the intrinsic column density $N_{\text{H}}(r)$.
$M_{\text{mol}}(r)$ has been recalculated
with the best-fit $\alpha_{\text{CO}}$.
The radial profiles all follow the
Equations~(\ref{eq:M_r})~$-$~(\ref{eq:NH_r}).

\section{Observed and modelled CO SLEDs
for the whole galaxy sample}
\label{sec:appendix}

\begin{figure}
    \includegraphics[width=\columnwidth]{
    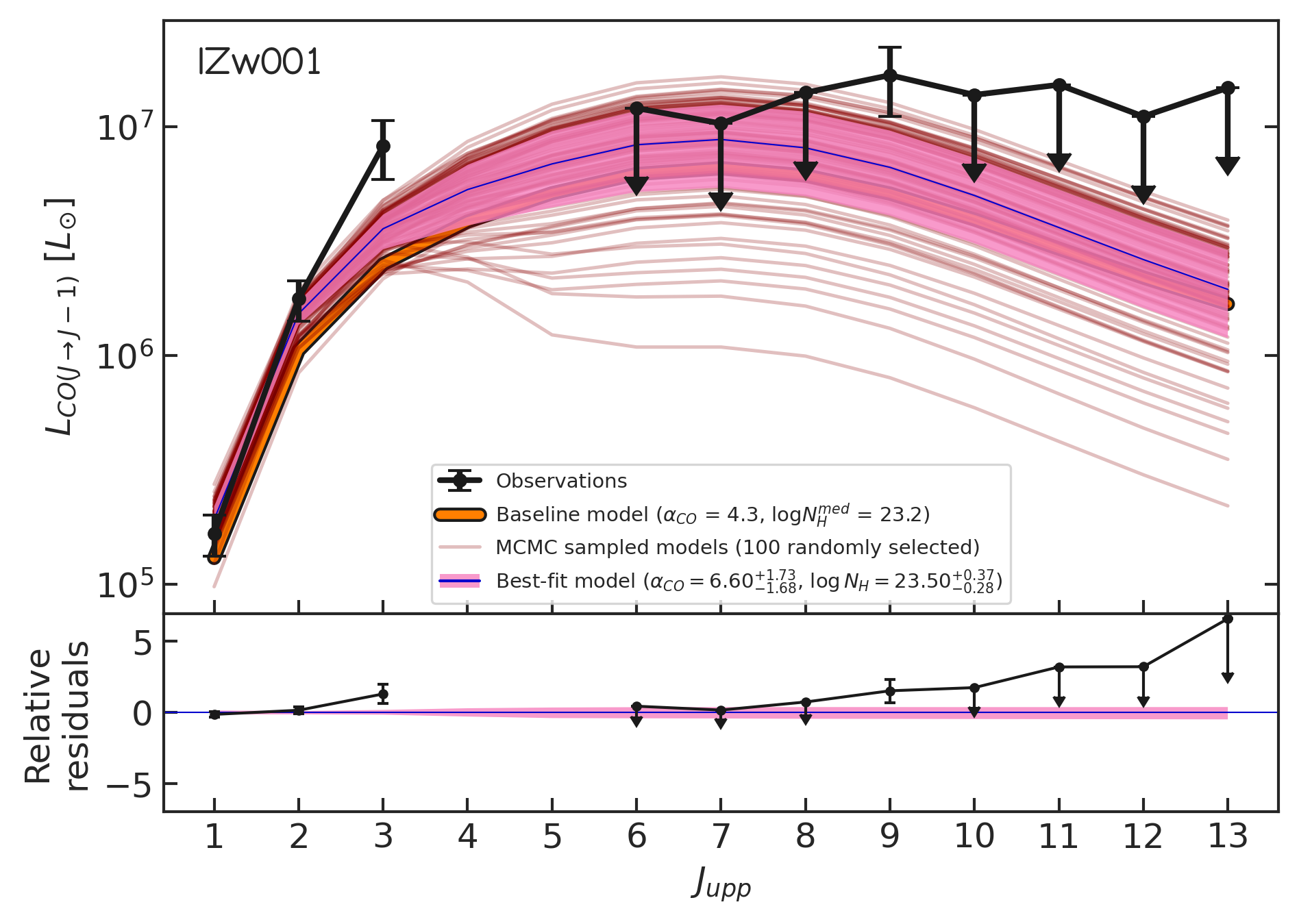}
    \caption{I Zw 1 (UGC 545) CO SLEDs.
    }
    \label{fig:izw1}
\end{figure}

\begin{figure}
    \includegraphics[width=\columnwidth]{
    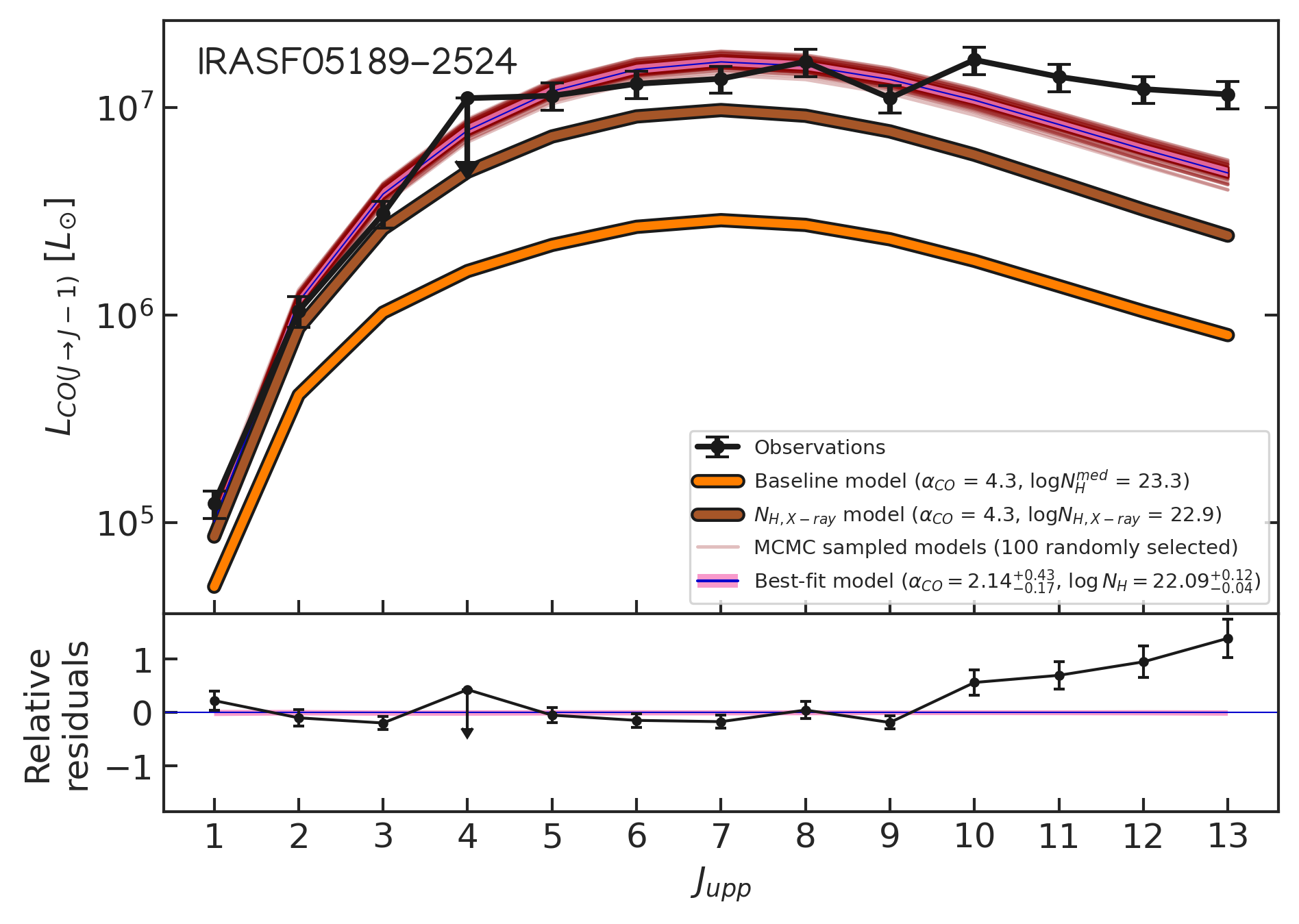}
    \caption{IRAS F05189$-$2524 CO SLEDs.}
    \label{fig:iras05189}
\end{figure}

\begin{figure}
    \includegraphics[width=\columnwidth]{
    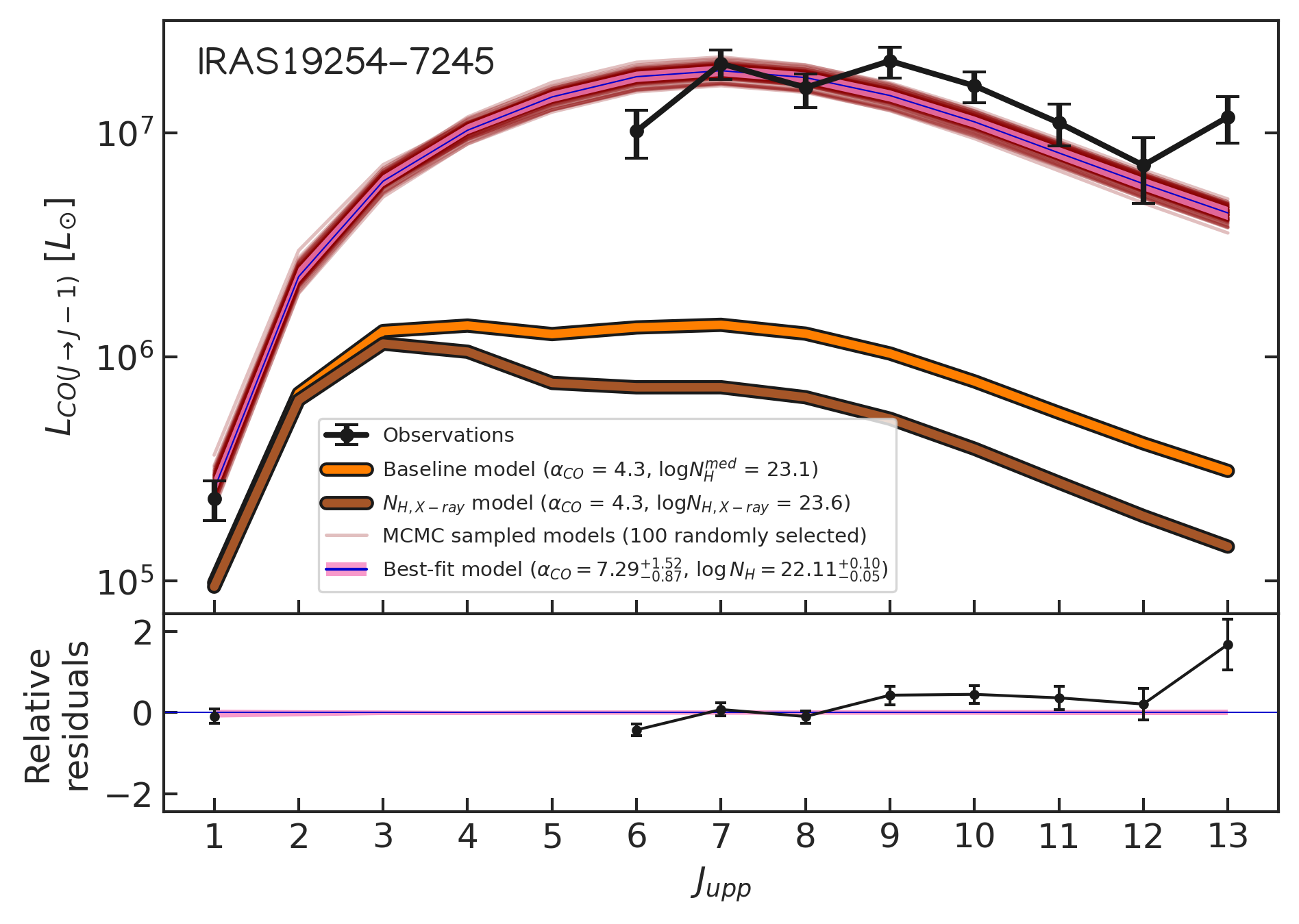}
    \caption{IRAS 19254$-$7245 (Superantennae) CO SLEDs.}
    \label{fig:iras19254}
\end{figure}

\begin{figure}
    \includegraphics[width=\columnwidth]{
    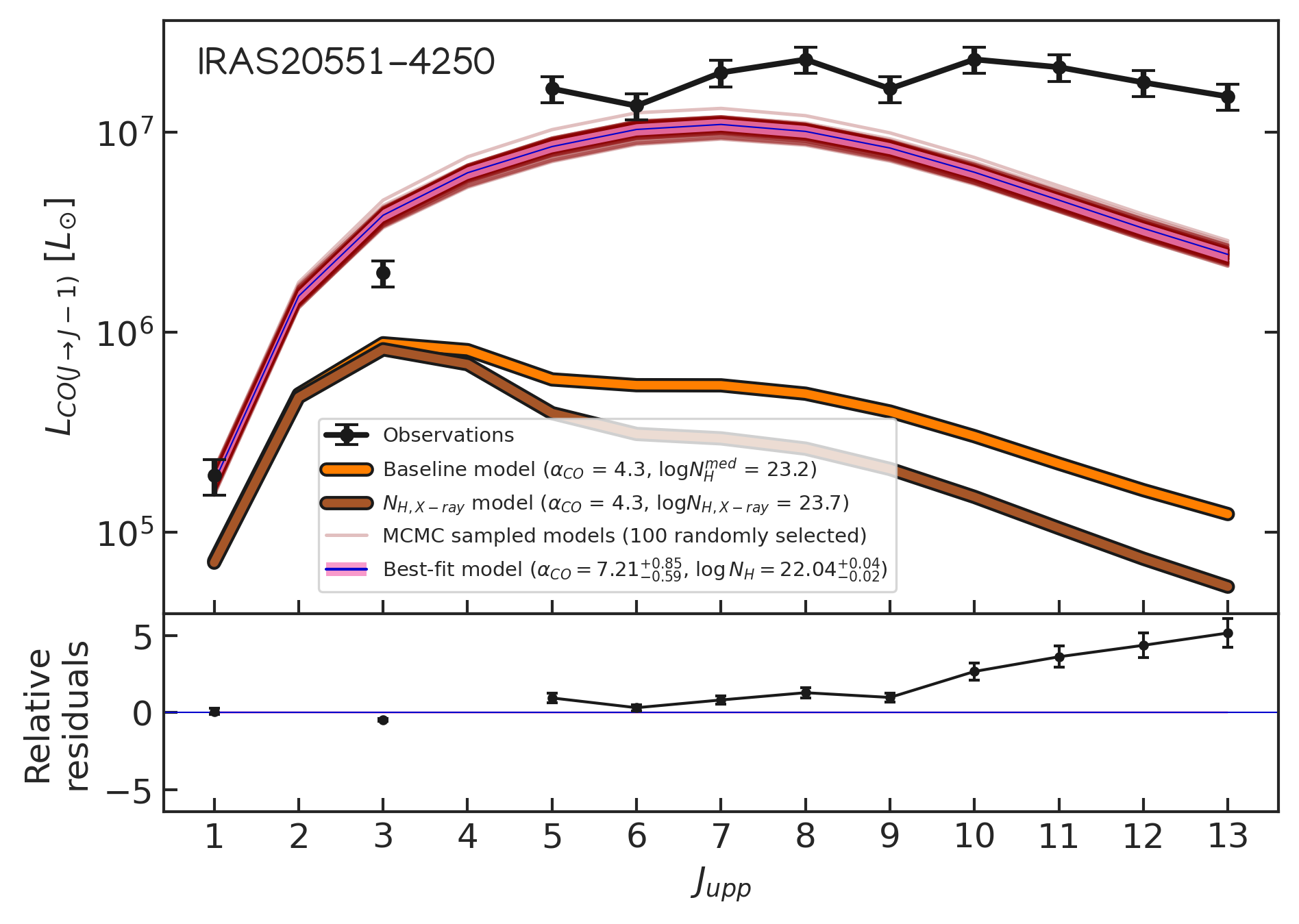}
    \caption{IRAS 20551$-$4250 (ESO 286$-$IG019) CO SLEDs.}
    \label{fig:iras20551}
\end{figure}

\begin{figure}
    \includegraphics[width=\columnwidth]{
    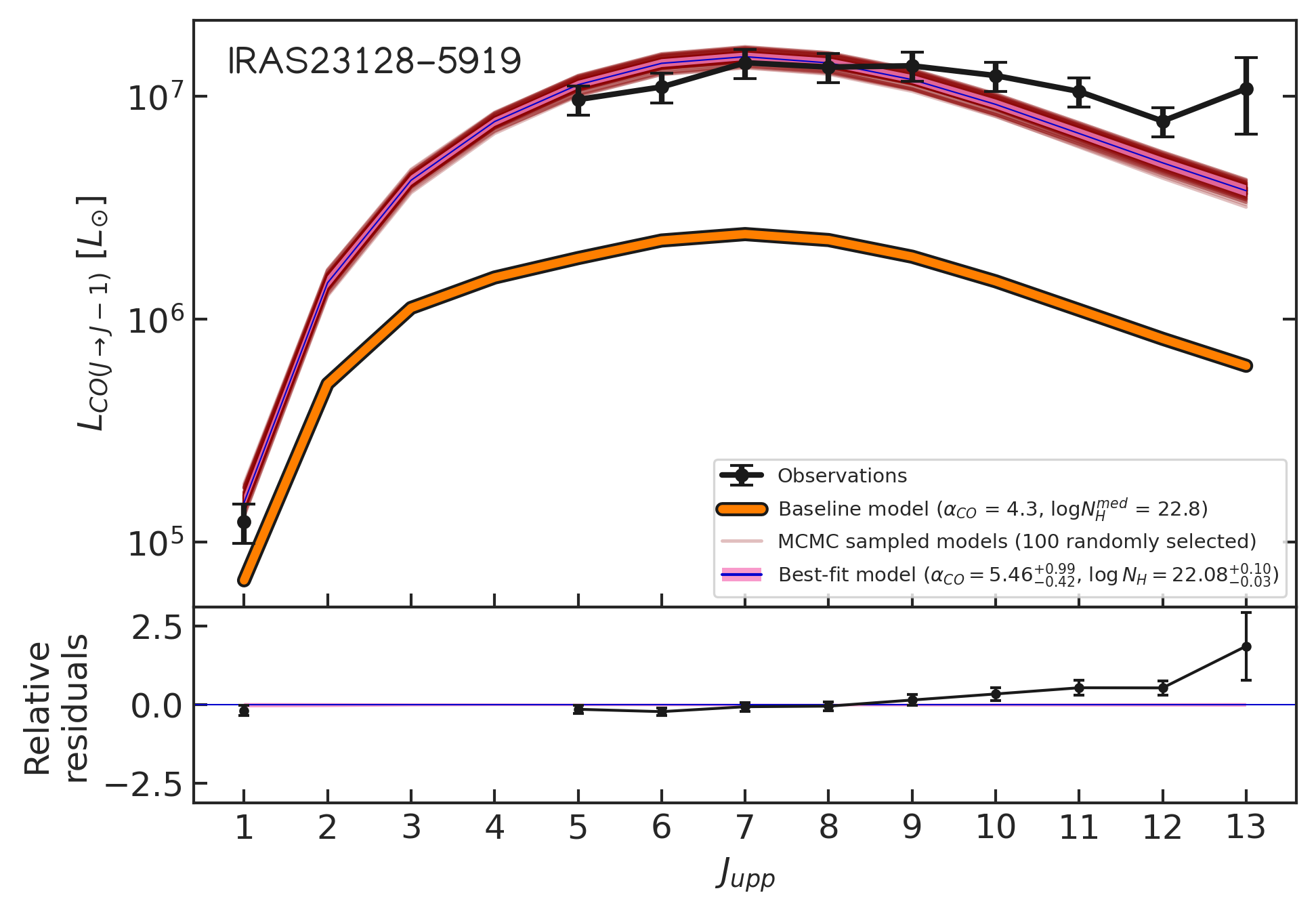}
    \caption{IRAS 23128$-$5919 (ESO 148$-$IG002) CO SLEDs.
    }
    \label{fig:iras23128}
\end{figure}

\begin{figure}
    \includegraphics[width=\columnwidth]{
    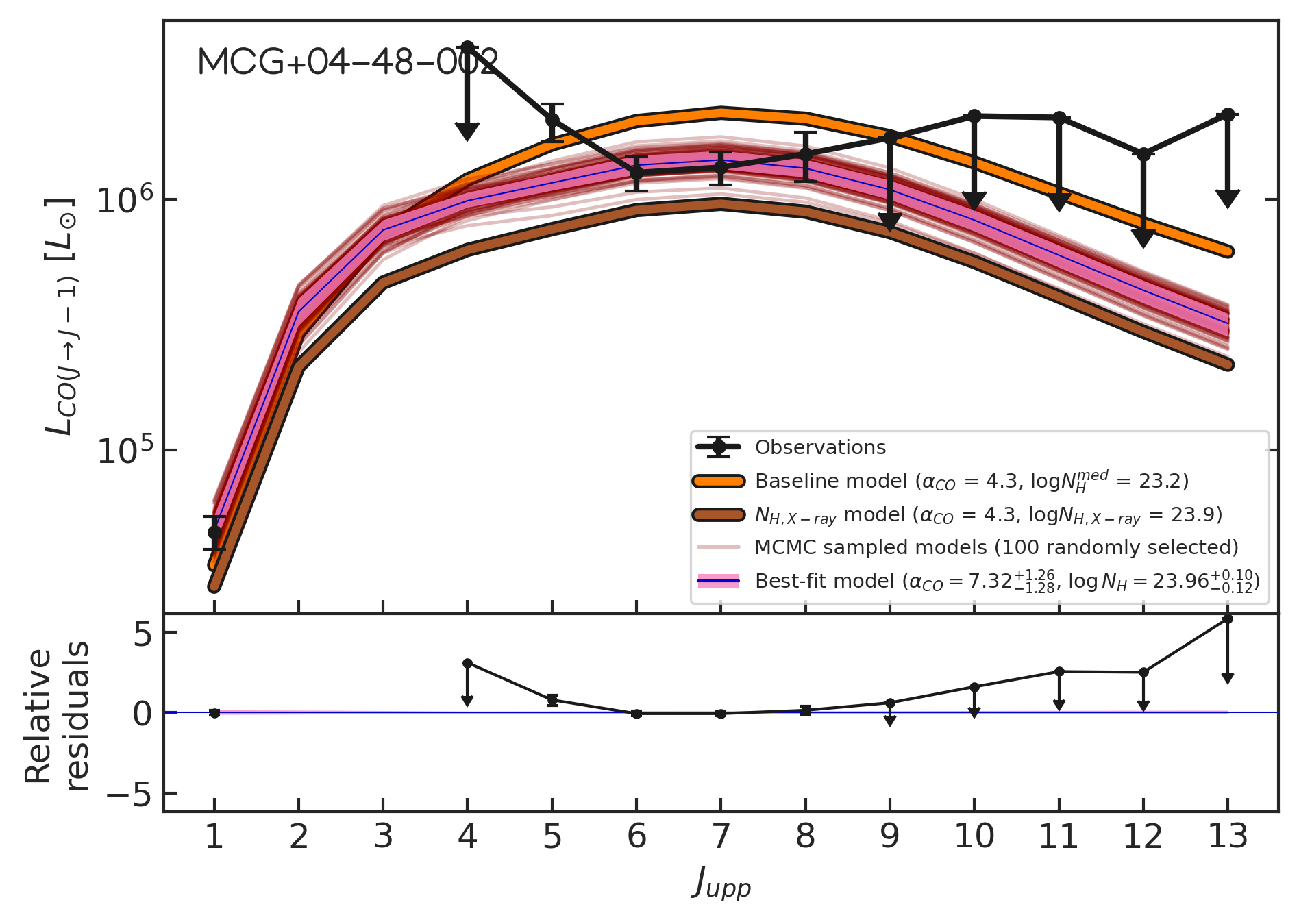}
    \caption{MCG +04$-$48$-$002 CO SLEDs.}
    \label{fig:mcg04}
\end{figure}

\begin{figure}
    \includegraphics[width=\columnwidth]{
    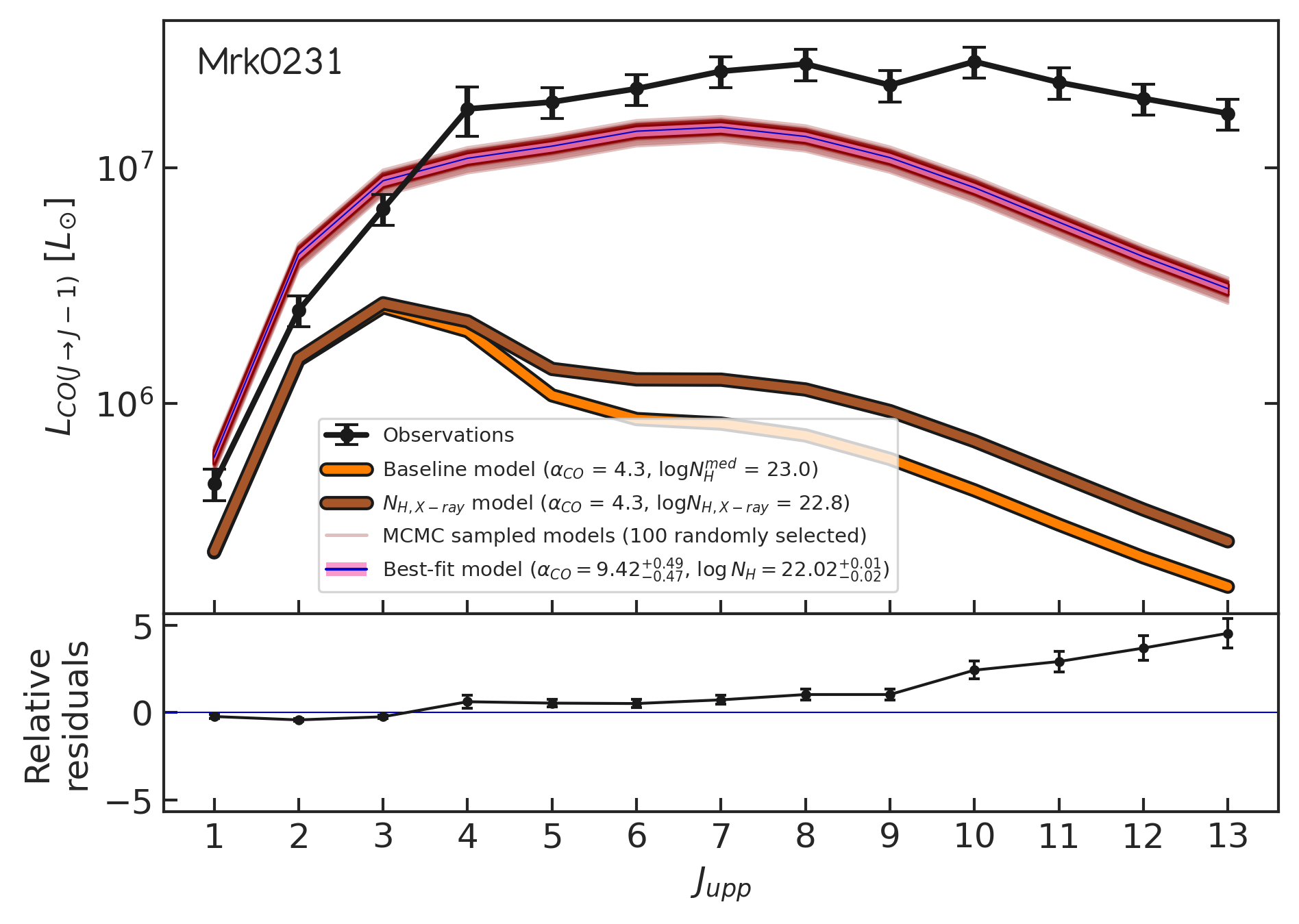}
    \caption{Mrk 231 CO SLEDs.}
    \label{fig:mrk231}
\end{figure}

\begin{figure}
    \includegraphics[width=\columnwidth]{
    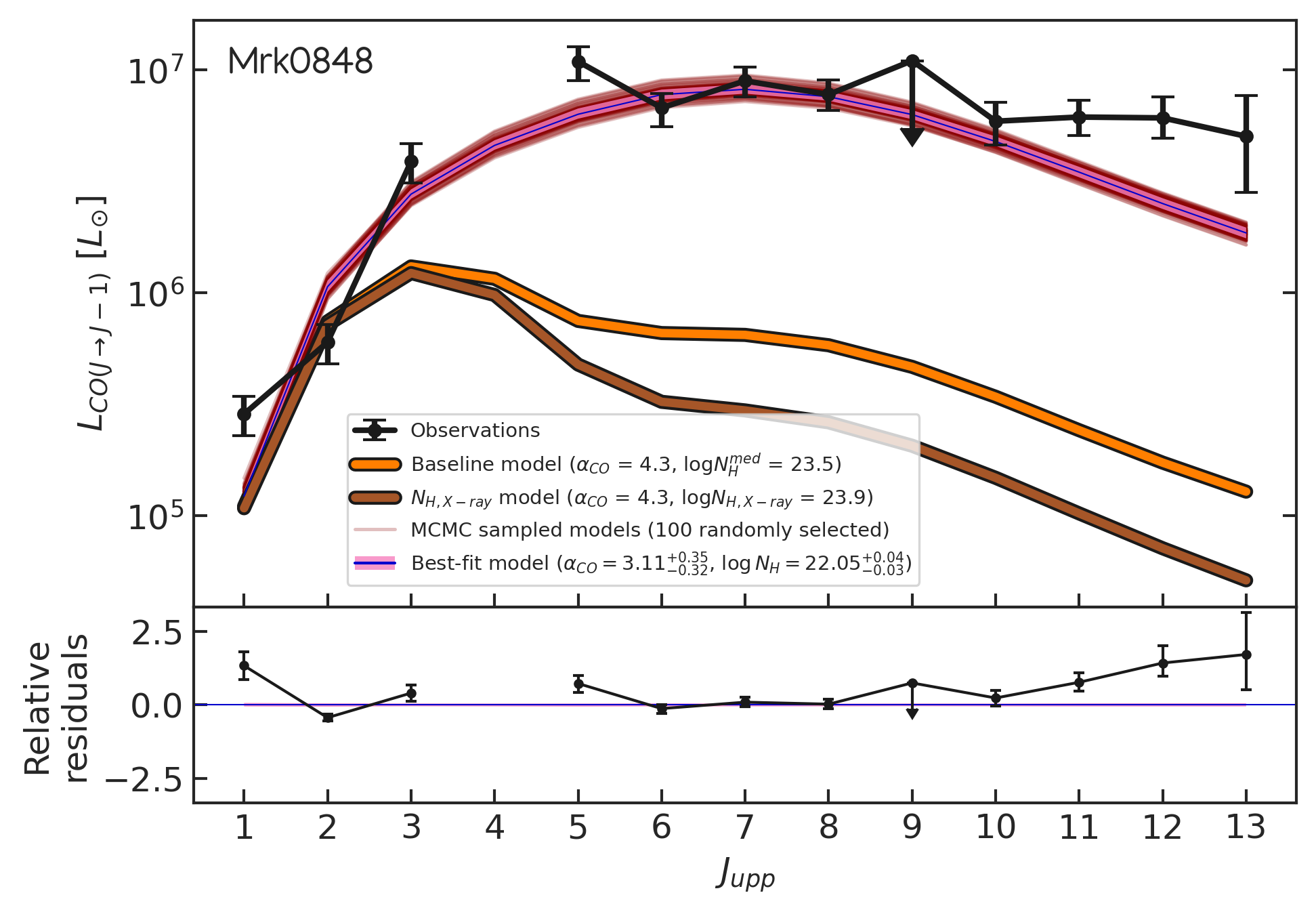}
    \caption{Mrk 848 (VV 705) CO SLEDs.}
    \label{fig:mrk848}
\end{figure}

\begin{figure}
    \includegraphics[width=\columnwidth]{
    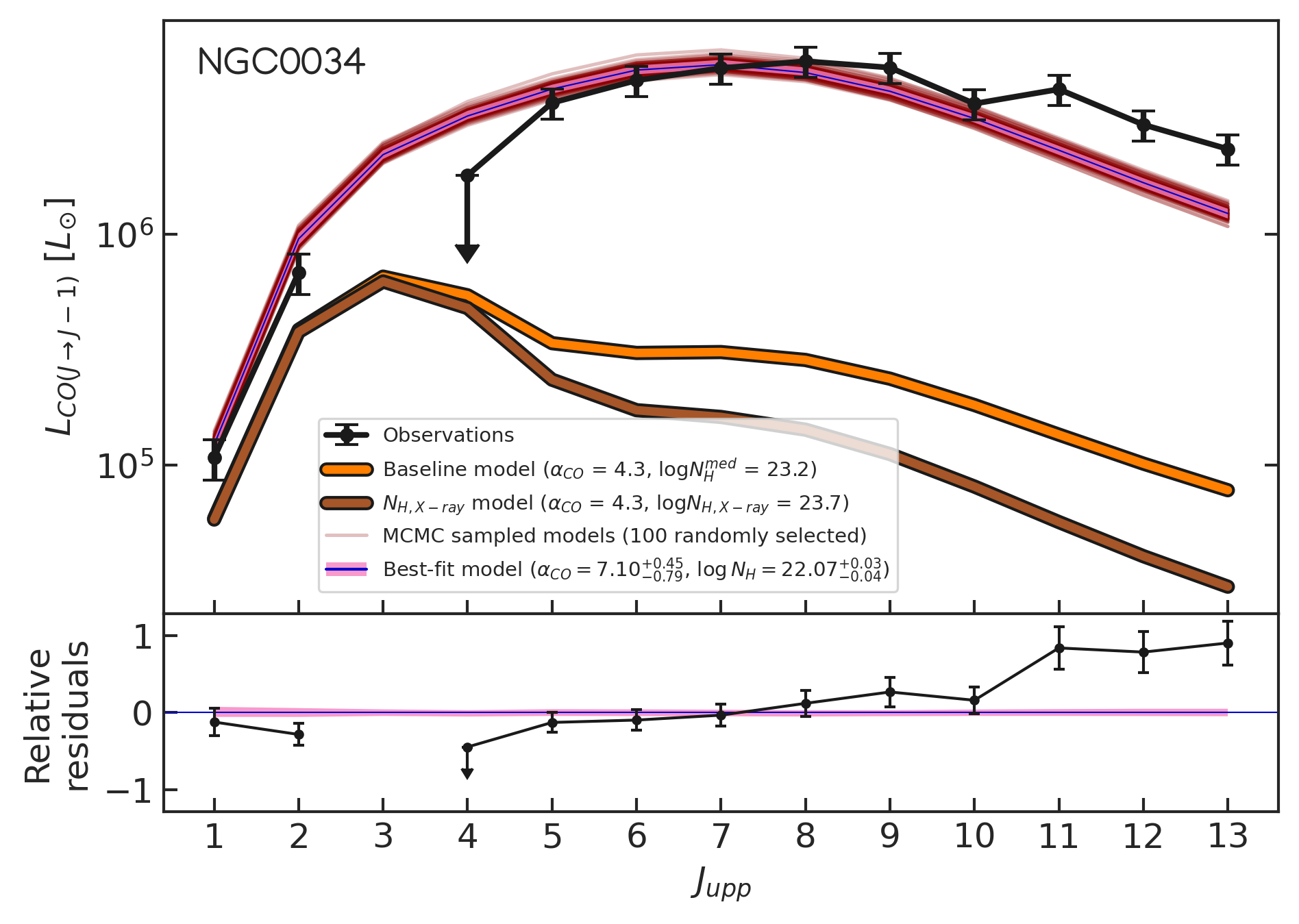}
    \caption{NGC 34 (Mrk 938) CO SLEDs.}
    \label{fig:ngc34}
\end{figure}

\begin{figure}
    \includegraphics[width=\columnwidth]{
    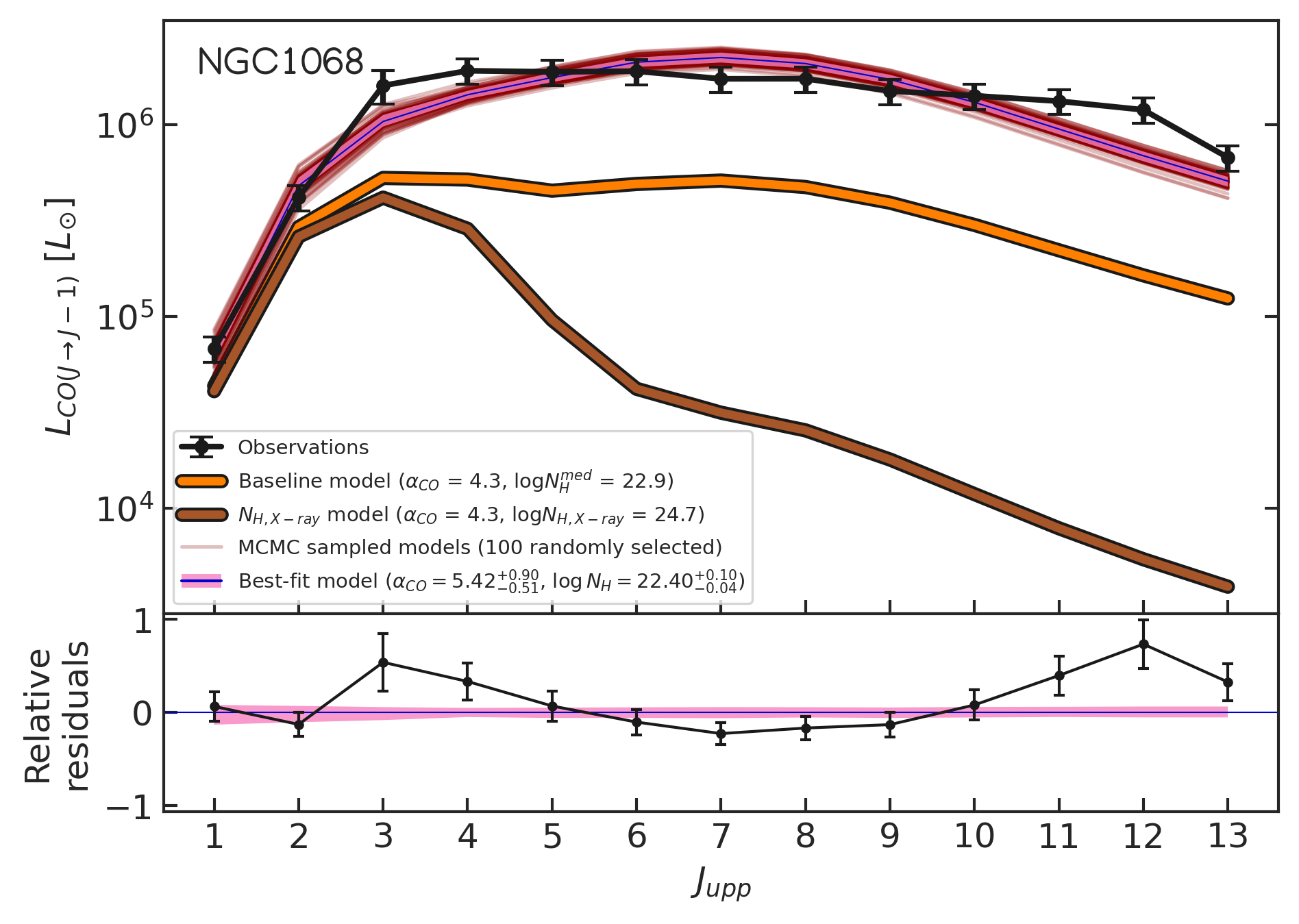}
    \caption{NGC 1068 (M77) CO SLEDs.}
    \label{fig:ngc1068}
\end{figure}

\begin{figure}
    \includegraphics[width=\columnwidth]{
    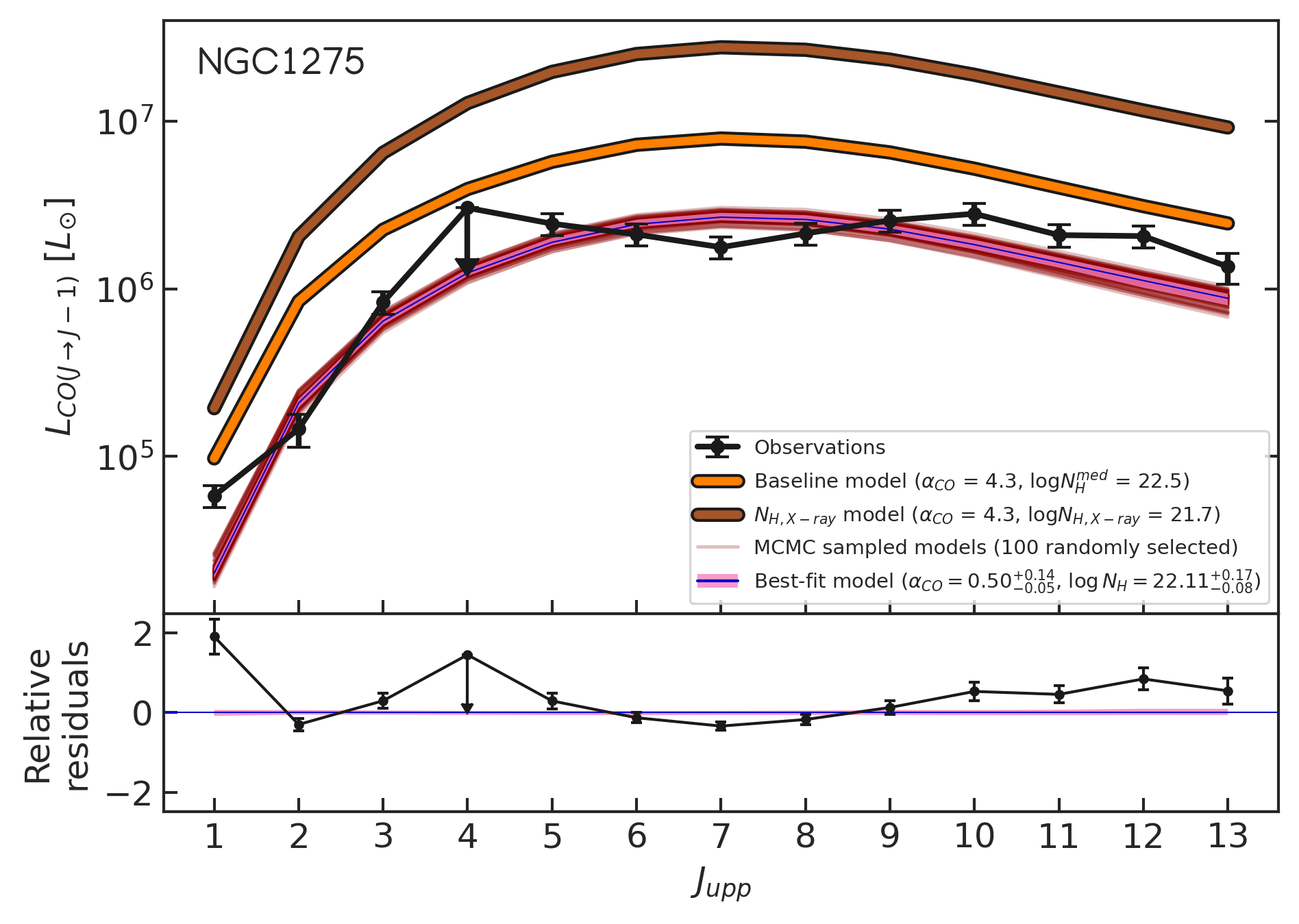}
    \caption{NGC 1275 (3C84, Perseus A) CO SLEDs.}
    \label{fig:ngc1275}
\end{figure}

\begin{figure}
    \includegraphics[width=\columnwidth]{
    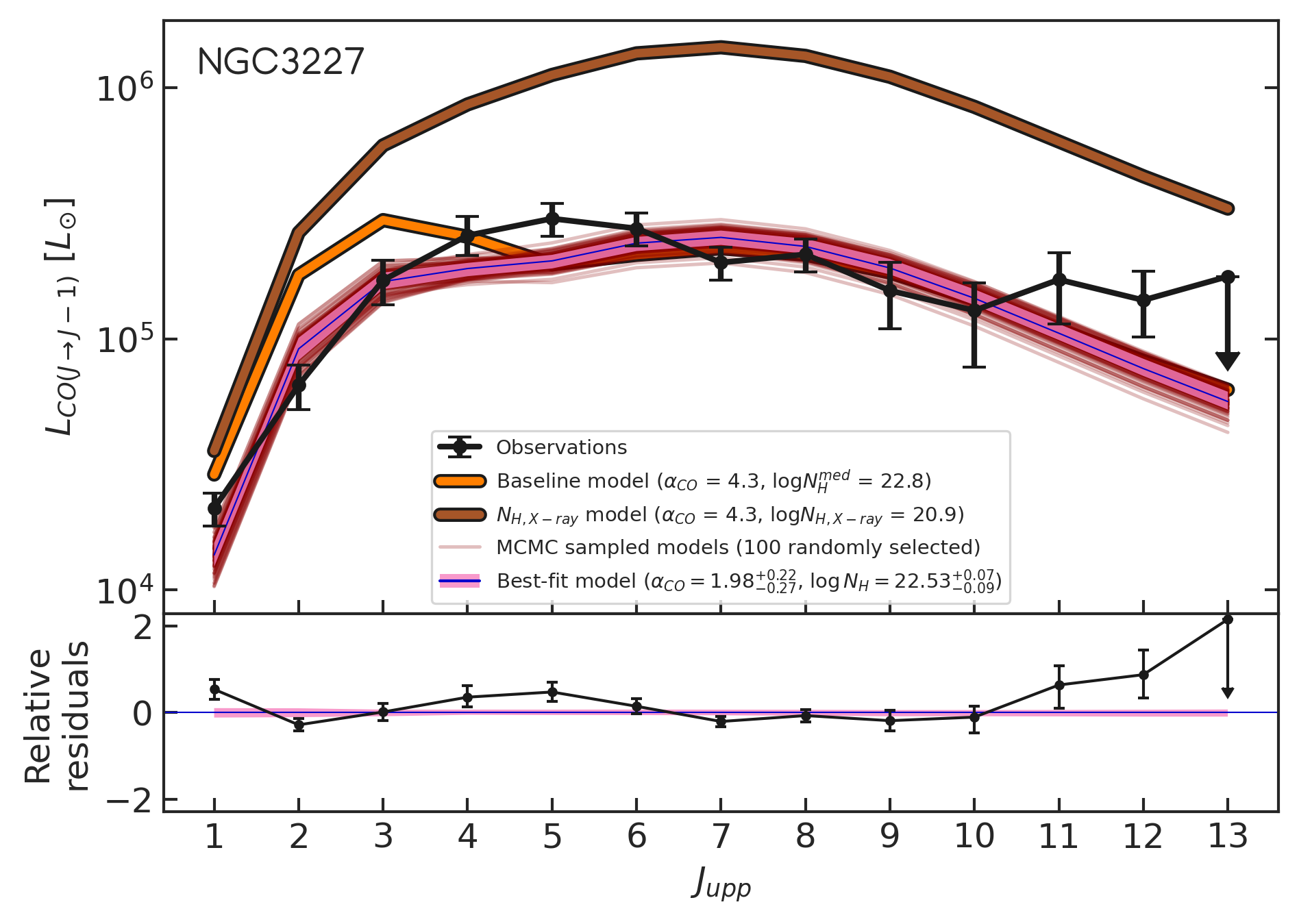}
    \caption{NGC 3227 CO SLEDs.}
    \label{fig:ngc3227}
\end{figure}

\begin{figure}
    \includegraphics[width=\columnwidth]{
    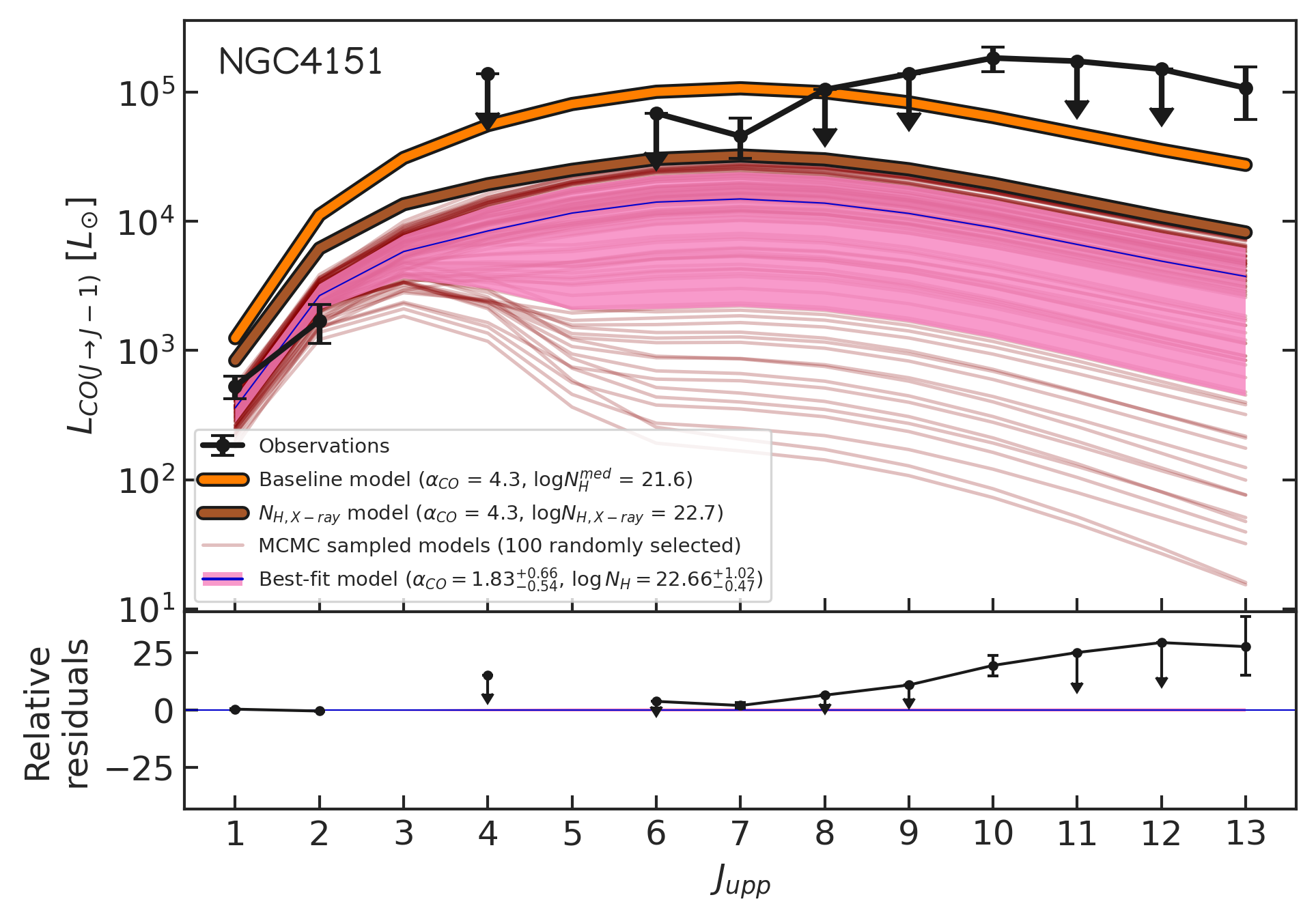}
    \caption{NGC 4151 CO SLEDs.}
    \label{fig:ngc4151}
\end{figure}

\begin{figure}
    \includegraphics[width=\columnwidth]{
    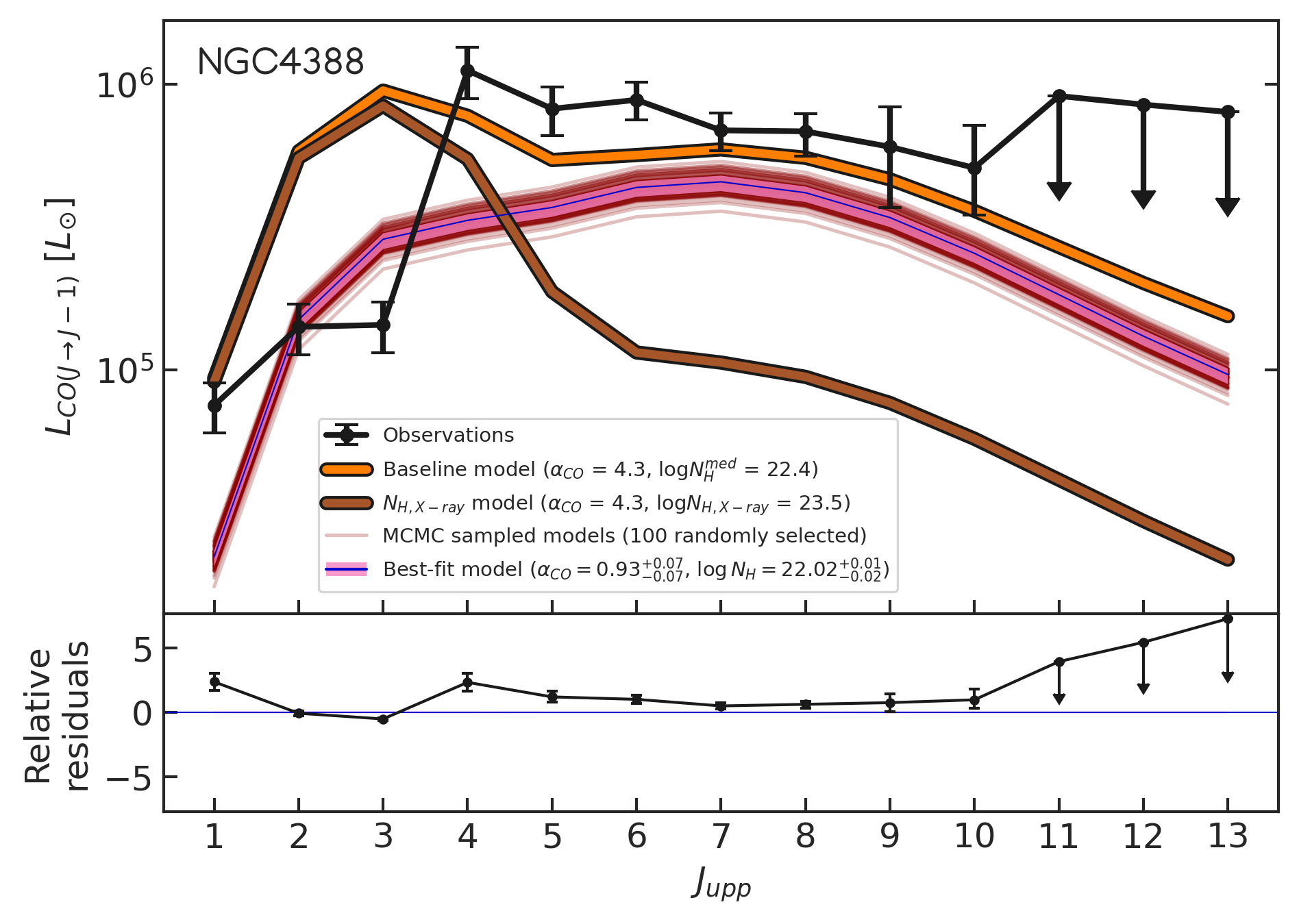}
    \caption{NGC 4388 CO SLEDs.}
    \label{fig:ngc4388}
\end{figure}

\begin{figure}
    \includegraphics[width=\columnwidth]{
    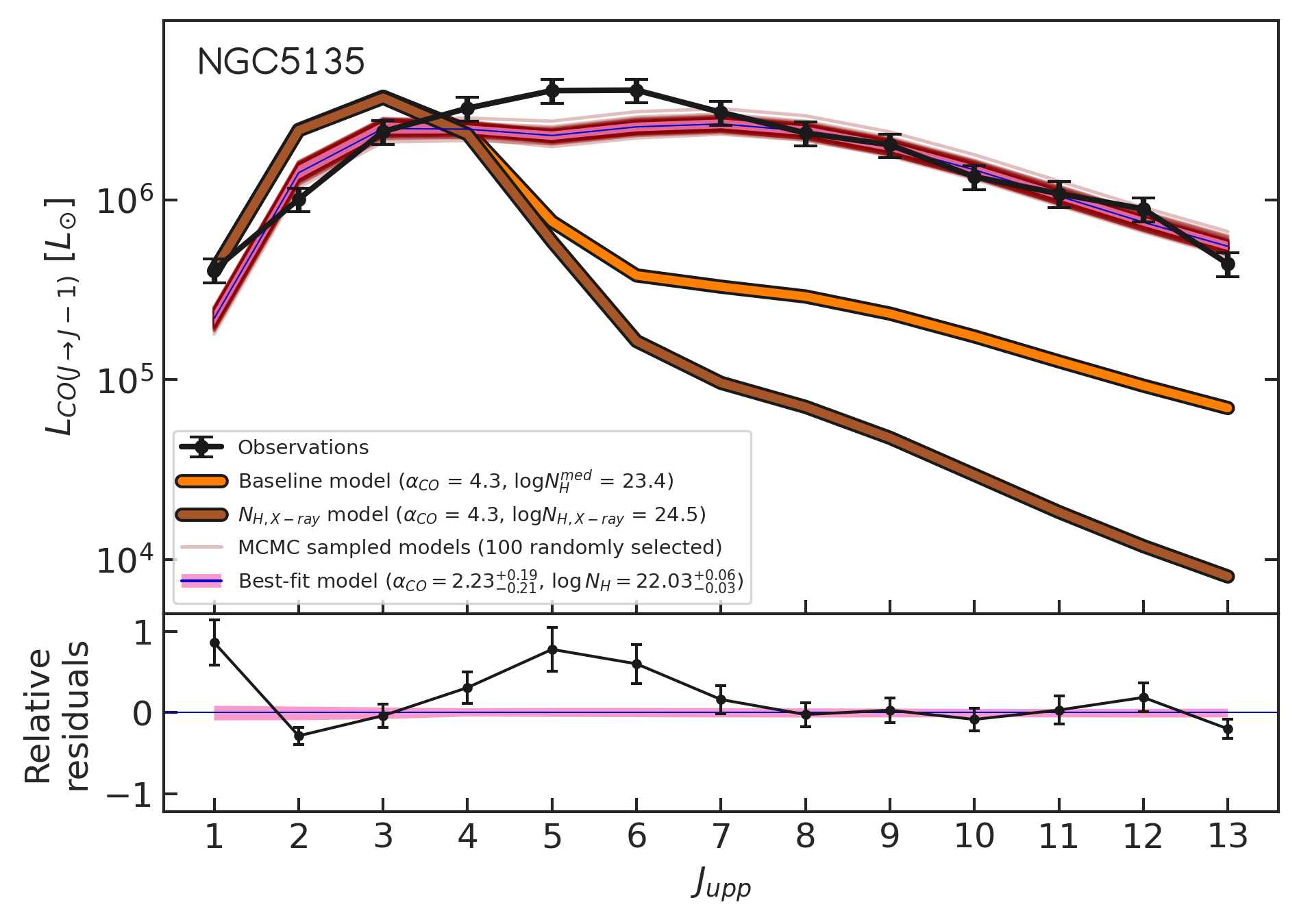}
    \caption{NGC 5135 CO SLEDs.}
    \label{fig:ngc5135}
\end{figure}

\begin{figure}
    \includegraphics[width=\columnwidth]{
    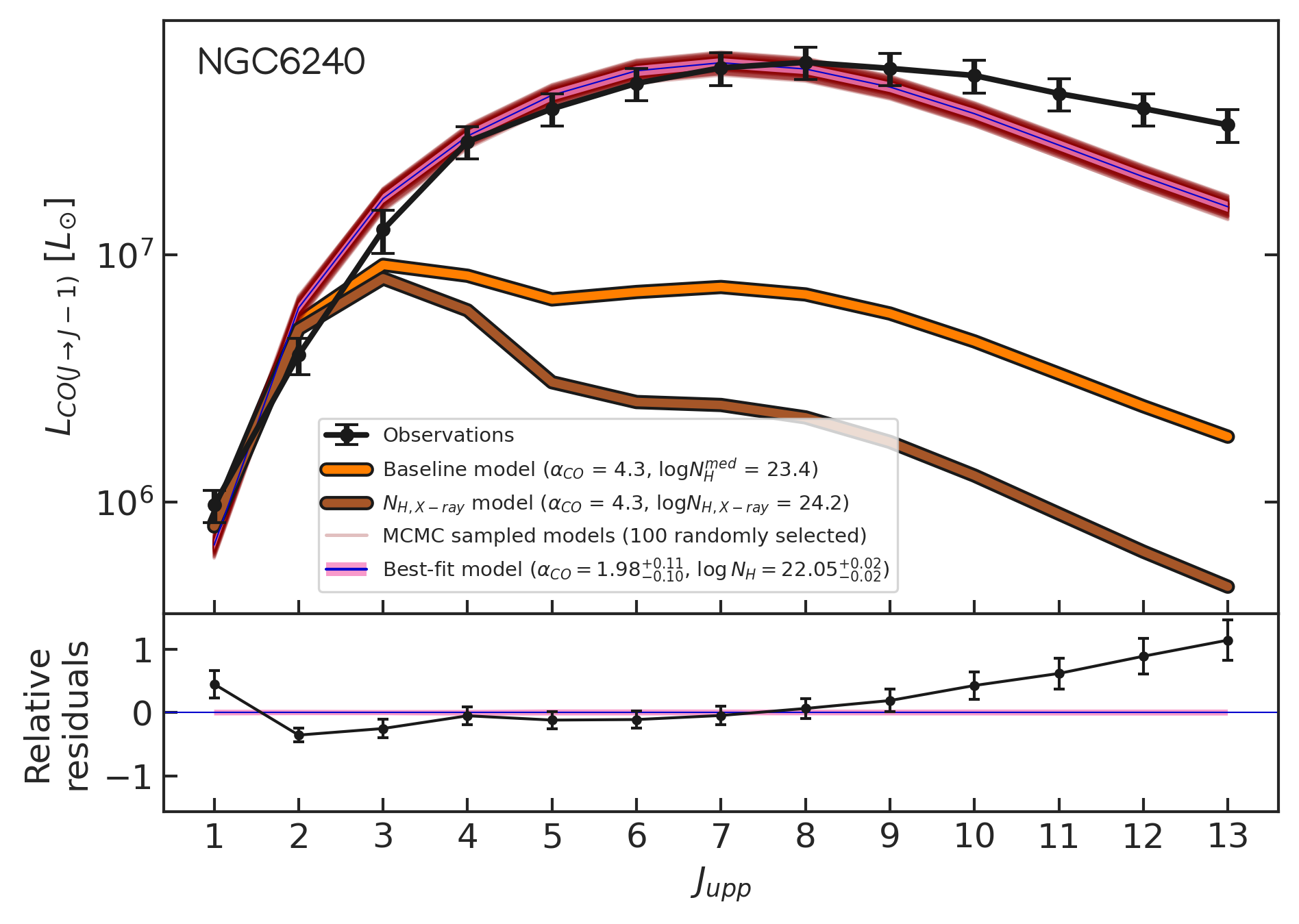}
    \caption{NGC 6240 CO SLEDs.}
    \label{fig:ngc6240}
\end{figure}

\begin{figure}
    \includegraphics[width=\columnwidth]{
    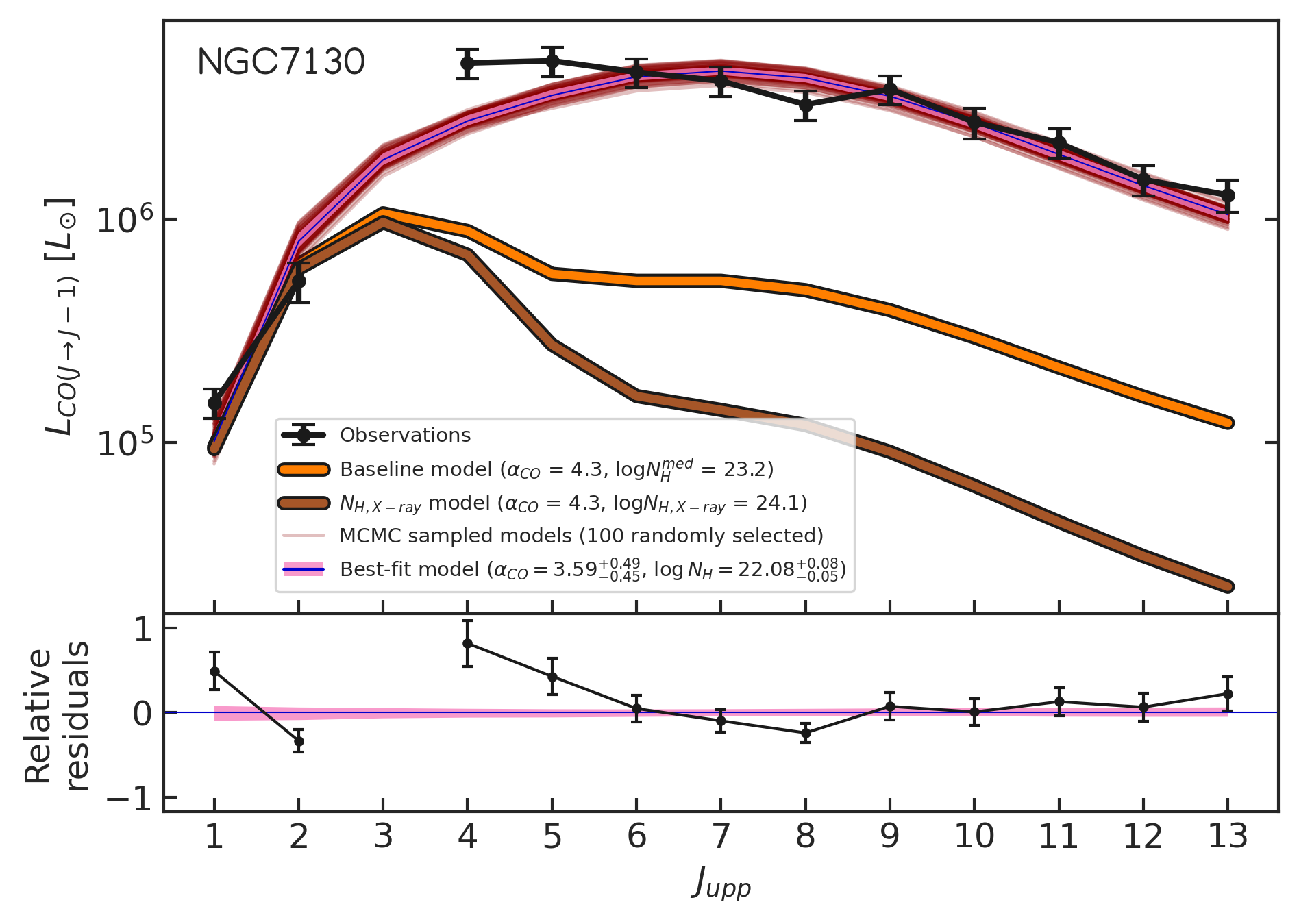}
    \caption{NGC 7130 CO SLEDs.}
    \label{fig:ngc7130}
\end{figure}

\begin{figure}
    \includegraphics[width=\columnwidth]{
    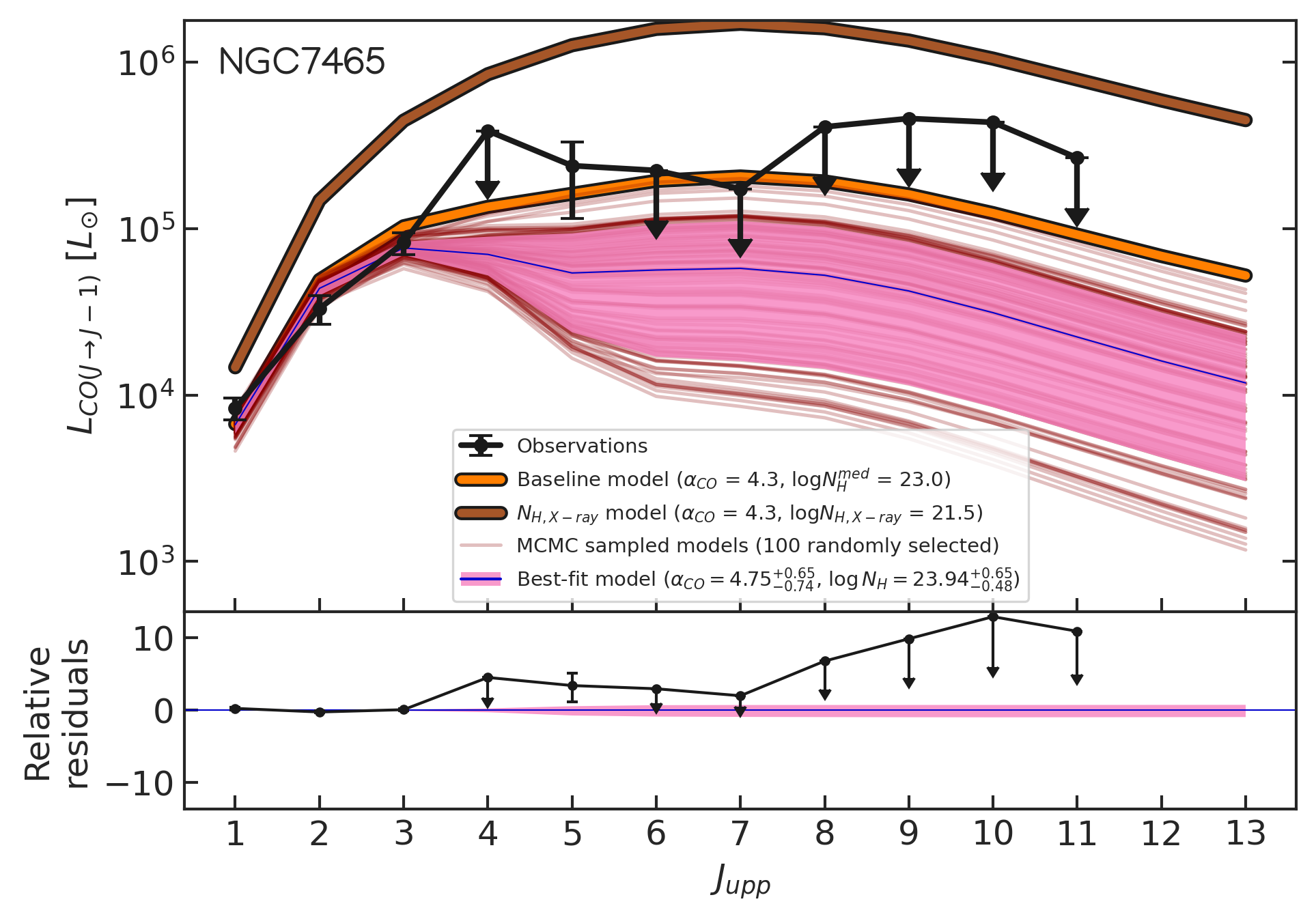}
    \caption{NGC 7465 CO SLEDs.}
    \label{fig:ngc7465}
\end{figure}

\begin{figure}
    \includegraphics[width=\columnwidth]{
    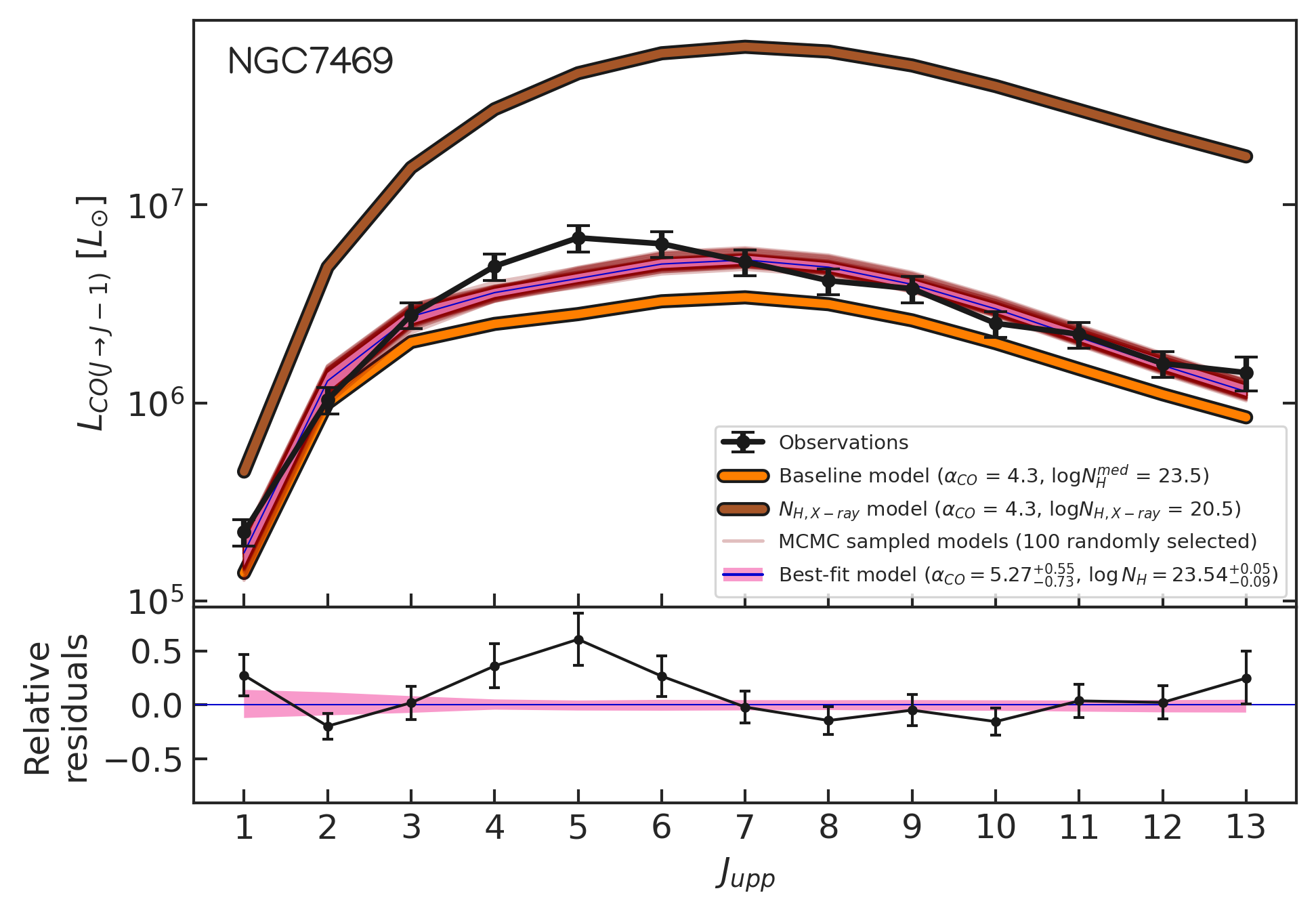}
    \caption{NGC 7469 CO SLEDs.}
    \label{fig:ngc7469}
\end{figure}

\begin{figure}
    \includegraphics[width=\columnwidth]{
    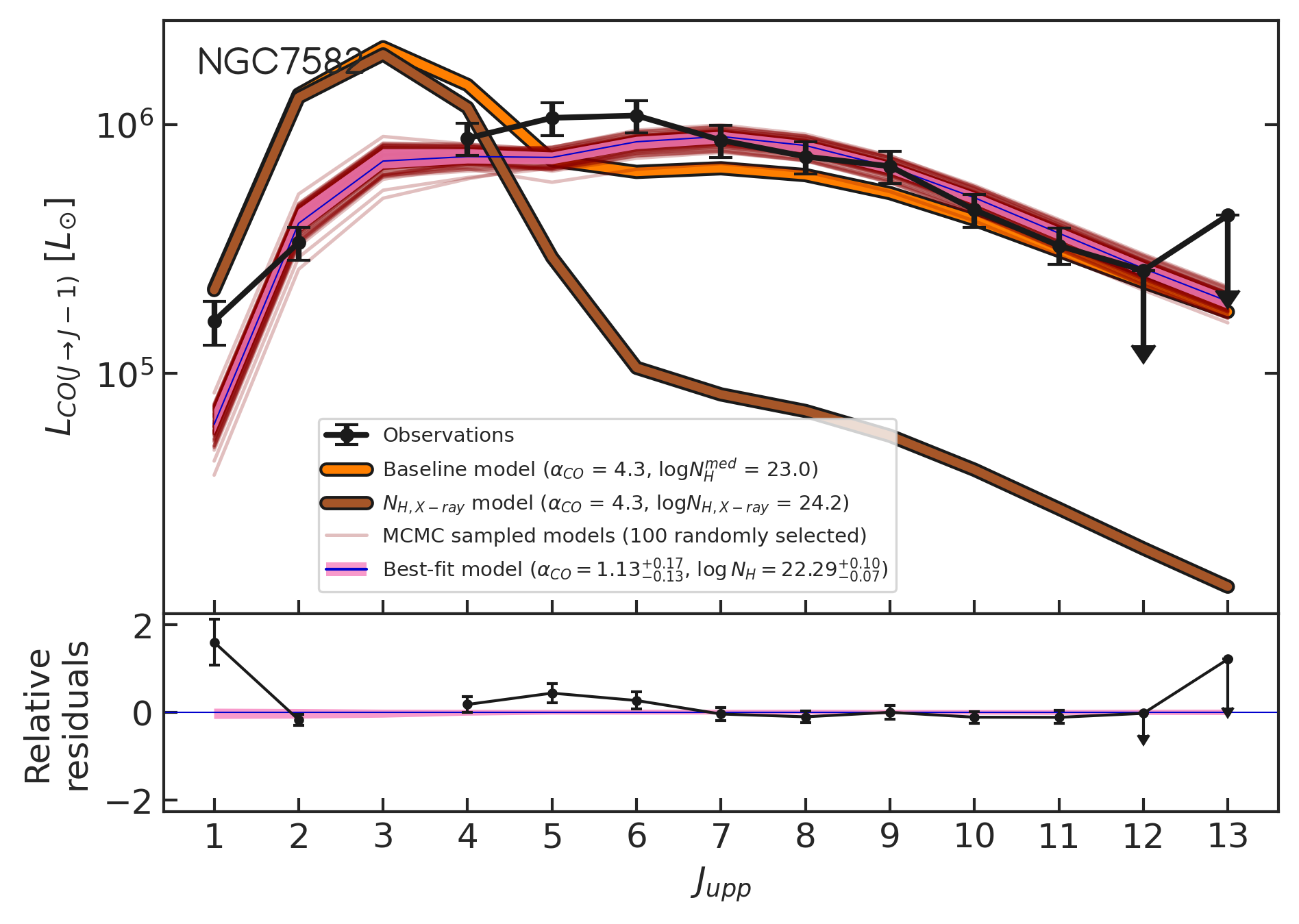}
    \caption{NGC 7582 CO SLEDs.}
    \label{fig:ngc7582}
\end{figure}

\begin{figure}
    \includegraphics[width=\columnwidth]{
    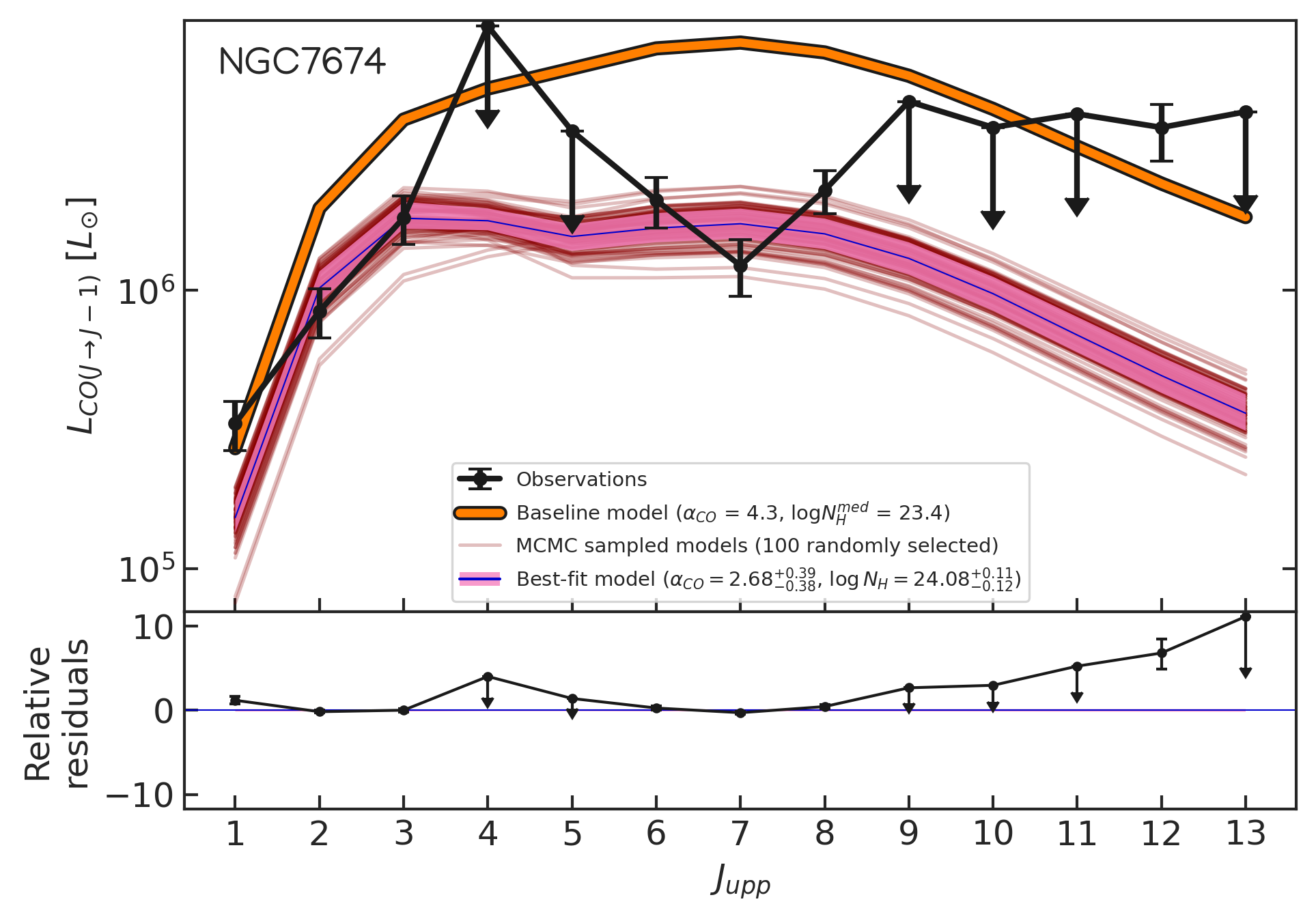}
    \caption{NGC 7674 (Mrk 533) CO SLEDs.
    }
    \label{fig:ngc7674}
\end{figure}

\begin{figure}
    \includegraphics[width=\columnwidth]{
    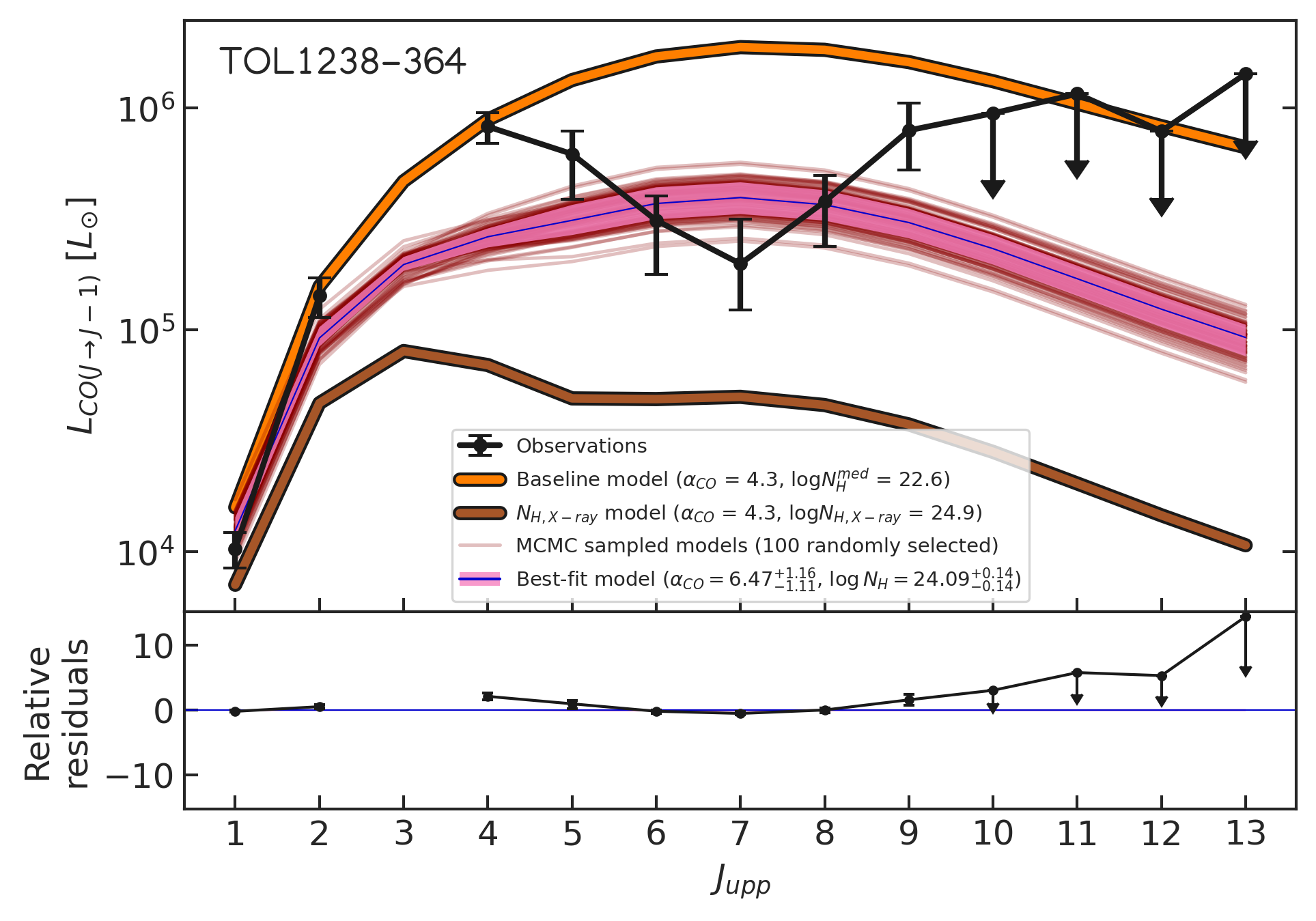}
    \caption{TOL 1238-364 (IC 3639) CO SLEDs.}
    \label{fig:tol1238}
\end{figure}

\begin{figure}
    \includegraphics[width=\columnwidth]{
    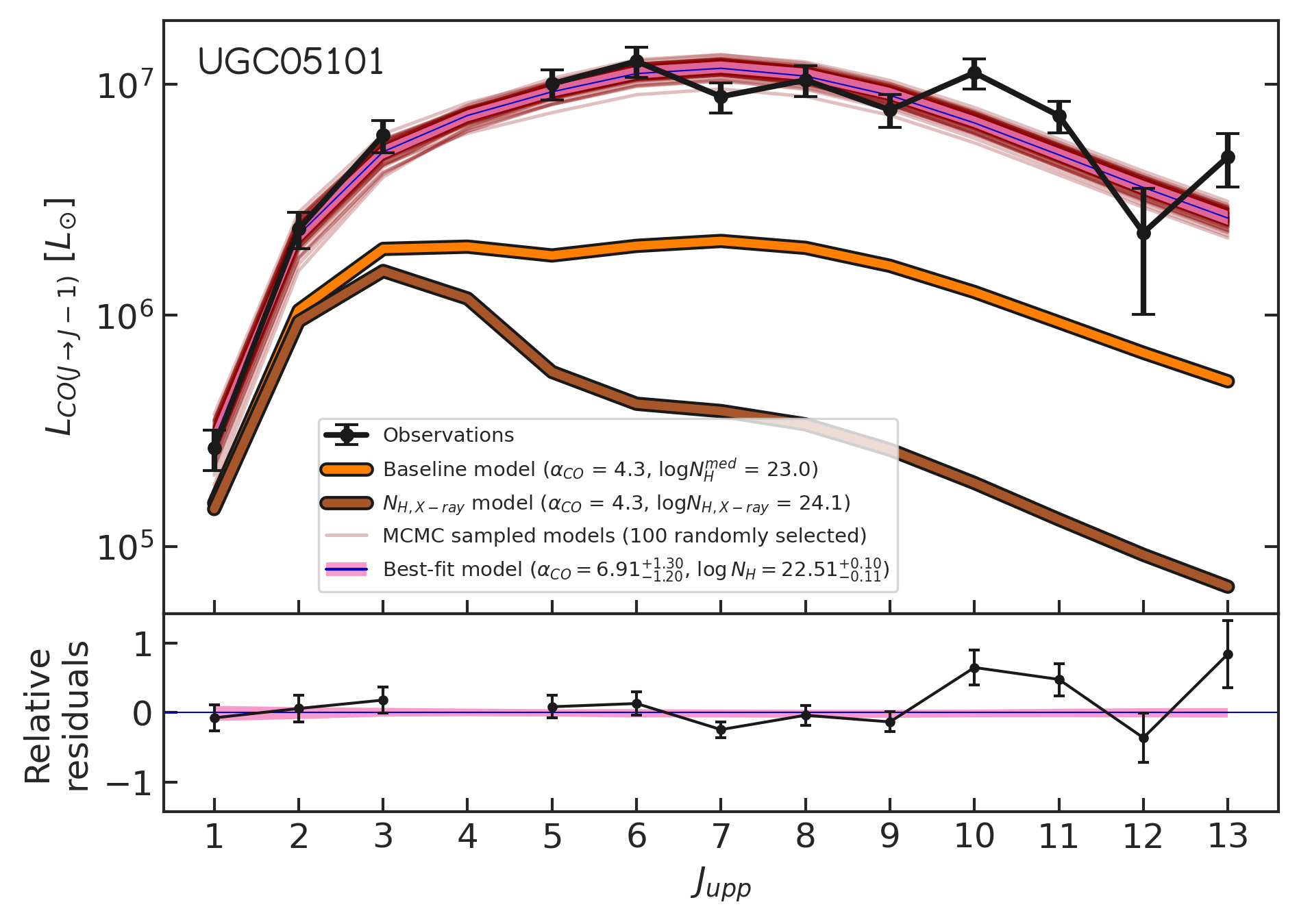}
    \caption{UGC 5101 CO SLEDs.}
    \label{fig:ugc5101}
\end{figure}

Figures~\ref{fig:izw1}~-~\ref{fig:ugc5101}
show the "Baseline model",
and "$N_{\text{H, X-ray}}$ model" for each galaxy
of the sample with orange and brown lines, respectively.
The observed CO SLED is plotted with a black line,
with vertical bars representing measurement errors,
and downward arrows indicating censored data points
(i.e. upper limits).
We randomly select,
from the posterior distributions of
the model parameters $\alpha_{\text{CO}}$
and $\log N_{\text{H}}$,
100 different modelled CO SLEDs, plotted
in red. The blue lines and pink shadings,
are the modelled CO SLEDs with the median
and $1\sigma$ spread of the model parameters
posterior distributions, respectively;
we call "Best-fit model" the one with
the median values of $\alpha_{\text{CO}}$
and $\log N_{\text{H}}$.
For each galaxy, the bottom panels of
Figures~\ref{fig:izw1}~-~\ref{fig:ugc5101}
show in black the relative residuals,
calculated as the difference between observations
and best fit model, divided by the latter;
the best-fit model is the blue line (fixed at 0),
with the $1\sigma$ spread in pink shading.

\newpage

\bsp	
\label{lastpage}
\end{document}